\documentclass[pra,aps,showpacs,amssymb,amsmath,amstext,amsfont,noshowkeys,groupedaddress,12pt]{revtex4}
\usepackage{graphicx}
\usepackage{accents}

\newcommand\zero[1]{\accentset{(0)}{#1}}
\newcommand\enn[1]{\accentset{(n)}{#1}}
\newcommand\starr[1]{\accentset{(\star)}{#1}}
\newcommand\emm[1]{\accentset{(\mathbf{m})}{#1}}
\newcommand\emmprime[1]{\accentset{(\mathbf{m}')}{#1}}

 \numberwithin{equation}{section}

\begin{document}
\title{Modified Semi-Classical Methods for Nonlinear Quantum Oscillations Problems}
\author{Vincent Moncrief}
\affiliation{Department of Physics and Department of Mathematics, \\ Yale University, P.O. Box 208120, New Haven, CT 06520, USA. \\ E-mail address: vincent.moncrief@yale.edu}
\author{Antonella Marini}
\affiliation{Department of Mathematics, \\ Yeshiva University, 500 West 185th Street, New York, NY 10033, USA. \\ and \\ Department of Mathematics, \\ University of L'Aquila, Via Vetoio, 67010 L'Aquila, AQ ITALY. \\ E-mail address: marini@yu.edu}
\author{Rachel Maitra}
\affiliation{Department of Physics, \\ Albion College, 611 E. Porter Street, Albion, MI 49224, USA. \\ E-mail address: rmaitra@albion.edu}
\date{\today}
\begin{abstract}
We develop a modified semi-classical approach to the approximate solution of Schr\"{o}dinger's equation for certain nonlinear quantum oscillations problems. In our approach, at lowest order, the Hamilton-Jacobi equation of the conventional semi-classical formalism is replaced by an inverted-potential-vanishing-energy variant thereof. With suitable smoothness, convexity and coercivity properties imposed on its potential energy function, we prove, using methods drawn from the calculus of variations together with the (Banach space) implicit function theorem, the existence of a global, smooth `fundamental solution' to this equation. Higher order quantum corrections thereto, for both ground and excited states, can then be computed through the integration of associated systems of linear transport equations, derived from Schr\"{o}dinger's equation, and formal expansions for the corresponding energy eigenvalues obtained therefrom by imposing the natural demand for smoothness on the (successively computed) quantum corrections to the eigenfunctions. For the special case of linear oscillators our expansions naturally truncate, reproducing the well-known exact solutions for the energy eigenfunctions and eigenvalues.

As an explicit application of our methods to computable nonlinear problems, we calculate a number of terms in the corresponding expansions for the one-dimensional anharmonic oscillators of quartic, sectic, octic, and dectic types and compare the results obtained with those of conventional Rayleigh/Schr\"{o}dinger perturbation theory. To the orders considered (and, conjecturally, to all orders) our eigenvalue expansions agree with those of Rayleigh/Schr\"{o}dinger theory whereas our wave functions more accurately capture the more-rapid-than-gaussian decay known to hold for the exact solutions to these problems. For the quartic oscillator in particular our results strongly suggest that both the ground state energy eigenvalue expansion and its associated wave function expansion are Borel summable to yield natural candidates for the actual exact ground state solution and its energy.

Our techniques for proving the existence of the crucial `fundamental solution' to the relevant (inverted-potential-vanishing energy) Hamilton-Jacobi equation have the important property of admitting interesting infinite dimensional generalizations. In a project paralleling the present one we shall show how this basic construction can be carried out for the Yang-Mills equations in Minkowski spacetime.
\end{abstract}
\pacs{02.30.Mv, 02.30.Xx, 03.65.Sq}
\maketitle

\section{Introduction}
\label{sec:introduction}

In this article we develop a modified semi-classical approach to the approximate solution of certain nonlinear quantum oscillations problems. Quantum systems of nonlinear oscillators have of course long been studied, from the semi-classical viewpoint as well many others, in the mathematical physics literature. Our aim here however is not simply to provide alternative proofs of already known results but instead to develop mathematical methods that can ultimately be applied to certain systems of quantized fields. We are pursuing that development in parallel to the present one and, in a companion paper, will show how some of the fundamental constructions of this article can, in fact, also be realized for the much more technically demanding case of nonabelian gauge fields \cite{Marini}. Since the study of finite dimensional systems, however, allows one to sidestep the intricate complications of regularization and renormalization, one can push the analysis of these to a deeper level than is readily attainable for field theoretic problems. For this reason we have been motivated to carry out the present, parallel study of quantum mechanical systems from the analogous, modified semi-classical viewpoint.

To some extent our work on these finite dimensional systems overlaps that already extensively developed in the microlocal analysis literature \cite{Dimassi1999} but it also differs from this work in fundamental ways that seem crucial for our ultimate, intended application of these methods to quantum field theory. But even at the finite dimensional level our approach offers some tangible advantages over that previously developed by unifying and globalizing several of the fundamental steps in the analysis and by doing so in a way that, in effect, lays the foundation for its eventual extension to infinite dimensional systems.

Finite systems of quantum oscillators are, of course, often considered as rough approximations to quantized fields (through, for example, Hamiltonian lattice discretization) and an important class of these is characterized by Schr\"{o}dinger operators of the form
\begin{equation}\label{eq:101}
\hat{H} = -\frac{\hbar^{2}}{2m} {}^{(n)}\!\Delta + \frac{1}{2}m \sum_{i=1}^{n} \omega_{i}^{2} (x^{i})^{2} + A (x^{1}, \ldots , x^{n}).
\end{equation}
Here \({}^{(n)}\!\Delta\) is the Laplacian on \(\mathbb{R}^{n}\), \({}^{(n)}\!\Delta = \sum_{i=1}^{n} \frac{\partial^{2}}{\partial x^{i2}}\), the `frequencies' \(\left\lbrace\omega_{i} | i \in [1, \ldots ,n]\right\rbrace\) are real and strictly positive and the potential function \(A : \mathbb{R}^{n} \rightarrow \mathbb{R}\) is smooth and incorporates the nonlinearities in the sense that its Taylor expansion about the origin in \(\mathbb{R}^{n}\) is required to begin at third order, i.e., to satisfy
\begin{equation}\label{eq:102}
\begin{split}
A(0, \ldots ,0)  &= \frac{\partial A(0, \ldots ,0)}{\partial x^{i}} = \frac{\partial^{2} A(0, \ldots ,0)}{\partial x^{i}\partial x^{j}}\\
&= 0\quad \forall\; i,j \in [1, \ldots ,n].
\end{split}
\end{equation}
We shall further require that the total potential energy function \(V : \mathbb{R}^{n} \rightarrow \mathbb{R}\) defined by
\begin{equation}\label{eq:103}
V(x^{1}, \ldots , x^{n}) := \frac{1}{2}m \sum_{i=1}^{n} \omega_{i}^{2} (x^{i})^{2} + A(x^{1}, \ldots , x^{n})
\end{equation}
satisfy \(V \geq 0\) on \(\mathbb{R}^{n}\) and have a unique, critical point corresponding to a vanishing global minimum at the origin so that
\begin{equation}\label{eq:104}
\begin{split}
V(\mathbf{x}) &:= V(x^{1}, \ldots ,x^{n}) > V(0, \ldots ,0) = 0\\
\forall\; \mathbf{x} &= (x^{1}, \ldots , x^{n}) \in \mathbb{R}^{n} \backslash (0, \ldots ,0).
\end{split}
\end{equation}
Later we shall impose a certain \textit{convexity} condition on \(V\) to insure the uniqueness and smoothness of our basic constructions and, in the event that \(A(\mathbf{x})\) has indefinite sign, a certain \textit{coercivity} condition bounding its behavior from below. Finally we shall require that the frequencies \(\lbrace\omega_{i}\rbrace\), characterizing the quadratic term in \(V(\mathbf{x})\) satisfy a convenient (but inessential) `non-resonance' condition that will simplify the analysis of quantum excited states.

Our approach begins by seeking a ground state wave function of the form
\begin{equation}\label{eq:105}
\zero{\psi}_{\hbar}(\mathbf{x}) = N_{\hbar} e^{-S_{\hbar}(\mathbf{x})/\hbar}
\end{equation}
where \(N_{\hbar}\) is a normalization constant and in which \(S_{\hbar}(\mathbf{x})\) is real-valued and admits a formal power series expansion in \(\hbar\) which we write as  \begin{equation}\label{eq:106}
S_{\hbar}(\mathbf{x}) \simeq S_{(0)}(\mathbf{x}) + \hbar S_{(1)}(\mathbf{x}) + \frac{\hbar^{2}}{2!} S_{(2)}(\mathbf{x}) + \dots + \frac{\hbar^{n}}{n!} S_{(n)}(\mathbf{x}) + \dots
\end{equation}
We expand the corresponding ground state energy eigenvalue \(\zero{E}_{\hbar}\) in the analogous way, writing
\begin{equation}\label{eq:107}
\zero{E}_{\hbar} \simeq \hbar \left(\zero{\mathcal{E}}_{(0)} + \hbar \zero{\mathcal{E}}_{(1)} + \frac{\hbar^{2}}{2!} \zero{\mathcal{E}}_{(2)} + \dots + \frac{\hbar^{n}}{n!} \zero{\mathcal{E}}_{(n)} + \dots\right)
\end{equation}
and substitute these ans\"{a}tze into the time-independent Schr\"{o}dinger equation
\begin{equation}\label{eq:108}
\hat{H} \zero{\psi}_{\hbar} = \zero{E}_{\hbar} \zero{\psi}_{\hbar}
\end{equation}
requiring the latter to hold, order by order, in powers of Planck's constant.

At leading order our formulation immediately generates the `inverted-potential-zero-energy' (or `ipze' for brevity) Hamilton-Jacobi equation,
\begin{equation}\label{eq:109}
\frac{1}{2m} \nabla S_{(0)} \cdot \nabla S_{(0)} - V = 0,
\end{equation}
for the function \(S_{(0)}(\mathbf{x})\). Under the convexity and coercivity hypotheses alluded to above (and given precisely by inequalities (\ref{eq:320}) and (\ref{eq:322}) below) we shall prove in Sect.~(\ref{sec:existence-solution}), using methods drawn from the calculus of variations, the existence and smoothness of a globally defined `fundamental' solution to Eq.~(\ref{eq:109}). The higher order `quantum corrections' to \(S_{(0)}(\mathbf{x})\) (i.e., the functions \(S_{(k)}(\mathbf{x})\) for \(k = 1, 2, \dots\)) can then be computed through the integration of a set of `transport equations' for these quantities along the integral curves of the gradient (semi-) flow generated by \(S_{(0)}(\mathbf{x})\). The natural demand for smoothness of these quantum corrections will force the (heretofore, undetermined) energy coefficients \(\lbrace\zero{\mathcal{E}}_{(0)}, \zero{\mathcal{E}}_{(1)}, \zero{\mathcal{E}}_{(2)}, \dots\rbrace\) all to take on specific, computable values.

Excited states will then be studied by substituting the ansatz
\begin{equation}\label{eq:110}
\starr{\psi}_{\hbar}(\mathbf{x}) = \starr{\phi}_{\hbar}(\mathbf{x}) e^{-S_{\hbar}(\mathbf{x})/\hbar}
\end{equation}
into the time-independent Schr\"{o}dinger equation
\begin{equation}\label{eq:111}
\hat{H} \starr{\psi}_{\hbar} = \starr{E}_{\hbar} \starr{\psi}_{\hbar}
\end{equation}
and formally expanding the wave function \(\starr{\phi}_{\hbar}\) and energy eigenvalue \(\starr{E}_{\hbar}\) in powers of \(\hbar\) as before
\begin{align}
\starr{\phi}_{\hbar} &\simeq \starr{\phi}_{(0)} + \hbar\starr{\phi}_{(1)} + \frac{\hbar^{2}}{2!} \starr{\phi}_{(2)} + \dots ,\label{eq:112}\\
\starr{E}_{\hbar} &\simeq \hbar\starr{\mathcal{E}}_{\hbar} = \hbar \left(\starr{\mathcal{E}}_{(0)} +  \hbar\starr{\mathcal{E}}_{(1)} + \frac{\hbar^{2}}{2!}\starr{\mathcal{E}}_{(2)} + \dots\right)\label{eq:113}
\end{align}
while retaining the same `universal' factor \(e^{-S_{\hbar}(\mathbf{x})/\hbar}\) determined by the ground state calculations.

It will prove to be convenient to reexpress the equations for an excited state in terms of the energy `gap' \(\Delta\starr{E}_{\hbar} = \hbar\Delta\starr{\mathcal{E}}_{\hbar}\) and its expansion coefficients defined via
\begin{equation}\label{eq:114}
\begin{split}
\Delta\starr{\mathcal{E}}_{\hbar} &:= \starr{\mathcal{E}}_{\hbar} - \zero{\mathcal{E}}_{\hbar}\\
 &= \left(\Delta\starr{\mathcal{E}}_{(0)} + \hbar \Delta\starr{\mathcal{E}}_{(1)} + \frac{\hbar^{2}}{2!} \Delta\starr{\mathcal{E}}_{(2)} + \dots + \frac{\hbar^{n}}{n!} \Delta\starr{\mathcal{E}}_{(n)} + \dots\right)
\end{split}
\end{equation}
where \(\Delta\starr{\mathcal{E}}_{(i)} := \starr{\mathcal{E}}_{(i)} - \zero{\mathcal{E}}_{(i)}\). Under a convenient, but inessential `non resonance' (or, more properly, `non-degeneracy') condition upon the frequencies \(\lbrace\omega_{i}\rbrace\) (defined precisely in (\ref{eq:244}) below) we shall show, by a sequence of arguments given in Sections (\ref{subsec:excited}) and (\ref{subsec:transport-excited-state}), that the globally smooth solutions for \(\starr{\phi}_{(0)}\) can each be naturally characterized by a collection of non-negative integers \(\mathbf{m} := (m_{1}, \ldots , m_{n})\) and, threafter refine the notation by replacing \(\starr{\phi}_{(0)}\) with \(\emm{\phi}_{(0)}\). The corresponding, lowest order energy gap coefficient will prove to be
\begin{equation}\label{eq:115}
\Delta \emm{\mathcal{E}}_{(0)} = \sum_{i=1}^{n} m_{i}\omega_{i}
\end{equation}
so that \(\Delta \emm{E}_{(0)} = \hbar \Delta \emm{\mathcal{E}}_{(0)} = \sum_{i=1}^{n} m_{i}\hbar\omega_{i}\) will coincide with the energy gap of a collection of pure, harmonic oscillators, the \(i\)-th one excited to its \(m_{i}\)-th energy level.

Higher order corrections, \(\left\lbrace\emm{\phi}_{(i)} | i = 1, 2, \dots\right\rbrace\), to the excited state wave functions can then be sequentially computed through the systematic integration of an appropriate set of linear transport equations derived from Schr\"{o}dinger's equation. The corresponding, higher order energy gap coefficients, \(\left\lbrace\Delta\emm{\mathcal{E}}_{(i)} | i = 1, 2, \dots\right\rbrace\), will each be uniquely determined by the natural demand for regularity of the functions \(\left\lbrace\emm{\phi}_{(i)}\right\rbrace\). A subtlety  of this excited state analysis is that one must first develop formal power series expansions for the solutions to the relevant transport equations and then smoothly modify these to generate the actual, globally defined smooth functions \(\left\lbrace\emm{\phi}_{(i)}(\mathbf{x}) | i = 1, 2, \dots\right\rbrace\). The formal expansions however will already suffice to determine the energy gap coefficients \(\left\lbrace\Delta\emm{\mathcal{E}}_{(i)} | i = 1, 2, \dots\right\rbrace\) which will remain unaffected by the subsequent smoothing operations needed to complete the construction of the wave functions.

It will become clear from our detailed analysis that all of the energy coefficients (for both ground and excited states) are uniquely determined from the Taylor expansion coefficients (about the origin in \(\mathbb{R}^{n}\)) of the potential energy function \(V(\mathbf{x})\). Thus for \textit{non-analytic} potential functions (many of which, of course, share the same formal Taylor expansion) the corresponding energy coefficients could not be expected to have better than an asymptotic validity as \(\hbar \rightarrow 0\). In this limit the wave functions become more and more sharply peaked about the origin and the detailed behavior of the potential energy away from the origin becomes accordingly less and less relevant to the determination of the energy spectrum. For analytic potentials on the other hand (which are of course uniquely determined by their convergent Taylor expansions) one expects more favorable behavior and we shall review, at the end of Sect.~(\ref{sect:integration-transport}), the precise sense in which formal expansions, of the types given above in Eqs.~(\ref{eq:106}--\ref{eq:107}) for the ground states and Eqs.~(\ref{eq:112}--\ref{eq:114}) for excited states, provide actual asymptotic approximations to solutions of  Schr\"{o}dinger's equation. The sharper error estimates derivable for analytic potentials will validate quantitatively the intuition sketched above that such methods should indeed be more favorable in the analytic case.

In Sect.~(\ref{sec:computable-examples}) we shall turn from general arguments to explicit calculations by applying our ideas concretely to the one-dimensional examples provided by quartic, sectic, octic and dectic anharmonic oscillators. For each of these models we have carried out the calculation of the ground state wave function and its energy eigenvalue to order \(\hbar^{25}\) and compared the results to those of conventional Rayleigh/Schr\"{o}dinger perturbation theory. In each case studied (and conjecturally, much more generally) we find that our eigenvalue expansions agree with the conventional ones and we conjecture that this agreement extends to all orders. For the oscillators considered it is well-known that the usual perturbative expansions for the ground state eigenvalues diverge but are nevertheless Borel summable to yield the \textit{exact} ground state energies [3,4]. If the conjectured agreement of our expansions with the conventional ones proves valid to all orders then of course the Borel summability conclusions would apply to our results as well.

For the case of the quartic oscillator in particular we present some rather striking evidence that the formal series (\ref{eq:106}) determining the ground state wave function (\ref{eq:105}) may itself be uniformly Borel summable and thus to yield a  natural candidate for the \textit{exact} ground state solution. For the quartic oscillator we also consider excited states, for an arbitrary excitation level, but only up through order \(\hbar^{3}\). Here too the eigenvalue expansions agree precisely with those of conventional perturbation theory and again we conjecture that this agreement should persist to all orders. For both ground and excited states, even though our eigenvalue expansions agree (to the order computed at least) with the conventional ones, our wave functions capture, even at lowest order, the more-rapid-than-gaussian decay known to be valid for such anharmonic oscillators. By contrast the conventional theory which approximates these solutions by expansions in harmonic oscillator wave functions, could only hope to recapture this rapid decay in the limit that these series are somehow fully summed. On the other hand we suggest a procedure whereby our wave functions could be expanded (in the oscillator coupling constant) and truncated in such a way as to (conjecturally) reproduce the approximate wave functions of the conventional theory.

To describe the ways in which our methods differ fundamentally from those of the microlocal analysis literature will proceed more easily after we have presented our results in detail. We therefore postpone this comparison until the concluding section wherein we also discuss the (closely related) program for extending our ideas to the infinite dimensional setting of quantum field theory.

\section{Finite Dimensional Oscillator Systems}
\label{sec:finite}

\subsection{Ground State Preliminaries}
\label{subsec:ground}

Substituting the ansatz (1.5) into the time-independent Schr\"{o}dinger equation for the ground state (1.8), expanding \(S_{\hbar}\) and \(\zero{E}_{\hbar}\) as in (1.6) and (1.7) respectively and requiring that the resulting formula hold order by order in \(\hbar\) leads to the following sequence of equations for the unknowns \(\left\lbrace S_{(0)}, S_{(1)}, S_{(2)}, \dotsc , \zero{\mathcal{E}}_{(0)}, \zero{\mathcal{E}}_{(1)}, \zero{\mathcal{E}}_{(2)}\right\rbrace\):
\begin{subequations}\label{eq:201}
\begin{align}
\frac{1}{2m} \nabla S_{(0)} \cdot \nabla S_{(0)} - V &= 0,\label{eq:201a}\\
-\frac{1}{m} \nabla S_{(0)} \cdot \nabla S_{(1)} + \frac{1}{2m} {}^{(n)}\!\Delta S_{(0)} &= \zero{\mathcal{E}}_{(0)},\label{eq:201b}\\
-\frac{1}{m} \nabla S_{(0)} \cdot \nabla S_{(2)} - \frac{1}{m} \nabla S_{(1)} \cdot \nabla S_{(1)} + \frac{1}{m} {}^{(n)}\!\Delta S_{(1)} &= 2 \zero{\mathcal{E}}_{(1)},\label{eq:201c}\\
-\frac{1}{m} \nabla S_{(0)} \cdot \nabla S_{(3)} - \frac{3}{m} \nabla S_{(1)} \cdot \nabla S_{(2)} + \frac{3}{2m} {}^{(n)}\!\Delta S_{(2)} &= 3 \zero{\mathcal{E}}_{(2)},\label{eq:201d}
\end{align}
\end{subequations}
and, for arbitrary \(k \geq 2\)
\begin{equation}\label{eq:202}
-\frac{1}{m} \nabla S_{(0)} \cdot \nabla S_{(k)} - \frac{1}{2m} \sum_{j=1}^{k-1} \frac{k!}{j! (k-j)!} \nabla S_{(j)} \cdot \nabla S_{(k-j)} + \frac{k}{2m} {}^{(n)}\!\Delta S_{(k-1)} = k \zero{\mathcal{E}}_{(k-1)}.
\end{equation}
These consist of the `ipze' Hamilton-Jacobi equation for \(S_{(0)}\) previously defined (\ref{eq:201a}) and a sequence of `transport' equations (\ref{eq:201b}--\ref{eq:202}) for the higher order quantum corrections \(\left\lbrace S_{(1)}, S_{(2)}, \dotsc , \zero{\mathcal{E}}_{(0)}, \zero{\mathcal{E}}_{(1)}, \dots\right\rbrace\).

As already mentioned, our strategy for solving this system will be first to show that Eq.~(\ref{eq:201a}) has a canonical, smooth, globally defined `fundamental' solution \(S_{(0)}\). Equipped with this solution we shall then show that the subsequent transport equations can, sequentially, be integrated to generate the quantum corrections \(\left\lbrace S_{(1)}, S_{(2)}, \dotsc , \right\rbrace\) to \(S_{(0)}\) and that the natural demand for global regularity of these functions forces the (heretofore, unknown) energy coefficients \(\left\lbrace\zero{\mathcal{E}}_{(0)}, \zero{\mathcal{E}}_{(1)}, \zero{\mathcal{E}}_{(2)}, \dotsc\right\rbrace\) all to take on specific, computable values.

Since \(S_{(0)}\) will play a central role in all of our constructions (including those for the excited states) let us first describe, intuitively, how we propose to generate it. As already noted, equation (\ref{eq:201a}) is the zero-energy Hamilton-Jacobi equation for the associated (inverted potential) mechanics problem whose Lagrangian \(L_{ip}\) and (conserved) energy \(E_{ip}\) are given by
\begin{equation}\label{eq:203}
\begin{split}
L_{ip}& (x^{1}, \dotsc  , x^{n}, \dot{x}^{1}, \dotsc , \dot{x}^{n})\\
&:= \frac{m}{2} \sum_{i=1}^{n} (\dot{x}^{i})^{2} - V_{ip} (x^{1}, \dotsc , x^{n})\\
&= \frac{m}{2} \sum_{i=1}^{n} (\dot{x}^{i})^{2} + V (x^{1}, \dotsc , x^{n}),
\end{split}
\end{equation}
and
\begin{equation}\label{eq:204}
\begin{split}
E_{ip} & (x^{1}, \dotsc , x^{n}, \dot{x}^{1}, \dotsc , \dot{x}^{n})\\
&:= \frac{m}{2} \sum_{i=1}^{n} (\dot{x}^{i})^{2} + V_{ip} (x^{1}, \dotsc , x^{n})\\
&= \frac{m}{2} \sum_{i=1}^{n} (\dot{x}^{i})^{2} - V (x^{1}, \dotsc , x^{n})
\end{split}
\end{equation}
wherein we have, for convenience, introduced the inverted potential function \(V_{ip} (x^{1}, \dotsc , x^{n})\) defined explicitly by
\begin{equation}\label{eq:205}
V_{ip} (x^{1}, \dotsc , x^{n}) = -V (x^{1}, \dotsc , x^{n}).
\end{equation}
The Hamiltonian corresponding to \(L_{ip}\) is given by
\begin{equation}\label{eq:206}
\begin{split}
H_{ip} & (x^{1}, \dotsc , x^{n}, p_{1}, \dotsc , p_{n})\\
&:= \frac{1}{2m} \sum_{i=1}^{n} (p_{i})^{2} + V_{ip} (x^{1}, \dotsc , x^{n})\\
&= \frac{1}{2m} \sum_{i=1}^{n} (p_{i})^{2} - V (x^{1}, \dotsc , x^{n})
\end{split}
\end{equation}
so that the associated (zero-energy) Hamilton-Jacobi equation (for an `action' function \(S_{(0)}\)) results from setting
\begin{equation}\label{eq:207}
H_{ip} \left(x^{1}, \dotsc , x^{n}, \frac{\partial S_{(0)}}{\partial x^{1}}, \dotsc , \frac{\partial S_{(0)}}{\partial x^{n}}\right) = 0
\end{equation}
and coincides with (\ref{eq:201a}).

In view of the conditions we have imposed on \(V (x^{1}, \dotsc , x^{n})\) (\textit{c.f.} 1.3, 1.4) the graph of \(V_{ip} (x^{1}, \dotsc , x^{n})\) is that of a potential `hill' (rather than `valley' or `well') whose summit is the unique, (vanishing) global maximum lying at the origin in \(\mathbb{R}^{n}\):
\begin{equation}\label{eq:208}
\begin{split}
& V_{ip}  (x^{1}, \dotsc , x^{n}) < V_{ip} (0, \dotsc , 0) = 0\\
& \forall\; (x^{1}, \dotsc , x^{n}) \in \mathbb{R}^{n} \backslash (0, \dotsc , 0).
\end{split}
\end{equation}
There is a canonical family of (vanishing ip-energy) solutions to the Euler Lagrange equations to attempt to construct for this problem. Suppose that, for arbitrary \((\zero{x}^{\, 1}, \dotsc , \zero{x}^{\, n}) \in \mathbb{R}^{n}\) specified at time \(t = 0\), one could find complementary initial data \((\zero{v}^{1}, \dotsc , \zero{v}^{\, n}) \in \mathbb{R}^{n}\) such that the solution curve \((x^{1}(t), \dotsc , x^{n}(t))\) determined by
\begin{equation}\label{eq:209}
(x^{1}(0), \dotsc , x^{n}(0)) = (\zero{x}^{\, 1}, \dotsc , \zero{x}^{\, n})
\end{equation}
and
\begin{equation}\label{eq:210}
\left( \frac{dx^{1}(0)}{dt}, \dotsc , \frac{dx^{n}(0)}{dt}\right) = (\zero{v}^{\, 1}, \dotsc , \zero{v}^{\, n})
\end{equation}
satisfied
\begin{equation}\label{eq:211}
\begin{split}
& E_{ip} \left(x^{1}(t), \dotsc , x^{n}(t); \frac{dx^{1}(t)}{dt}, \dotsc , \frac{dx^{n}(t)}{dt}\right)\\
& = 0
\end{split}
\end{equation}
and
\begin{equation}\label{eq:212}
\lim_{t \rightarrow -\infty}{\left(x^{1}(t), \dotsc , x^{n}(t); \frac{dx^{1}(t)}{dt}, \dotsc , \frac{dx^{n}(t)}{dt}\right)} = (0, \dotsc , 0; 0, \dotsc , 0).
\end{equation}
In other words suppose that, for any initial position \((\zero{x}^{\, 1}, \dotsc , \zero{x}^{\, n})\), one could find a (vanishing ip-energy) solution curve that tended asymptotically, as \(t \searrow - \infty\), to the potential summit lying at the origin in \(\mathbb{R}^{n}\). Such a solution would necessarily, in view of its vanishing ip-energy, also have asymptotically vanishing velocity. If such a curve existed for each \((\zero{x}^{\, 1}, \dotsc , \zero{x}^{\, n}) \in \mathbb{R}^{n}\) and were uniquely determined by this data then the collection of such curves would comprise the `canonical' family that we are interested in.

We shall prove below, using the direct method of the calculus of variations and  additional convexity and coercivity hypotheses on the potential energy function \(V (x^{1}, \dotsc , x^{n})\), that such a solution curve does indeed exist for arbitrary \((\zero{x}^{\, 1}, \dotsc , \zero{x}^{\, n}) \in \mathbb{R}^{n}\) and that this curve is always uniquely determined by the given data. It will follow from the proof that the `action' integral
\begin{equation}\label{eq:213}
\mathcal{S} (\zero{x}^{\, 1}, \dotsc \zero{x}^{\, n}) := \int_{-\infty}^{0} dt\; L_{ip} \left(x^{1}(t), \dotsc , x^{n}(t), \frac{dx^{1}(t)}{dt}, \dotsc , \frac{dx^{n}(t)}{dt}\right)
\end{equation}
converges on each such curve and that (again using the convexity hypothesis) the resulting function \(\mathcal{S}: \mathbb{R}^{n} \rightarrow \mathbb{R}\) is smooth (i.e., \(C^{\infty}\)) and satisfies the ipze Hamilton-Jacobi equation globally on \(\mathbb{R}^{n}\). We shall also find that the gradient of \(S\) generates the complementary initial data \((\zero{v}^{\, 1}, \dotsc , \zero{v}^{\, n})\), mentioned above, via the formula
\begin{equation}\label{eq:214}
\begin{split}
(\zero{p}_{1}, \dotsc , \zero{p}_{n}) & := \left(\frac{\partial\mathcal{S}}{\partial x^{1}}, \dotsc , \frac{\partial\mathcal{S}}{\partial x^{n}}\right) (\zero{x}^{\, 1}, \dotsc , \zero{x}^{\, n})\\
&= (m\zero{v}^{\, 1}, \dotsc ,  m\zero{v}^{\, n})
\end{split}
\end{equation}
and, more generally, that the canonical family of solutions curves described above corresponds to the gradient `semi-flow' of \(\mathcal{S}\) defined via
\begin{equation}\label{eq:215}
m \frac{dx^{i}(t)}{dt} = \frac{\partial\mathcal{S}}{\partial x^{i}} (x^{1}(t), \dotsc , x^{n}(t))
\end{equation}
where \(i = 1, \dotsc , n\), \(x^{i}(0) = \zero{x}^{i}\). It will be natural at this point to identify \(\mathcal{S}\) with the fundamental solution, \(S_{(0)}\), to the ipze Hamilton-Jacobi equation that we have been seeking.

We use the term `semi-flow' for that generated through Eq.~(\ref{eq:215}) since it will turn out that each integral curve will typically only exist on an interval of the form \((-\infty, \varepsilon)\) for some \(\varepsilon > 0\) (which depends upon the curve under study). An important feature of these curves is that none of them (except the trivial one having \((\zero{x}^{\, 1}, \dotsc , \zero{x}^{\, n}) = (0, \dotsc , 0)\)) can achieve the potential energy summit at a finite time \(t^{*}\) but instead must only approach this point asymptotically as \(t \searrow -\infty\). The well-known reason for this is that if, at any time \(t^{*}\), one had \((x^{1}(t^{*}), \dotsc , x^{n}(t^{*})) = (0, \dotsc , 0)\) then, by the vanishing of its ip-energy, one would also have \(\left(\frac{dx^{1}(t^{*})}{dt}, \dotsc , \frac{dx^{n}(t^{*})}{dt}\right) = (0, \dotsc , 0)\) at this instant. But this is `initial' (at time \(t^{*}\)) data for the trivial solution \((x^{1}(t), \dotsc , x^{n}(t)) = (0, \dotsc , 0)\quad \forall\; t \in \mathbb{R}\) which sits at the summit for all \(t \in \mathbb{R}\) and thus, by uniqueness of solutions to the Euler Lagrange equations, could not have had the (non-trivial) initial (at time 0) conditions assumed for it. The semi-infinite character of the intervals of existence for these integral solution curves will play a crucial role in the subsequent analysis of the transport equations (\ref{eq:201b}--\ref{eq:202}) which, along the integral curves of the gradient of \(S_{(0)} = \mathcal{S}\), can now be written
\begin{subequations}\label{eq:216}
\begin{align}
\frac{dS_{(1)}}{dt} (x^{1}(t), \dotsc , x^{n}(t)) &= \left(\frac{1}{2m} {}^{(n)}\!\Delta S_{(0)} - \zero{\mathcal{E}}_{(0)}\right) (x^{1}(t), \dotsc , x^{n}(t)),\label{eq:216a}\\
\frac{dS_{(2)}}{dt} (x^{1}(t), \dotsc , x^{n}(t)) &= \left(\frac{1}{m} {}^{(n)}\!\Delta S_{(1)} - \frac{1}{m} \nabla S_{(1)} \cdot \nabla S_{(1)} - 2\zero{\mathcal{E}}_{(1)}\right) (x^{1}(t), \dotsc , x^{n}(t)), \label{eq:216b}
\end{align}
\end{subequations}
and, for general \(k \geq 2\),
\begin{equation}\label{eq:217}
\begin{split}
& \frac{dS_{(k)}}{dt} (x^{1}(t), \dotsc , x^{n}(t))\\
&= \left(\frac{k}{2m} {}^{(n)}\!\Delta S_{(k-1)} - \frac{1}{2m} \sum_{j=1}^{k-1} \frac{k!}{j!(k-j)!} \nabla S_{(j)} \cdot \nabla S_{(k-j)} - k\zero{\mathcal{E}}_{(k-1)}\right) (x^{1}(t), \dotsc , x^{n}(t)).
\end{split}
\end{equation}
The integrals of the right hand sides of these equations, over the semi-infinite intervals \((-\infty,0] \ni t\), will prove to converge, yielding smooth, sequentially defined formulas for the quantities \(\left\lbrace S_{(i)} (\zero{x}^{\, 1}, \dotsc , \zero{x}^{\, n}) - S_{(i)} (0, \dotsc , 0)\; \vert\; i = 1,2, \ldots\right\rbrace\) if and only if the energy coefficients \(\left\lbrace\zero{\mathcal{E}}_{(0)}, \zero{\mathcal{E}}_{(1)}, \ldots\right\rbrace\) are successively defined by
\begin{subequations}\label{eq:218}
\begin{align}
\zero{\mathcal{E}}_{(0)} &= \left\lbrack\frac{1}{2m} {}^{(n)}\!\Delta S_{(0)}\right\rbrack (0, \dotsc , 0), \label{eq:218a}\\
\zero{\mathcal{E}}_{(1)} &= \left\lbrack\frac{1}{2m} {}^{(n)}\!\Delta S_{(1)} - \frac{1}{2m} \nabla S_{(1)} \cdot \nabla S_{(1)}\right\rbrack (0, \dotsc , 0), \ldots \label{eq:218b}\\
\zero{\mathcal{E}}_{(k-1)} &= \left\lbrack\frac{1}{2m} {}^{(n)}\!\Delta S_{(k-1)} - \frac{1}{2m} \sum_{j=1}^{k-1} \frac{(k-1)!}{j! (k-j)!} \nabla S_{(j)} \cdot \nabla S_{(k-j)}\right\rbrack (0, \dotsc , 0). \label{eq:218c}
\end{align}
\end{subequations}
This choice for the \(\zero{\mathcal{E}}_{(i)}\)'s, together with (arbitrary) choices for the constants of integration \(\lbrace S_{(i)} (0, \dotsc , 0)\; \vert\; i = 1, 2, \ldots\rbrace\) (which, however, must, for each \(i\), be chosen independently of the integral curve in the canonical collection) will thus result in a well-defined sequence \(\lbrace S_{(1)}, S_{(2)} \ldots\rbrace (x^{1}, \dotsc , x^{n})\) of smooth quantum corrections to \(S_{(0)} (x^{1}, \dotsc , x^{n})\). Since we have allowed for a normalization constant, \(N_{\hbar}\), in the formula for the ground state wave function (1.5) one could always choose the integration constants \(\lbrace S_{(i)} (0, \dotsc ,0)\; \vert\; i = 1, 2, \ldots\rbrace\) to vanish without any essential loss of generality.

Though, by construction, our canonical solution curves for the ip-problem will always exist on the interval \((-\infty,0]\) they will typically only, if continued in the positive time direction, extend to an interval of the form \((-\infty,\varepsilon)\) for some \(\varepsilon > 0\) (where \(\varepsilon\) depends upon the curve in question). This follows from the more-rapid-than-quadratic decay of \(V_{ip} (x^{1}, \dotsc , x^{n})\) whose corresponding, repulsive force drives the solution curves to infinity in a finite time. For this reason the gradient `flow' generated by \(S_{(0)}\) is only a `semi-flow' as we have already mentioned.

\subsection{Excited State Preliminaries}
\label{subsec:excited}

To investigate excited states we substitute the ansatz
\begin{equation}\label{eq:219}
\starr{\psi}_{\hbar} = \starr{\phi}_{\hbar} e^{-S_{\hbar}/\hbar}
\end{equation}
into the (time-independent) Schr\"{o}dinger equation
\begin{equation}\label{eq:220}
\begin{split}
\hat{H} \starr{\psi}_{\hbar} &= -\frac{\hbar^{2}}{2m} {}^{(n)}\!\Delta\starr{\psi}_{\hbar} + V\starr{\psi}_{\hbar}\\
 &= \starr{E}_{\hbar} \starr{\psi}_{\hbar}
\end{split}
\end{equation}
and formally expand the wave function \(\starr{\phi}_{\hbar}\) and energy eigenvalue \(\starr{E}_{\hbar}\) in powers of \(\hbar\) as before
\begin{subequations}\label{eq:221}
\begin{align}
\starr{\phi}_{\hbar} &= \starr{\phi}_{(0)} + \frac{\hbar}{1!} \starr{\phi}_{(1)} + \frac{\hbar^{2}}{2!} \starr{\phi}_{(2)} + \dots \label{eq:221a}\\
\starr{E}_{\hbar} &:= \hbar\starr{\mathcal{E}}_{\hbar} = \hbar \left(\starr{\mathcal{E}}_{(0)} + \frac{\hbar}{1!} \starr{\mathcal{E}}_{(1)} + \frac{\hbar^{2}}{2!} \starr{\mathcal{E}}_{(2)} + \dots\right). \label{eq:221b}
\end{align}
\end{subequations}
The function \(S_{\hbar}\) is here chosen to \textit{coincide} with that previously defined for the ground state and thus admits a formal expansion of the same type
\begin{equation}\label{eq:222}
S_{\hbar} = S_{(0)} + \frac{\hbar}{1!}S_{(1)} + \frac{\hbar^{2}}{2!}S_{(2)} + \dots .
\end{equation}
Since, however, each of the coefficients \(\lbrace S_{(i)}\; |\; i = 0, 1, 2, \dots\rbrace\) has already been computed at this point and does not vary with the excited state under consideration, we refrain from attaching a superfluous, overhead \((\ast)\) thereto.

With the ansatz (\ref{eq:219}) Schr\"{o}dinger's equation (\ref{eq:220}) now takes the form
\begin{equation}\label{eq:223}
\left\lbrace\frac{\hbar}{2m} {}^{(n)}\!\Delta S_{\hbar} - \frac{1}{2m} \nabla S_{\hbar} \cdot \nabla S_{\hbar} + V - \hbar\starr{\mathcal{E}}_{\hbar}\right\rbrace\; \starr{\phi}_{\hbar} + \hbar \left(\frac{\nabla S_{\hbar}}{m}\right) \cdot \nabla\starr{\phi}_{\hbar} - \frac{\hbar^{2}}{2m} {}^{(n)}\!\Delta\starr{\phi}_{\hbar} = 0.
\end{equation}
But since the ground state wave function
\begin{equation}\label{eq:224}
\zero{\psi}_{\hbar} = \zero{N}_{\hbar} e^{-S_{\hbar}/\hbar}
\end{equation}
satisfies
\begin{equation}\label{eq:225}
\frac{\hbar}{2m} {}^{(n)}\!\Delta S_{\hbar} - \frac{1}{2m} \nabla S_{\hbar} \cdot \nabla S_{\hbar} + V = \hbar\zero{\mathcal{E}}_{\hbar}
\end{equation}
(with, by hypothesis, the same \(S_{\hbar}\)) we can reexpress Eq.~(\ref{eq:223}) in the simplified form
\begin{equation}\label{eq:226}
\left(\frac{\nabla S_{\hbar}}{m}\right) \cdot \nabla\starr{\phi}_{\hbar} - \frac{\hbar}{2m} {}^{(n)}\!\Delta\starr{\phi}_{\hbar} = \Delta\starr{\mathcal{E}}_{\hbar}\starr{\phi}_{\hbar}
\end{equation}
where \(\Delta\starr{\mathcal{E}}_{\hbar}\) determines the `energy gap',
\begin{equation}\label{eq:227}
\begin{split}
\Delta\starr{\mathcal{E}}_{\hbar} &:= \starr{\mathcal{E}}_{\hbar} - \zero{\mathcal{E}}_{\hbar}\\
&= \left(\Delta\starr{\mathcal{E}}_{(0)} + \frac{\hbar}{1!} \Delta\starr{\mathcal{E}}_{(1)} + \frac{\hbar^{2}}{2!} \Delta\starr{\mathcal{E}}_{(2)} + \dots\right)
\end{split}
\end{equation}
(with \(\Delta\starr{\mathcal{E}}_{(i)} = \starr{\mathcal{E}}_{(i)} - \zero{\mathcal{E}}_{(i)}\)), between the ground and excited states.

Substituting the foregoing expansions into Eq.~(\ref{eq:226}) and requiring the latter to hold order by order in \(\hbar\) leads to the following sequence of equations for the unknowns \(\left\lbrace\starr{\phi}_{(0)}, \starr{\phi}_{(1)}, \starr{\phi}_{(2)}, \dots ; \Delta\starr{\mathcal{E}}_{(0)}, \Delta\starr{\mathcal{E}}_{(1)}, \Delta\starr{\mathcal{E}}_{(2)},\dots\right\rbrace\):
\begin{subequations}\label{eq:228}
\begin{align}
\left(\frac{\nabla S_{(0)}}{m}\right) \cdot \nabla\starr{\phi}_{(0)} - \Delta\starr{\mathcal{E}}_{(0)}\starr{\phi}_{(0)} &= 0,\label{eq:228a}\\
\left(\frac{\nabla S_{(0)}}{m}\right) \cdot \nabla\starr{\phi}_{(1)} - \Delta\starr{\mathcal{E}}_{(0)}\starr{\phi}_{(1)} &= \Delta\starr{\mathcal{E}}_{(1)}\starr{\phi}_{(0)} + \left(\frac{1}{2m}\right) {}^{(n)}\!\Delta\starr{\phi}_{(0)} - \frac{\nabla S_{(1)}}{m} \cdot \nabla\starr{\phi}_{(0)},\label{eq:228b}\\
\left(\frac{\nabla S_{(0)}}{m}\right) \cdot \nabla\starr{\phi}_{(2)} - \Delta\starr{\mathcal{E}}_{(0)}\starr{\phi}_{(2)} &= \Delta\starr{\mathcal{E}}_{(2)}\starr{\phi}_{(0)} + 2\Delta\starr{\mathcal{E}}_{(1)}\starr{\phi}_{(1)} + \left(\frac{1}{m}\right) {}^{(n)}\!\Delta\starr{\phi}_{(1)}\nonumber\\
 &\qquad - \left(\frac{\nabla S_{(2)}}{m}\right) \cdot \nabla\starr{\phi}_{(0)} - 2\left(\frac{\nabla S_{(1)}}{m}\right) \cdot \nabla\starr{\phi}_{(1)}, \label{eq:228c}
\end{align}
\end{subequations}
and, for arbitrary \(k \geq 1\),
\begin{equation}\label{eq:229}
\begin{split}
\left(\frac{\nabla S_{(0)}}{m}\right) \cdot \nabla\starr{\phi}_{(k)} - \Delta\starr{\mathcal{E}}_{(0)}\starr{\phi}_{(k)} &= \frac{k}{2m} {}^{(n)}\!\Delta\starr{\phi}_{(k-1)}\\
 &~ + \sum_{j=1}^{k} \frac{k!}{j!(k-j)!} \left[\Delta\starr{\mathcal{E}}_{(j)}\starr{\phi}_{(k-j)} - \left(\frac{\nabla S_{(j)}}{m}\right) \cdot \nabla\starr{\phi}_{(k-j)}\right].
\end{split}
\end{equation}
In view of the presence of the constant term \(\Delta\starr{\mathcal{E}}_{(0)}\) occurring in the `transport operator'
\begin{equation}\label{eq:230}
\starr{\mathcal{L}} := \left(\frac{\nabla S_{(0)}}{m}\right) \cdot \nabla - \Delta\starr{\mathcal{E}}_{(0)}
\end{equation}
that characterizes the left hand sides of Eqs.~(\ref{eq:228}--\ref{eq:229}), these transport equations require a somewhat different treatment from that described in the previous section for the ground state. Fortunately, however, there are well-developed methods for solving such systems and we shall review these below in Sect.~(\ref{subsec:transport-excited-state}).

The aforementioned methods begin with the development of formal power series solutions to the relevant equations and then proceed (in non-analytic cases) to `correct' these formal expansions in a systematic way so as to produce actual smooth solutions. Since these formal expansions, however, are straightforward to develop and since they already yield, by themselves, an algorithm for the computation of the energy gap coefficients \(\left\lbrace\Delta\starr{\mathcal{E}}_{(0)}, \Delta\starr{\mathcal{E}}_{(1)}, \Delta\starr{\mathcal{E}}_{(2)}, \dots\right\rbrace\) we shall review them here, postponing the refinements needed to complete the determination of the smooth wave function coefficients \(\left\lbrace\starr{\phi}_{(0)}, \starr{\phi}_{(1)}, \starr{\phi}_{(2)}, \dots\right\rbrace\) until Sect.~(\ref{subsec:transport-excited-state}) below.

We shall prove in Sect.~(\ref{subsec:smoothness}) that our fundamental solution, \(S_{(0)}(\mathbf{x})\), to the ipze Hamilton-Jacobi equation has the properties
\begin{equation}\label{eq:231}
S_{(0)}(\mathbf{x}) = \frac{1}{2} m \sum_{i=1}^{n} \omega_{i}(x^{i})^{2} + O(|\mathbf{x}|^{3})
\end{equation}
and
\begin{equation}\label{eq:232}
\partial_{j} S_{(0)}(\mathbf{x}) = m\omega_{j}x^{j} + O(|\mathbf{x}|^{2})
\end{equation}
(\textit{c.f.}, Eqs.~(\ref{eq:368}) and (\ref{eq:382}) respectively). Thus, writing
\begin{equation}\label{eq:233}
\mathcal{S}_{(0)} (\mathbf{x}) = \frac{1}{2} m \sum_{i=1}^{n} \omega_{i}(x^{i})^{2} + m \Sigma_{(0)}(\mathbf{x})
\end{equation}
where
\begin{equation}\label{eq:234}
\Sigma_{(0)}(\mathbf{x}) = O(|\mathbf{x}|^{3}), \qquad \partial_{j}\Sigma_{(0)}(\mathbf{x}) = O(|\mathbf{x}|^{2})
\end{equation}
we split the operator \(\starr{\mathcal{L}}\) defined above into linear
\begin{equation}\label{eq:235}
\starr{\mathcal{L}}_{0} = \left(\sum_{i=1}^{n} \omega_{i}x^{i} \frac{\partial}{\partial x^{i}}\right) - \Delta\starr{\mathcal{E}}_{(0)}
\end{equation}
and remaining, higher order terms
\begin{equation}\label{eq:236}
\starr{\mathcal{L}}_{R} := \sum_{i=1}^{n} \frac{\partial\Sigma_{(0)}}{\partial x^{i}} \frac{\partial}{\partial x^{i}}
\end{equation}
so that \(\starr{\mathcal{L}} = \starr{\mathcal{L}}_{0} + \starr{\mathcal{L}}_{R}\).

As a first step toward solving Eq.~(\ref{eq:228a}), which now reads
\begin{equation}\label{eq:237}
\starr{\mathcal{L}}\;\:\starr{\phi}_{(0)} = 0,
\end{equation}
let us begin with the simpler equation
\begin{equation}\label{eq:238}
\starr{\mathcal{L}}_{0}\starr{f} = 0.
\end{equation}
This (linear, separable) equation is easily seen to have smooth, non-trivial solutions within the space of homogeneous polynomials in the \(n\) variables \(\lbrace x^{1}, \ldots , x^{n}\rbrace\) if and only if \(\Delta\starr{\mathcal{E}}_{(0)}\) has the form
\begin{equation}\label{eq:239}
\Delta\starr{\mathcal{E}}_{(0)} = \Delta\emm{\mathcal{E}}_{(0)} := \sum_{i=1}^{n} m_{i}\omega_{i}
\end{equation}
where the \(\lbrace m_{i}\rbrace\) are non-negative integers such that
\begin{equation}\label{eq:240}
|\mathbf{m}| := \sum_{i=1}^{n} m_{i} \geq 1.
\end{equation}
A corresponding solution, \(\emm{f}_{(0)}(\mathbf{x})\), to
\begin{equation}\label{eq:241}
\emm{\mathcal{L}}_{0}\emm{f}_{(0)} = 0
\end{equation}
with
\begin{equation}\label{eq:242}
\begin{split}
\emm{\mathcal{L}}_{0} &:= \sum_{i=1}^{n} \omega_{i}x^{i} \frac{\partial}{\partial x^{i}} - \Delta\emm{\mathcal{E}}_{(0)}\\
 &= \sum_{i=1}^{n} \omega_{i}x^{i} \frac{\partial}{\partial x^{i}} - \sum_{i=1}^{n} m_{i}\omega_{i}
\end{split}
\end{equation}
will then be given by
\begin{equation}\label{eq:243}
\begin{split}
\emm{f}_{(0)}(\mathbf{x}) &= (x^{i})^{m_{1}} (x^{2})^{m_{2}} \dots (x^{n})^{m_{n}}\\
&:= \mathbf{x}^{\mathbf{m}}
\end{split}
\end{equation}
and, of course, by arbitrary constant multiples thereof.

However, the solution (\ref{eq:243}) to (\ref{eq:241}) will not be unique (up to the trivial, multiplicative constant) unless
\begin{equation}\label{eq:244}
\Delta\emmprime{\mathcal{E}}_{(0)} := \sum_{i=1}^{n} m'_{i}\omega_{i} = \Delta\emm{\mathcal{E}}_{(0)} = \sum_{i=1}^{n} m_{i}\omega_{i}
\end{equation}
implies \(m'_{i} = m_{i}\; \forall\; i \in [1, \ldots , n]\) since, otherwise
\begin{equation}\label{eq:245}
\emmprime{f}_{(0)}(\mathbf{x}) := (x^{1})^{m'_{1}} \ldots (x^{n})^{m'_{n}}
\end{equation}
or, more generally, arbitrary linear combinations of \(\emm{f}_{(0)}\) and \(\emmprime{f}_{(0)}\) would provide additional, independent solutions to Eq.~(\ref{eq:241}).

While this non-degeneracy for the energy gap coefficient \(\Delta\emm{\mathcal{E}}_{(0)}\) (for a fixed, chosen \(\mathbf{m} = (m_{1}, \ldots , m_{n})\)) is, strictly speaking, not needed for our constructions, its breakdown would necessitate the extension of our power series expansions in (integral) powers of \(\hbar\) to allow for half-integral powers thereof and thus somewhat complicate the analysis (\textit{c.f.}, the discussion given in Sect.~(\ref{sec:existence-solution}) of Ref.~\cite{Dimassi1999} and further references on this issue cited therein). To sidestep such complications (at least temporarily) we shall henceforth impose the aforementioned non-degeneracy requirement as a restriction upon the `frequencies' \(\lbrace\omega_{i}\rbrace\) to be considered. For non-degeneracy to hold, not just for some particular choice of the `quantum numbers' \(m = \lbrace m_{1}, \ldots , m_{n}\rbrace\), but for all allowed choices, we shall need to assume that the equation
\begin{equation}\label{eq:246}
\sum_{i=1}^{n} \ell_{i}\omega_{i} = 0
\end{equation}
has no, nontrivial solutions for arbitrary \(\ell = \lbrace\ell_{1}, \ldots , \ell_{n}\rbrace \in \mathbb{Z}^{n}\).

We can now proceed to construct a (formal, power series) solution to
\begin{equation}\label{eq:247}
\emm{\mathcal{L}}\;\: \emm{f} = 0
\end{equation}
of the form
\begin{equation}\label{eq:248}
\emm{f} = \sum_{k=|m|}^{\infty} \emm{f}_{k}
\end{equation}
 where each \(\emm{f}_{k}\) belongs to the space, \(\mathcal{P}_{hom}^{k}\), of homogeneous polynomials of degree \(k\) in the \(n\) variables \(\mathbf{x} = \lbrace x^{1}, \ldots , x^{n}\rbrace\). Note that the monomials
\begin{equation}\label{eq:249}
\mathbf{x}^{\mathbf{k}} := (x^{1})^{\mathbf{k}_{1}} \dots (x^{n})^{\mathbf{k}_{n}},
\end{equation}
(with \(\mathbf{k} = (k_{1}, \ldots , k_{n})\), \(|\mathbf{k}| = \sum_{i=1}^{n} k_{i}\) and \(k_{i} \in \mathbb{N} \cup \lbrace 0\rbrace \forall\; i \in [1, \ldots , n]\)), provide a basis for the eigenvectors of \(\emm{\mathcal{L}}_{0}\) restricted to \(\mathcal{P}_{hom}^{k}\) with corresponding eigenvalues given by
\begin{equation}\label{eq:250}
\emm{\mathcal{L}}_{0}\mathbf{x}^{\mathbf{k}} = \left(\sum_{i=1}^{n} (k_{i} - m_{i})\omega_{i}\right)\mathbf{x}^{\mathbf{k}}
\end{equation}
which, by virtue of our non-degeneracy condition, are non-vanishing whenever \(|\mathbf{k}| \neq |\mathbf{m}|\). Thus, in particular, \(\emm{\mathcal{L}}_{0}\) is a bijection on \(\mathcal{P}_{hom}^{k}\) provided that \(k \neq |\mathbf{m}|\).

Returning to Eq.~(\ref{eq:247}) and setting
\begin{equation}\label{eq:251}
\emm{f}_{|\mathbf{m}|} = (x^{1})^{m_{1}} \dots (x^{n})^{m_{n}} := \mathbf{x}^{\mathbf{m}}
\end{equation}
one finds that, in the sense of formal power series,
\begin{equation}\label{eq:252}
\emm{\mathcal{L}}\;\: \emm{f}_{|\mathbf{m}|} = \sum_{k=|m|+1}^{\infty} g_{k}
\end{equation}
where each \(g_{k} \in \mathcal{P}_{hom}^{k}\). Taking \(\emm{f}_{|\mathbf{m}|+1}\) to be the (unique in \(\mathcal{P}_{hom}^{|\mathbf{m}|+1}\)) solution to
\begin{equation}\label{eq:253}
\emm{\mathcal{L}}_{0}\;\: \emm{f}_{|\mathbf{m}|+1} = -g_{|m|+1}
\end{equation}
one then finds that
\begin{equation}\label{eq:254}
\emm{\mathcal{L}} \left(\emm{f}_{|\mathbf{m}|} + \emm{f}_{|\mathbf{m}|+1}\right) = \sum_{k=|m|+2}^{\infty} h_{k}
\end{equation}
for certain \(h_{k} \in \mathcal{P}_{hom}^{k}\). Thus, taking \(\emm{f}_{|\mathbf{m}|+2}\) to be the unique solution (in \(\mathcal{P}_{hom}^{|m|+2}\)) to
\begin{equation}\label{eq:255}
\emm{\mathcal{L}}_{0}\;\: \emm{f}_{|\mathbf{m}|+2} = -h_{|m|+2},
\end{equation}
one proceeds, in this way, to construct the complete formal series solution (\ref{eq:248}) to (\ref{eq:247}) that we have been seeking.

Clearly inhomogeneous equations can be treated in the same way provided the inhomogeneity contains no (formal expansion) term proportional to \(\emm{f}_{|\mathbf{m}|} = (x^{1})^{m_{1}} \dots (x^{n})^{m_{n}}\) (the kernel of \(\emm{\mathcal{L}}_{0}\)). One thus arrives at the
\begin{quote}
\textit{\textbf{Proposition} (3.4 of Ref.~\cite{Dimassi1999}):\hfill\break
For every formal power series \(g\) at \(\mathbf{x} = \mathbf{0}\) there is a unique constant \(\emm{\lambda}_{(g)}\) such that
\begin{equation*}
\emm{\mathcal{L}}\;\: \emm{f} = g - \emm{\lambda}(g)\;\: \emm{f}_{|\mathbf{m}|}
\end{equation*}
has a solution in the sense of formal power series. This solution is unique up to a multiple of \(\emm{f}_{|\mathbf{m}|}\).}
\end{quote}

Applying the foregoing proposition it is now clear how to solve the sequence of transport equations (\ref{eq:228}--\ref{eq:229}) in the sense of finding formal power series expansions for the  \(\lbrace\emm{\phi}_{(k)}\; |\; k = 0, 1, \dots\rbrace\). Except perhaps in the case of analytic potential functions one cannot expect these formal series expansions to converge but we shall see later, in Sect.~(\ref{subsec:transport-excited-state}), how they can all be `corrected' to yield genuine, smooth solutions to Eqs.~(\ref{eq:228}--\ref{eq:229}).

One has however, already at the formal level, succeeded to determine the energy gap coefficients \(\lbrace\Delta\emm{\mathcal{E}}_{(0)}, \Delta\emm{\mathcal{E}}_{(1)}, \Delta\emm{\mathcal{E}}_{(2)}, \dots \rbrace\) in the sense that, once  \(\Delta\emm{\mathcal{E}}_{(0)}\) has been chosen, all the subsequent coefficients \(\Delta\emm{\mathcal{E}}_{(k)}\), for \(k = 1, 2, \dots\), are then uniquely determined and remain unmodified by the subsequent refinements needed to construct the smooth functions \(\lbrace\emm{\phi}_{k}\; |\; k = 0, 1, 2, \dots\rbrace\). This is a straightforward consequence of the foregoing proposition and the structure of Eqs.~(\ref{eq:228}--\ref{eq:229}) wherein each successive \(\Delta\emm{\mathcal{E}}_{(k)}\) provides precisely the constant, \(\emm{\lambda}(g)\), needed for that step in the argument.

Even though formal expansion methods were not needed for the \textit{ground state} analysis they could nevertheless have been used to develop formal series solutions to Eqs.~(\ref{eq:201}--\ref{eq:202}) and thereby to determine the energy  coefficients \(\lbrace\zero{\mathcal{E}}_{(0)}, \zero{\mathcal{E}}_{(1)}, \dots\rbrace\) defined through Eq.~(\ref{eq:218}). Note, furthermore, that even the nonlinear equation (\ref{eq:201a}) can be solved formally by setting
\begin{equation}\label{eq:256}
\tilde{S}_{(0)}(\mathbf{x}) = \frac{1}{2} m \sum_{i=1}^{n} \omega_{i}(x^{i})^{2} + \sum_{k=3}^{\infty} s_{k}(\mathbf{x})
\end{equation}
where \(s_{k}(\mathbf{x}) \in \mathcal{P}_{hom}^{k}\) and recalling that, in view of our hypotheses for the potential energy function, the latter admits a formal expansion of the form
\begin{equation}\label{eq:257}
\tilde{V}(\mathbf{x}) = \frac{1}{2} m \sum_{i=1}^{n} \omega_{i}^{2}(x^{i})^{2} + \sum_{k=3}^{\infty} v_{k}(\mathbf{x})
\end{equation}
where \(v_{k} \in \mathcal{P}_{hom}^{k}\). Substituting (\ref{eq:256}) and (\ref{eq:257}) into Eq.~(\ref{eq:201a}) it is straightforward to verify that each successive \(s_{k}(\mathbf{x})\) is uniquely determined by an argument that closely parallels that sketched above for the excited states. Using the formal expansion \(\tilde{S}_{(0)}(\mathbf{x})\) in place of the exact solution \(S_{(0)}(\mathbf{x})\) in Eqs.~(\ref{eq:201b}--\ref{eq:202}) it is easy to see that these can each be sequentially solved (formally) by the same techniques developed for the excited states and the energy coefficients \(\lbrace\zero{\mathcal{E}}_{(k)}\; |\; k = 0, 1, \dots\rbrace\) thereby determined along with formal expansions \(\lbrace\tilde{S}_{(k)}(\mathbf{x})\; |\; k = 0, 1, 2, \dots\rbrace\) for the actual smooth functions \(\lbrace S_{(k)}(\mathbf{x})\rbrace\).

The point of these remarks is that, for both ground and excited states, the energy and energy gap coefficients together with corresponding formal expansions for the wave function coefficients can all be computed independently of any exact construction of the solutions to the relevant equations. It is not difficult to see that all of these elements are determined from the (formal, in the non-analytic case) Taylor expansion coefficients of the potential energy function \(V(x)\), i.e., from the collection \(\left\lbrace C_{\alpha_{1}, \dots , \alpha_{n}} \; |\; \alpha_{i} \in \mathbb{N} \cup \lbrace 0\rbrace \mbox{ for } i \in [1, \dots , n]\right\rbrace\) with
\begin{equation}\label{eq:258}
C_{\alpha_{1}, \dots , \alpha_{n}} := \left.\frac{\partial^{|\alpha|} V(\mathbf{x})}{\partial x^{1\; \alpha_{1}} \dots \partial x^{n\; \alpha_{n}}}\right|_{\mathbf{x} = \mathbf{0}}
\end{equation}
It follows that, for non-analytic potentials (many of which share the same Taylor expansion coefficients but not the same energy spectrum), the computable energy and energy gap coefficients, for example, could not be expected to have anything more than an asymptotic significance in the limit as \(\hbar\) tends to zero. For analytic potential energies though, it is known that the situation is significantly more favorable and we shall review what is known about the precise sense in which such constructions provide approximate solutions to Schr\"{o}dinger's equation below at the end of Sect.~(\ref{sect:integration-transport}). In Sect.~(\ref{sec:computable-examples}), which is devoted to the study of certain (analytic) anharmonic oscillators in one dimension, we shall discuss some evidence that constructions of the type under study, combined with suitable Borel resummations, may even yield exact results.

We conclude this section with a brief discussion of an alternative technique for solving the basic equation
\begin{equation}\label{eq:259}
\starr{\mathcal{L}}\;\: \starr{\phi}_{(0)} := \left[\left(\frac{\nabla S_{(0)}}{m}\right) \cdot \nabla - \Delta\starr{\mathcal{E}}_{(0)}\right] \starr{\phi}_{(0)} = 0
\end{equation}
through an application of the well-known Sternberg linearization theorem [5]. For a class of smooth vector fields including those of the type
\begin{equation}\label{eq:260}
\frac{\nabla S_{(0)}}{m} \cdot \nabla = \sum_{i=1}^{n} \left(\omega_{i}x^{i} \frac{\partial}{\partial x^{i}} + \partial_{i} \Sigma_{(0)} \frac{\partial}{\partial x^{i}}\right)
\end{equation}
(with \(\partial_{j}\Sigma_{(0)}(\mathbf{x}) = O(|\mathbf{x}|^{2})\)), Sternberg proved that, if the `frequencies' \(\lbrace\omega_{i}\rbrace\) satisfy a certain `non-resonance condition' (which is implied by our non-degeneracy condition (\ref{eq:246})), then there is a (local) diffeomorphism,
\begin{equation}\label{eq:261}
\begin{split}
\mathbf{x} &\longmapsto \mathbf{y} = \mu(\mathbf{x}),\\
y^{i} &= \mu^{i} (x^{1}, \dots , x^{n})\; \forall\; i \in [1, \dots , n],
\end{split}
\end{equation}
defined on a neighborhood of the origin in \(\mathbb{R}^{n}\) such that, in terms of the new coordinates \(\lbrace y^{1}, \dots , y^{n}\rbrace\) the corresponding vector field takes the purely linear form
\begin{equation}\label{eq:262}
\frac{\nabla S_{(0)}}{m} \cdot \nabla = \sum_{i=1}^{n} \omega_{i}y^{i} \frac{\partial}{\partial y^{i}}.
\end{equation}
Using special features of the Hamilton-Jacobi (semi-) flows generated by our particular vector fields we shall show, in the Appendix, that Sternberg's \(\mu\) can always be extended to a global diffeomorphism mapping \(\mathbb{R}^{n} = \left\lbrace(x^{1}, \dots , x^{n})\; |\; x^{i} \in \mathbb{R}\right\rbrace\) to a corresponding star-shaped domain in the space of the \(\lbrace y^{1}, \dots , y^{n}\rbrace\) coordinates. That the image of \(\mathbb{R}^{n}\) under this extended \(\mu\) does not normally exhaust another copy of \(\mathbb{R}^{n}\), but only a star-shaped domain therein, follows from the fact, already noted, that the integral curves of \(\frac{\nabla S_{(0)}}{m}\) typically only persist for semi-infinite time intervals of the type (\(-\infty, t^{\ast}(\gamma)\)) whereas these curves, expressed in the new coordinates, each have the explicit form
\begin{equation}\label{eq:263}
y_{\gamma}^{i}(t) = y_{\gamma}^{i}(0) e^{\omega_{i}t}
\end{equation}
for suitable constants \(\lbrace (y_{\gamma}^{i}(0))\; |\; i \in [1, \dots , n]\rbrace\). The boundary of the star-shaped domain in question is thus defined by the collection of ideal endpoints with coordinates
\begin{equation}\label{eq:264}
\starr{y}_{\gamma}^{\: i} = y_{\gamma}^{i}(0) e^{\omega_{i}t^{\ast}(\gamma)}.
\end{equation}

In terms of the new coordinates the regular solutions to
\begin{equation}\label{eq:265}
\starr{\mathcal{L}}\;\: \starr{\phi}_{(0)} = \sum_{i=1}^{n} \omega_{i}y^{i} \frac{\partial}{\partial y^{i}} \starr{\phi}_{(0)} - \Delta\starr{\mathcal{E}}_{(0)}\;\: \starr{\phi}_{(0)} = 0
\end{equation}
are given simply by (constant multiples of) the monomials
\begin{equation}\label{eq:266}
\emm{\phi}_{(0)}(\mathbf{y}) := (y^{1})^{m_{1}} \dots (y^{n})^{m_{n}}
\end{equation}
with corresponding eigenvalues
\begin{equation}\label{eq:267}
\Delta\emm{\mathcal{E}}_{(0)} = \sum_{i=1}^{n} m_{i}\omega_{i}
\end{equation}
as before. Note that the \(\lbrace\emm{\phi}_{(0)}(\mathbf{y})\rbrace\) are normally (i.e., for genuinely \textit{nonlinear} oscillators) bounded by virtue of the bounded nature of the range of the \(y\)-coordinates.

By contrast, for the case of \textit{purely linear} oscillators (for which the original \(x\)-coordinates are already of Sternberg type but with unbounded range) the corresponding monomials, \(\emm{\phi}_{(0)}(\mathbf{x}) = (x^{1})^{m_{1}} \dots (x^{n})^{m_{n}},\) provide the highest order (in \(\mathbf{x}\)) terms in the usual product of Hermite polynomials that characterizes the excited state wave functions and it is straightforward to verify that the constructions of this section merely fill in the lower order terms of these polynomials and then terminate, thus reproducing the well-known exact results for \textit{linear} oscillators. More precisely, one finds, by direct calculation, that all of the higher corrections to
\begin{equation}\label{eq:268}
S_{(0)}^{\mathrm{linear}}(\mathbf{x}) := \frac{1}{2} m \sum_{i=1}^{n} \omega_{i} (x^{i})^{2}
\end{equation}
vanish, leaving the familiar gaussian exponential factor common to all the states, and that the series expansions (\ref{eq:221a}) and (\ref{eq:221b}) now terminate yielding
\begin{equation}\label{eq:269}
\emm{\phi}_{\hbar}(\mathbf{x}) = \emm{N}_{\hbar} H_{m_{1}} \left(\sqrt{\frac{m\omega_{1}}{\hbar}} x^{1}\right) \dots H_{m_{n}} \left(\sqrt{\frac{m\omega_{n}}{\hbar}} x^{n}\right)
\end{equation}
and
\begin{equation}\label{eq:270}
\emm{E}_{\hbar} = \sum_{i=1}^{n} \left(m_{i} +  \frac{1}{2}\right) \hbar\omega_{i}.
\end{equation}

\section{Existence and Smoothness of the Fundamental Solution}
\label{sec:existence-solution}

\subsection{Existence and Regularity of Minimizers}
\label{subsec:existence-minimizers}

A natural approach for seeking solutions to the inverted potential (ip) dynamics problem formulated above is to look for minimizers of the ip action functional
\begin{equation}\label{eq:301}
\begin{split}
\mathcal{I}_{ip} [\gamma] &:= \int_{-\infty}^{0} L_{ip} \left(x^{1}(t), \dotsc , x^{n}(t), \dot{x}^{1}(t), \dotsc , \dot{x}^{n}(t)\right)\; dt\\
&= \int_{-\infty}^{0} \left\lbrace\frac{1}{2}m \sum_{i=1}^{n} \left\lbrace\left(\dot{x}^{i}(t)\right)^{2} + \omega_{i}^{2} \left(x^{i}(t)\right)^{2}\right\rbrace + A \left(x^{1}(t), \dotsc , x^{n}(t)\right)\right\rbrace\; dt
\end{split}
\end{equation}
within the affine space of curves
\begin{equation}\label{eq:302}
\begin{split}
\mathcal{D}_{\mathbf{x}} &:= \left\lbrace\gamma \in H^{1} (I,\mathbb{R}^{n})\; \vert\; I = (-\infty,0],\right.\\
\gamma(t) &= \left(x^{1}(t), \dotsc , x^{n}(t)\right), \lim_{t \nearrow 0}{\gamma(t)} = \mathbf{x}\\
&= \left.\vphantom{\gamma\in H^{1} (I,\mathbb{R}^{n})\; \vert\; I = (-\infty,0),} (x^{1}, \dotsc ,x^{n}) \in \mathbb{R}^{n}\right\rbrace.
\end{split}
\end{equation}
Here \(H^{1} (I,\mathbb{R}^{n})\) is the Sobolev space of (distributional) curves in \(\mathbb{R}^{n}\) equipped with the norm
\begin{equation}\label{eq:303}
\begin{split}
\vert\!\vert\gamma (\cdot)\vert\!\vert_{H^{1} (I,\mathbb{R}^{n})} &= \left\lbrace\int_{-\infty}^{0} \sum_{i=1}^{n} \left\lbrack\left(\dot{x}^{i}(t)\right)^{2} + \omega_{i}^{2} \left( x^{i}(t)\right)^{2}\right\rbrack\; dt\right\rbrace^{1/2}\\
& < \infty
\end{split}
\end{equation}
and \(\mathbf{x} = (x^{1}, \dotsc , x^{n})\) is an arbitrary but fixed right endpoint lying in \(\mathbb{R}^{n}\). From the Sobolev embedding theorem for \(H^{s}\)-maps \cite{Blanchard1992,Buttazzo2005} one has that \(H^{1} (I,\mathbb{R}^{n})\) is continuously embedded in
\begin{equation}\label{eq:304}
\begin{split}
C_{b}^{0} (I,\mathbb{R}^{n}) &:= \left\lbrace\vphantom{\binom{\text{sup}}{t \in I} \sqrt{\sum_{i=1}^{n} \left( x^{i}(t)\right)^{2}} < \infty}\gamma\in C^{0} (I,\mathbb{R}^{n}) \vert\right.\\
\vert\!\vert\gamma (\cdot)\vert\!\vert_{L^{\infty} (I,\mathbb{R}^{n})} &= \left.\sup_{t \in I} \sqrt{\sum_{i=1}^{n} \left( x^{i}(t)\right)^{2}} < \infty\right\rbrace
\end{split}
\end{equation}
where \(C^{0} (I,\mathbb{R}^{n})\) is the space of continuous curves \(\gamma : I \rightarrow \mathbb{R}^{n}\), and furthermore that these curves automatically (as a consequence of having finite \(H^{1}\)norm) `vanish at infinity' in the sense that
\begin{equation}\label{eq:305}
\begin{split}
\lim_{t \rightarrow -\infty}{|\gamma (t)|} &= \lim_{t \rightarrow -\infty}{\sqrt{\sum_{i=1}^{\infty} \left(x^{i}(t)\right)^{2}}}\\
&= 0.
\end{split}
\end{equation}
Thus the curves in \(\mathcal{D}_{\mathbf{x}}\) have their (asymptotically attained) left endpoints at the origin in \(\mathbb{R}^{n}\) which, in our setup, coincides with the unique, global maximum of the inverted potential function \(V_{ip} (x^{1}, \dotsc , x^{n}) = -V (x^{1}, \dotsc , x^{n})\).

Strictly speaking the `curves' in \(H^{1} (I, \mathbb{R}^{n})\) are \textit{distributions} but, by virtue of the Sobolev embedding theorem cited above, each such distribution can be represented by a continuous curve which (by a slight abuse of notation) we also write as \(\gamma : I \rightarrow \mathbb{R}^{n}\). For this reason one can meaningfully speak of the \textit{values} of \(\gamma (t)\) (as points in \(\mathbb{R}^{n}\)) for any \(t \in I = (-\infty,0]\) and also impose the right-endpoint boundary condition (included in the definition of \(\mathcal{D}_{\mathbf{x}}\)) that
\begin{equation}\label{eq:306}
\lim_{t \nearrow 0}{\gamma (t)} = \mathbf{x} = (x^{1}, \dotsc , x^{n}) \in \mathbb{R}^{n}.
\end{equation}
For a more extensive discussion of the values and boundary values of curves in \(H^{1} (I,\mathbb{R}^{n})\), when \(I\) is a finite domain, see Section 2.1 of Ref.~\cite{Buttazzo2005}. That we are instead working on the semi-infinite domain \(I = (-\infty,0)\) is taken into account, as \(t \rightarrow -\infty\), by the `vanishing at infinity' result mentioned above.

The first two terms in the explicit integral formula for \(\mathcal{I}_{ip} [\gamma]\) are automatically finite for any curve \(\gamma \in \mathcal{D}_{\mathbf{x}}\) since these terms comprise simply a (positive constant) multiple of the squared \(H^{1} (I,\mathbb{R}^{n})\) norm of \(\gamma\). The integral of the remaining term in \(\mathcal{I}_{ip}[\gamma]\) is also finite since, by assumption, the potential function \(A (x^{1}, \dotsc , x^{n})\) is smooth and satisfies
\begin{equation}\label{eq:307}
A(0, \dotsc ,0) = \frac{\partial A(0, \dotsc ,0)}{\partial x^{i}} = \frac{\partial^{2} A(0, \dotsc ,0)}{\partial x^{i}\partial x^{i}} = 0
\end{equation}
\(\forall\; i,j \in \lbrack 1, \dotsc ,n\rbrack\). It follows that the function \(A(x^{1}, \dotsc ,x^{n})/\langle \mathbf{x,x}\rangle_{\omega}\), where \(\langle \mathbf{x,x}\rangle_{\omega} := \sum_{i=1}^{n} \omega_{i}^{2}(x^{i})^{2}\), is bounded on bounded subsets of \(\mathbb{R}^{n}\) and thus, since the curves \(\gamma \in \mathcal{D}_{\mathbf{x}}\) are each bounded, we have that
\begin{gather}\label{eq:308}
\left\vert\int_{-\infty}^{0} dt\; A(\gamma (t))\right\vert \leq \int_{-\infty}^{0} dt\; \left\vert\frac{A(\gamma (t))}{\left\langle\gamma (t),\gamma (t)\right\rangle_{\omega}}\right\vert \left\langle\gamma (t),\gamma (t)\right\rangle_{\omega}\\
\leq\; \mathrm{constant}\; \vert\!\vert\gamma (\cdot)\vert\!\vert_{L^{2}(I,\mathbb{R}^{n})}^{2} \leq\; \mathrm{constant}\; \vert\!\vert\gamma (\cdot)\vert\!\vert_{H^{1}(I,\mathbb{R}^{n})}^{2}.\nonumber
\end{gather}

Thus the ip action integral \(\mathcal{I}_{ip}[\gamma]\) is finite for any curve \(\gamma \in \mathcal{D}_{\mathbf{x}}\), for any \(\mathbf{x} \in \mathbb{R}^{n}\) and, in view of our requirement that
\begin{equation}\label{eq:309}
V(x^{1}, \dotsc ,x^{n}) = \frac{1}{2} \sum_{i=1}^{n} m\omega_{i}^{2} (x^{i})^{2} + A(x^{1}, \dotsc ,x^{n}) \geq 0,
\end{equation}
\(\mathcal{I}_{ip}[\gamma] \geq 0\) for any such curve as well. Summarizing the above we have that, for any \(\mathbf{x} \in \mathbb{R}^{n}\), each \(\gamma \in \mathcal{D}_{\mathbf{x}}\) can be represented by a continuous curve in \(C_{b}^{0} (I,\mathbb{R}^{n})\) (also called \(\gamma\)) for which the ip action functional is finite and bounded from below,
\begin{subequations}\label{eq:310}
\begin{equation}\label{eq:310a}
\mathcal{I}_{ip} : \mathcal{D}_{\mathbf{x}} \rightarrow \mathbb{R}^{+} = [0,\infty),
\end{equation}
and for which the boundary conditions
\begin{align}
\lim_{t \rightarrow -\infty}{\gamma (t)} &= (0, \dotsc , 0)\label{eq:310b}\\
\intertext{and}
\lim_{t \nearrow 0}{\gamma (t)} &= (x^{1}, \dotsc , x^{n}) = \mathbf{x} \in \mathbb{R}^{n}\label{eq:310c}
\end{align}
\end{subequations}
are `built in'.

If a minimizer for \(\mathcal{I}_{ip}\) exists for each \(\mathbf{x} \in \mathbb{R}^{n}\) we shall (tentatively) identify the sought-after `fundamental solution' \(S_{0}(\mathbf{x})\) by setting
\begin{equation}\label{eq:311}
S_{(0)}(\mathbf{x}) = \mathcal{I}_{ip}[\gamma_{\mathbf{x}}]
\end{equation}
where \(\gamma_{\mathbf{x}}\) is a minimizer corresponding to the chosen boundary point \(\mathbf{x}\).

Note that if a minimizing curve, corresponding to a given \(\mathbf{x}\), should exist but fail to be unique, the value of \(S_{(0)}(\mathbf{x})\) would nevertheless be well-defined. But such a failure of uniqueness of the minimizing curves, if it occurred, would strongly suggest a corresponding breakdown in the differentiability of the fundamental solution \(S_{(0)}(\mathbf{x})\).

The reason for this is that if \(S_{(0)}(\mathbf{x})\) were indeed a differentiable solution to the (ipze) Hamiltonian-Jacobi equation then its gradient at \(\mathbf{x}\) should yield the complementary, canonical `initial data', (\(\mathbf{x}, \mathbf{p} = \text{grad}\;{S_{(0)}(\mathbf{x})}\)), for the corresponding, minimizing solution curve `lifted' up to the phase space \(T^{*}\mathbb{R}^{n} \approx \mathbb{R}^{2n}\). But the gradient of a differentiable \(S_{(0)}(\mathbf{x})\) would not have the multiple values needed to accommodate multiple minimizing solution curves. To avoid this potential difficulty we shall later impose a certain `convexity' condition on the potential function \(V(x^{1}, \dotsc , x^{n})\) that will suffice to globally exclude the occurrence of multiple minimizers. Then, with some further work we shall be able to prove that \(S_{(0)} : \mathbb{R}^{n} \rightarrow \mathbb{R}\) is in fact a smooth (i.e., \(C^{\infty}\)) function and that it satisfies the ipze Hamilton-Jacobi equation globally on \(\mathbb{R}^{n}\).

For the purpose of appealing to some standard results in the calculus of variations it will be convenient to work with a slight reformulation of the problem defined above. If the potential function \(A (x^{1}, \dotsc , x^{n})\) were taken to vanish then it is straightforward to show that the curve \(\zero{\gamma}_{\mathbf{x}} \in \mathcal{D}_{\mathbf{x}}\) defined by
\begin{equation}\label{eq:312}
\zero{\gamma}_{\mathbf{x}}(t) = (x^{1}e^{\omega_{1}t}, \dotsc , x^{n}e^{\omega_{n}t})
\end{equation}
would be the (unique) minimizer for the corresponding linear problem having boundary data \(\mathbf{x} \in \mathbb{R}^{n}\). For the nonlinear problem, when \(A (x^{1}, \dotsc , x^{n})\) is reinstated, one can express any \(\gamma \in \mathcal{D}_{\mathbf{x}}\) as
\begin{equation}\label{eq:313}
\gamma = \zero{\gamma}_{\mathbf{x}} + u
\end{equation}
where \(u\) is a curve in the fixed (i.e., independent of \(\mathbf{x}\)) \textit{linear} space
\begin{equation}\label{eq:314}
\mathcal{D} := \left\lbrace u \in H^{1} (I,\mathbb{R}^{n})\; \vert\; I = (-\infty,0],\; \lim_{t \nearrow 0}{u(t)} = (0, \dotsc ,0)\right\rbrace
\end{equation}
characterized by vanishing endpoint data. For arbitrary \(\mathbf{x} \in \mathbb{R}^{n}\) we can thus seek minimizers \(u \in \mathcal{D}\) for the functional
\begin{equation}\label{eq:315}
\tilde{\mathcal{I}}_{ip,\mathbf{x}}[u] := \mathcal{I}_{ip}[\zero{\gamma}_{\mathbf{x}} + u]
\end{equation}
rather than (equivalently) seeking minimizes \(\gamma \in \mathcal{D}_{\mathbf{x}}\) for the original functional \(\mathcal{I}_{ip}\). Writing \(u : I \rightarrow \mathbb{R}^{n}\) in component form, \(t \longmapsto \left(u^{1}(t), \dotsc , u^{n}(t)\right)\) one can easily verify that
\begin{equation}\label{eq:316}
\begin{split}
\tilde{\mathcal{I}}_{ip,\mathbf{x}}[u] &= \frac{1}{2} m \sum_{i=1}^{n} \omega_{i} (x^{i})^{2}\\
&+ \int_{-\infty}^{0} \left\lbrace \frac{1}{2} m \sum_{i=1}^{n} \left\lbrack \left(\dot{u}^{i}(t)\right)^{2} + \omega_{i}^{2} \left(u^{i}(t)\right)^{2}\right\rbrack\right.\\
& \left.\vphantom{\frac{1}{2} m \sum_{i=1}^{n} \left\lbrace \left(\dot{U}^{i}(t)\right)^{2} + \omega_{i}^{2} \left(U^{i}(t)\right)^{2}\right\rbrace} + A \left(x^{1}(t), \dotsc , x^{n}(t)\right)\right\rbrace\; dt
\end{split}
\end{equation}
where \(x^{i}(t) = x^{i}e^{\omega_{i}t} + u^{i}(t)\). The first term on the right hand side of (\ref{eq:316}) is independent of \(u\) and can thus be disregarded in minimizing the action for fixed \(\mathbf{x}\). Terms that would be bilinear in \(x^{i}e^{\omega_{i}t}\) and \(u^{i}(t)\) are absent by virtue of the fact that \(\zero{\gamma}_{\mathbf{x}}(t) = (x^{1}e^{\omega_{1}t}, \dotsc , x^{n}e^{\omega_{n}t})\) is a critical curve (in fact a minimizer) for the linearized problem. To simplify the notation slightly let us define
\begin{equation}\label{eq:317}
\mathcal{S}_{\mathbf{x}}[u] := \tilde{\mathcal{I}}_{ip,\mathbf{x}}[u] = \mathcal{I}_{ip} [\zero{\gamma}_{\mathbf{x}} + u]
\end{equation}
and seek minimizers for \(\mathcal{S}_{\mathbf{x}}\) within this linear (Hilbert) space \(\mathcal{D}\).

Standard results for the `direct method' in the calculus of variations yield the following fundamental theorem [Theorem 1.2.5 of Ref.~\cite{Blanchard1992}]:
\begin{quote}
\textit{Let \(X\) be a reflexive Banach space and \(M \subset X\) a weakly (sequentially) closed subset. Let \(f:M \rightarrow \mathbb{R}\) be a coercive, weakly (sequentially) lower semicontinuous function on \(M\). Then \(m(f) = \inf_{x \in M}{f(x)}\) is finite and is attained at a point \(x_{0} \in M\); i.e., \(m(f) = f(x_{0})\).}
\end{quote}
For our problem we take \(M = X = \mathcal{D}\), the (reflexive) Hilbert space defined above, and seek to verify the coercivity and (weak, sequential) lower semicontinuity of \begin{equation}\label{eq:318}
\mathcal{S}_{\mathbf{x}} : \mathcal{D} \rightarrow \mathbb{R}.
\end{equation}
for any fixed \(\mathbf{x} \in \mathbb{R}^{n}\).

Coercivity of \(\mathcal{S}_{\mathbf{x}}\) on \(\mathcal{D}\) follows if \(\vert\!\vert u(\cdot)\vert\!\vert_{H^{1}(I,\mathbb{R}^{n})} \rightarrow \infty\) always implies \(\mathcal{S}_{\mathbf{x}}[u] \rightarrow \infty\). But \(\mathcal{S}_{\mathbf{x}}[u]\) can be written as
\begin{equation}\label{eq:319}
\begin{split}
\mathcal{S}_{\mathbf{x}}[u] &= \frac{1}{2} m \sum_{i=1}^{n} \omega_{i}(x^{i})^{2}\\
&+ \frac{1}{2} m \vert\!\vert u(\cdot)\vert\!\vert^{2}_{H^{1} (I,\mathbb{R}^{n})} + \int_{-\infty}^{0} A \left(x^{1}(t), \dotsc , x^{n}(t)\right)\; dt
\end{split}
\end{equation}
where \(x^{i}(t) = x^{i}e^{\omega_{i}t} + u^{i}(t)\). Clearly a sufficient condition that ensures coercivity (but allows some negativity in the function \(A (x^{1}, \dotsc , x^{n})\)) is the requirement that
\begin{equation}\label{eq:320}
A(x^{1}, \dotsc ,x^{n}) \geq -\frac{1}{2} m \sum_{i=1}^{n} \lambda_{i}^{2} (x^{i})^{2}
\end{equation}
for some constants \(\lbrace\lambda_{i}\rbrace\) such that \(\lambda_{i}^{2} < \omega_{i}^{2}\; \forall\; i \in \lbrace 1, \dotsc , n\rbrace\). Simple examples where this inequality could be arranged to hold nontrivially include polynomials involving indefinite terms of less than the maximal order. We shall not attempt here to characterize the most general coercive choice for \(A\) but simply assume that it has been chosen to satisfy this condition. For Fr\'{e}chet differentiable functionals (such as \(\mathcal{S}_{\mathbf{x}}\)) a useful criterion for coercivity is provided by Lemma 2.5.2 of Ref.~\cite{Blanchard1992}.

If, as is true in our case, the functional in question is twice continuously Fr\'{e}chet differentiable on its domain Banach space, then a sufficient condition for its weak lower semicontinuity (derived in Lemma 2.5.1 of Ref.~\cite{Blanchard1992}) is the (non-strict) positivity of its second Fr\'{e}chet derivative (i.e., \(D^{2} f(x)\; (h,h) \geq 0\) in the notation of Ref.~\cite{Blanchard1992}) at an arbitrary point in the domain of \(f\). For our problem the second Fr\'{e}chet derivative of \(\mathcal{S}_{\mathbf{x}}\) is given explicitly by
\begin{equation}\label{eq:321}
\begin{split}
D^{2}\mathcal{S}_{\mathbf{x}}[u] & \cdot (h,h)\\
&= \int_{-\infty}^{0} dt\; \left\lbrace m \sum_{i=1}^{n} \left\lbrack\left( \dot{h}^{i}(t)\right)^{2} + \omega_{i}^{2} \left( h^{i}(t)\right)^{2}\right\rbrack\right.\\
& \left.\vphantom{m \sum_{i=1}^{n} \left\lbrack\left( \dot{h}^{i}(t)\right)^{2} + \omega_{i}^{2} \left( h^{i}(t)\right)^{2}\right\rbrack} + \sum_{i,j=1}^{n} \frac{\partial^{2}A}{\partial x^{i}\partial x^{j}} \left(x^{1}(t), \dotsc , x^{n}(t)\right)\; h^{i}(t) h^{j}(t)\right\rbrace
\end{split}
\end{equation}
where \(u \in \mathcal{D}\) and \(h \in T_{u}\mathcal{D} \approx \mathcal{D}\) are arbitrary elements of \(\mathcal{D}\) and its tangent space at \(u\) respectively and, as before, \(x^{i}(t) = x^{i}e^{\omega_{i}t} + u^{i}(t)\). Clearly, a sufficient condition to ensure the positivity of this expression is the pointwise \textit{convexity} requirement
\begin{equation}\label{eq:322}
\begin{split}
\sum_{i,j=1}^{n} & \frac{\partial^{2} V}{\partial x^{i}\partial x^{j}}\; (x^{1},\dotsc , x^{n}) \xi^{i}\xi^{j}\\
&= \sum_{i=1}^{n} m\omega_{i}^{2} (\xi^{i})^{2} + \sum_{i,j=1}^{n} \frac{\partial^{2} A}{\partial x^{i}\partial x^{j}} (x^{1}, \dotsc ,x^{n}) \xi^{i}\xi^{j}\\
& \geq 0
\end{split}
\end{equation}
\(\forall\; \mathbf{x} = (x^{1}, \dotsc ,x^{n}) \in \mathbb{R}^{n}\) and all \(\xi = (\xi^{1}, \dotsc ,\xi^{n}) \in \mathbb{R}^{n}\). We shall henceforth assume that \(A (x^{1}, \dotsc ,x^{n})\) satisfies this condition as well as that needed for coercivity. It follows from the fundamental theorem quoted above that a minimizer for \(\mathcal{S}_{\mathbf{x}}\) always exists.

Uniqueness of the minimizer will follow if we can show that \(\mathcal{S}_{\mathbf{x}}\) (which is globally defined on the convex space \(\mathcal{D}\)) is strictly convex, i.e., that for any \(u,v \in \mathcal{D}\), \(u \neq v\), \(0 < \lambda < 1 \Rightarrow\)
\begin{equation}\label{eq323}
\mathcal{S}_{\mathbf{x}} (\lambda u + (1 - \lambda) v) < \lambda\mathcal{S}_{\mathbf{x}}(u) + (1 - \lambda) \mathcal{S}_{\mathbf{x}} (v)
\end{equation}
(c.f., Theorem 1.1.3 of Ref.~\cite{Blanchard1992}). From Theorem 2.6.1 of this same reference however one knows that, for a continuously Fr\'{e}chet differentiable functional (such as \(\mathcal{S}_{\mathbf{x}}: \mathcal{D} \rightarrow \mathbb{R}\)), strict convexity is equivalent to strict monotonicity of the Fr\'{e}chet derivative which, for our problem, corresponds to the following (strict) inequality,
\begin{equation}\label{eq:324}
D\mathcal{S}_{\mathbf{x}}(u) \cdot (u - v) - D\mathcal{S}_{\mathbf{x}} (v) \cdot (u - v) > 0
\end{equation}
holding for all \(u,v \in \mathcal{D}\) whenever \(u \neq v\). Written out explicitly this is equivalent to the requirement that
\begin{equation}\label{eq:325}
\begin{split}
\int_{-\infty}^{0} dt\; & \left\lbrace m \sum_{i=1}^{n} \left(\dot{u}^{i}(t) - \dot{v}^{i}(t)\right)^{2}\right.\\
&+ \left.\vphantom{m \sum_{i=1}^{n} \left(\dot{U}^{i}(t) - \dot{V}^{i}(t)\right)^{2}} \sum_{i=1}^{n} \left(\frac{\partial V}{\partial x^{i}} \left(\zero{\gamma}_{\mathbf{x}}(t) + u(t)\right) - \frac{\partial V}{\partial x^{i}} \left(\zero{\gamma}_{\mathbf{x}}(t) + v(t)\right)\right)\; \left(u^{i}(t) - v^{i}(t)\right)\right\rbrace\\
& > 0
\end{split}
\end{equation}
whenever \(u - v \neq 0\). Since elements of \(\mathcal{D}\) vanish at both endpoints the first integral is strictly positive for \(u \neq v\) and thus a sufficient condition for the strict convexity of \(\mathcal{S}_{\mathbf{x}}\) is that
\begin{equation}\label{eq:326}
\sum_{i=1}^{n} \left(\frac{\partial V}{\partial x^{i}} (\eta + \alpha) - \frac{\partial V}{\partial x^{i}} (\eta + \beta)\right)\; (\alpha^{i} - \beta^{i}) \geq 0
\end{equation}
hold for arbitrary \(\eta, \alpha, \beta \in \mathbb{R}^{n}\). But, noting that
\begin{equation}\label{eq:327}
\begin{split}
\sum_{i=1}^{n} & \left\lbrack\frac{\partial V}{\partial x^{i}} (\eta + \alpha) - \frac{\partial V}{\partial x^{i}} (\eta + \beta)\right\rbrack\; (\alpha^{i} - \beta^{i}),\\
&= \sum_{i=1}^{n} (\alpha^{i} - \beta^{i})\; \int_{0}^{1} d\lambda\; \frac{d}{d\lambda} \left(\frac{\partial V}{\partial x^{i}} (\eta + \lambda\alpha + (1 - \lambda)\beta)\right)\\
&= \sum_{i,j=1}^{n} (\alpha^{i} - \beta^{i})\; \int_{0}^{1} d\lambda\; \frac{\partial^{2}V}{\partial x^{i}\partial x^{j}} \left(\eta + \lambda\alpha + (1 - \lambda)\beta\right) (\alpha^{j} - \beta^{j})\\
&= \int_{0}^{1} d\lambda \sum_{i,j = 1}^{n} \frac{\partial^{2}V}{\partial x^{i}\partial x^{j}} \left(\eta + \lambda\alpha + (1 - \lambda)\beta\right) (\alpha^{i} - \beta^{i}) (\alpha^{j} - \beta^{j})
\end{split}
\end{equation}
we see that this follows automatically from our previous condition on \(V\) that
\begin{equation}\label{eq:328}
\sum_{i,j=1}^{n} \frac{\partial V}{\partial x^{i}\partial x^{j}} (x^{1}, \dotsc ,x^{n}) \xi^{i}\xi^{j} \geq 0
\end{equation}
\(\forall\; (x^{1}, \dotsc ,x^{n}) \in \mathbb{R}^{n}\) and all \((\xi^{1}, \dotsc , \xi^{n}) \in \mathbb{R}^{n}\). Thus the foregoing sufficient condition for the weak lower semicontinuity of \(\mathcal{S}_{\mathbf{x}}\) guarantees as well the uniqueness of the minimizer for arbitrarily chosen \(\mathbf{x} \in \mathbb{R}^{n}\). For this reason we can write, without ambiguity, \(u_{\mathbf{x}}\) for the unique minimizer of \(\mathcal{S}_{\mathbf{x}}\) and thus define a real-valued function \(S_{(0)}\) on \(\mathbb{R}^{n}\) by setting
\begin{equation}\label{eq:329}
S_{(0)}(\mathbf{x}) := \mathcal{S}_{\mathbf{x}}(u_{\mathbf{x}}).
\end{equation}
\(S_{(0)}\) is, of course, our candidate for the sought-after fundamental solution to the ipze Hamilton-Jacobi equation but, at this point, we don't even know if \(S_{(0)}\) is differentiable. We shall prove below that it is actually smooth and that it does indeed solve the Hamilton-Jacobi equation but, to lay the groundwork for this, we first need to show that the minimizing curves \(\left\lbrace u_{\mathbf{x}} \in \mathcal{D} | \mathbf{x} \in \mathbb{R}^{n}\right\rbrace\) are smooth solutions of the (ip) Euler Lagrange equations, each having vanishing (ip) energy.

We already know that each minimizer \(u_{\mathbf{x}}\) can be represented by a bounded, continuous curve on \(I = (-\infty,0)\) that, furthermore, has the asymptotic behavior
\begin{subequations}\label{eq:330}
\begin{align}
\lim_{t\rightarrow -\infty} u_{\mathbf{x}} &= (0, \dotsc ,0)\label{eq:330a}\\
\lim_{t\nearrow 0} u_{\mathbf{x}} &= (x^{1}, \dotsc ,x^{n})\label{eq:330b}
\end{align}
\end{subequations}
but, as an element of \(H^{1}(I,\mathbb{R}^{n})\), it need not, a priori, be smooth enough to satisfy the Euler Lagrange equations,
\begin{equation}\label{eq:331}
m \frac{d^{2}x^{i}}{dt^{2}} = -\frac{\partial V_{ip}}{\partial x^{i}} = \frac{\partial V}{\partial x^{i}}.
\end{equation}
The machinery needed to prove that our minimizers are indeed actual smooth curves on \(I\) (i.e., elements of \(C^{\infty} (I,\mathbb{R}^{n})\)) that further do satisfy the Euler Lagrange equations is provided, in technical detail, by the argument in Section 4.1 of Ref.~\cite{Buttazzo2005}. The basic tool needed for this is Theorem 4.1 of the foregoing reference but, since several of the hypotheses of this theorem are not satisfied by our problem, we first need to show how the proof given in this reference can be modified to apply to the setup dealt with here.

First of all the foregoing reference deals only with minimizers defined on bounded intervals of the type \(I = (a,b)\) whereas we need to handle the case for which \(a \rightarrow -\infty\). But, taking \(b = 0\), we can evaluate our minimizer \(u_{\mathbf{x}}\) at any point \(a \in (-\infty,0)\) and restrict the domain of definition of \(u_{\mathbf{x}}\) to the subinterval \(I_{a} = (a,0)\) whereon \(u_{\mathbf{x}}\) has the boundary values given by
\begin{subequations}\label{eq:332}
\begin{align}
\lim_{t\searrow a}{u_{\mathbf{x}}(t)} &= u_{\mathbf{x}} (a) := \mathbf{y}\label{eq:332a}\\
\lim_{t\nearrow 0}{u_{\mathbf{x}}(t)} &= \mathbf{x}.\label{eq:332b}
\end{align}
\end{subequations}
It is now easy to see that this curve must be the (unique) minimizer for the corresponding restricted variational integral
\begin{equation}\label{eq:333}
\mathcal{I}_{ip,(a,0)}[\gamma] := \int_{a}^{0} L_{ip} \left(x^{1}(t), \dotsc , x^{n}(t),\dot{x}^{1}(t), \dotsc ,\dot{x}^{n}(t)\right)\; dt
\end{equation}
when the curves \(\gamma \in H^{1} \left((a,0),\mathbb{R}^{n}\right)\) are constrained to have the boundary values
\begin{subequations}\label{eq:334}
\begin{align}
\lim_{t\searrow a}{\gamma(t)} &= \mathbf{y}\label{eq:334a}\\
\lim_{t\nearrow 0}{\gamma(t)} &= \mathbf{x}.\label{eq:334b}
\end{align}
\end{subequations}
If this were not the case we could replace the segment of \(u_{\mathbf{x}}\) along the subinterval \((a,0)\) by the true minimizer for this segment to get a different curve \(u_{\mathbf{x}}^{*} : (-\infty,0) \rightarrow \mathbb{R}^{n}\), still in \(H^{1}(I,\mathbb{R}^{n})\) with the original boundary values, for which the value of \(\mathcal{S}_{\mathbf{x}}\) is thereby decreased,
\begin{equation}\label{eq:335}
\mathcal{S}_{\mathbf{x}} [u_{\mathbf{x}}^{*}] < \mathcal{S}_{\mathbf{x}} [u_{\mathbf{x}}].
\end{equation}
But this is impossible since, by construction, \(u_{\mathbf{x}}\) was the actual minimizer for the original problem. Thus we can always `localize' our problem to subintervals of the form treated by the given theorem in Ref.~\cite{Buttazzo2005}.

But this same theorem has hypotheses that, in our context, constrain the growth of the potential function \(V (x^{1}, \dotsc ,x^{n})\) and its gradient \(\nabla V (x^{1}, \dotsc ,x^{n})\) for large \(|\mathbf{x}|\) (c.f., hypotheses (i) and (ii) of Theorem 4.1). Such constraints would normally not be satisfied by the potential functions we wish to consider. But since our minimizers are a priori bounded we are free to smoothly modify their associated potential energy functions outside of sufficiently large balls in \(\mathbb{R}^{n}\) so that the modified potentials satisfy the needed growth restrictions while the range of the original minimizer remains entirely within the unmodified domain. The original minimizing curve may now only provide a \textit{local} minimizer for the modified variational problem (since a `true' minimizer might need to traverse a modified portion of the domain of the potential function). But a \textit{local minimizer} is all that is required by the remaining hypotheses of the cited theorem.

In summary, with the modifications sketched above, one can apply Theorem 4.1 of Ref.~\cite{Buttazzo2005} to the correspondingly modified version of our variational problem and use it to conclude that each of our original minimizers, \(u_{\mathbf{x}} : (-\infty,0) \rightarrow \mathbb{R}^{n}\), is a globally smooth solution to the Euler Lagrange equations defined on \(I = (-\infty,0)\). In fact if the potential energy should happen to be real analytic, instead of merely smooth, then the minimizers themselves would be real analytic \cite{Buttazzo2005} instead of merely \(C^{\infty}\).

Any smooth solution to the Euler Lagrange equations has conserved (ip) energy but we shall show momentarily that this conserved energy actually vanishes for any minimizer \(u_{\mathbf{x}}\), i.e., that
\begin{equation}\label{eq:336}
E_{ip} \left(x^{1}(t),\dotsc ,x^{n}(t), \dot{x}^{1}(t), \dotsc ,\dot{x}^{n}(t)\right) = 0
\end{equation}
where \(x^{i}(t) = x^{i}e^{\omega_{i}t} + u_{\mathbf{x}}^{i}(t)\) with \(E_{ip}\) defined by Eq.~(\ref{eq:204}).

This additional regularity of the solution curves follows from the fact that, since the minimizers satisfy the Euler Lagrange equations and since these curves lie in \(H^{1}(I,\mathbb{R}^{n})\), one can apply an argument similar to that leading to inequality (\ref{eq:308}) to show that the force term in the Euler Lagrange equations (i.e., the right hand side of Eq.~(\ref{eq:331})) has finite \(L^{2}(I,\mathbb{R}^{n})\)-norm. Thus the acceleration is square integrable over \(I\) and hence the solution curves actually belong to \(H^{2}(I,\mathbb{R}^{n})\). From the `vanishing-at-infinity' result cited earlier it now follows that the velocity, \(\frac{d}{dt} \mathbf{x}(t) = \frac{d\zero{\gamma}_{\mathbf{x}}(t)}{dt} + \frac{du(t)}{dt}\), of any such minimizer vanishes as \(t \rightarrow -\infty\) and thus that the conserved ip energy of each such curve has the value zero.

By successively differentiating the Euler Lagrange equations with respect to \(t\) and applying H\"{o}lder's inequality and the Sobolev embedding theorem inductively to the higher order `forcing terms' generated thereby, one can easily show that every minimizing solution curve actually belongs to \(H^{s}(I,\mathbb{R}^{n})\) for arbitrary (integral) \(s \geq 0\). Thus the acceleration and all the (successively computed) higher \(t\) derivatives of an arbitrary minimizer vanish as \(t \rightarrow -\infty\), much as for the minimizer,
\begin{equation*}
\zero{\gamma}_{\mathbf{x}}(t) = \left(x^{1}e^{\omega_{1}t}, \dotsc , x^{n}e^{\omega_{n}t}\right),
\end{equation*}
of the corresponding linearized problem.

\subsection{Smoothness and Asymptotics of the Fundamental Solution}
\label{subsec:smoothness}

To show that \(S_{(0)} : \mathbb{R}^{n} \rightarrow \mathbb{R}\) (defined by (\ref{eq:329})) is smooth (i.e., \(C^{\infty}\)) we proceed by first showing that the minimizing solution curves \(u_{\mathbf{x}} : I \rightarrow \mathbb{R}^{n}\) depend smoothly upon \(\mathbf{x}\). It will then follow that
\begin{equation}\label{eq:337}
S_{(0)} (\mathbf{x}) := \mathcal{S}_{\mathbf{x}} (u_{\mathbf{x}}) = \mathcal{I}_{ip} [\gamma_{\mathbf{x}}]
\end{equation}
is defined by the composition of smooth maps and thus is smooth.

The space
\begin{equation}\label{eq:338}
\begin{split}
\mathcal{M}^{s} &:= \left\lbrace\zero{\gamma}_{\mathbf{x}} \in H^{s} (I,\mathbb{R}^{n}) \vert \mathbf{x} \in \mathbb{R}^{n}\right.,\\
\zero{\gamma}_{\mathbf{x}} &= \left.\vphantom{\zero{\gamma}_{\mathbf{x}} \in H^{S} (I,\mathbb{R}^{n}) \vert \mathbf{x} \in \mathbb{R}^{n}} (x^{1} e^{\omega_{1}t}, \dotsc ,x^{n} e^{\omega_{n}t}), s \geq 1\; \text{(integral)}\right\rbrace
\end{split}
\end{equation}
is an n-dimensional, closed subspace of \(H^{s} (I,\mathbb{R}^{n})\) comprised of the minimizers for the linearized problem. Let
\begin{equation}\label{eq:339}
\begin{split}
\mathcal{D}^{s} &:= \left\lbrace u \in H^{s} (I,\mathbb{R}^{n}) \vert \lim_{t \nearrow 0}{u(t)} = (0, \dotsc ,0)\right.,\\
& \left.\vphantom{\lbrace u \in H^{S} (I,\mathbb{R}^{n}) \vert \lim_{t \nearrow 0}{u(t)} = (0, \dotsc ,0)}s \geq 1\; \text{(integral)}\right\rbrace
\end{split}
\end{equation}
designate the space of \(H^{s}\)-curves on \(I = (-\infty,0]\) having vanishing boundary data. Now, for any \(s \geq 3\), define the Euler Lagrange map
\begin{equation}\label{eq:340}
\mathcal{E}^{s} : \mathcal{M}^{s} \times \mathcal{D}^{s} \rightarrow \Sigma^{s-2},
\end{equation}
where
\begin{equation}\label{eq:341}
\Sigma^{s-2} := \left\lbrace \sigma \in H^{s-2} (I,\mathbb{R}^{n})\right\rbrace,
\end{equation}
by
\begin{equation}\label{eq:342}
\begin{split}
\mathcal{E}^{s} (\zero{\gamma}_{\mathbf{x}},u)^{i}(t) &:= m \frac{d^{2}}{dt^{2}} (\zero{\gamma}_{\mathbf{x}}^{i}(t) + u^{i}(t)) - \frac{\partial V}{\partial x^{i}} (\zero{\gamma}_{\mathbf{x}}(t) + u(t))\\
&= m \frac{d^{2}u^{i}(t)}{dt^{2}} - \left\lbrace m \omega_{i}^{2} u^{i}(t) + \frac{\partial A}{\partial x^{i}} (\zero{\gamma}_{\mathbf{x}}(t) + u(t))\right\rbrace.
\end{split}
\end{equation}
Using the tools discussed above it is straightforward to verify that the right hand side of this formula is indeed an element of \(\Sigma^{s-2}\). In particular, note that, for the special case of polynomial potentials, the fact that the `forcing term' \(\frac{\partial A}{\partial x^{i}} (\zero{\gamma}_{\mathbf{x}}(\cdot) + u(\cdot))\) lies in \(H^{s}(I,\mathbb{R}^{n})\) (hence, a fortiori in \(H^{s-2}(I,\mathbb{R}^{n})\)) follows from the fact that \(H^{s}\)-maps in one-dimension from an algebra under pointwise multiplication for any \(s \geq 1\) \cite{Buttazzo2005b}. Using the same tools one also verifies that \(\mathcal{E}^{s}\) is a Fr\'{e}chet smooth (i.e., \(C^{\infty}\)) functional of its arguments.

From the main results of this section we know that the equation
\begin{equation}\label{eq:343}
\mathcal{E}^{s} (\zero{\gamma}_{\mathbf{x}},u) (\cdot) = 0
\end{equation}
has a unique solution, \(u_{\mathbf{x}} \in \mathcal{D}^{s} \subset \mathcal{D}\), for any \(\zero{\gamma}_{\mathbf{x}} \in \mathcal{M}^{s}\; \forall\; s \geq 1\). To show that \(u_{\mathbf{x}}\) depends smoothly on \(\zero{\gamma}_{\mathbf{x}}\) (hence smoothly on \(\mathbf{x}\) since \(\zero{\gamma}_{\mathbf{x}}\) is linear in \(\mathbf{x}\)) we employ the Banach space version of the implicit function theorem \cite{Abraham1983}. The Fr\'{e}chet derivative of \(\mathcal{E}^{s}(\zero{\gamma}_{\mathbf{x}},u)\) with respect to its second argument, evaluated at \(u = u_{\mathbf{x}}\), is given by
\begin{equation}\label{eq:344}
\begin{split}
&\left\lbrace D_{2} \mathcal{E}^{s} (\zero{\gamma}_{\mathbf{x}},u_{\mathbf{x}}) \cdot h(t)\right\rbrace^{i}\\
&= m \frac{d^{2}h^{i}(t)}{dt^{2}} - m \omega_{i}^{2}h^{i}(t)\\
&- \sum_{j=1}^{n} \frac{\partial^{2}A}{\partial x^{i}\partial x^{j}} (\zero{\gamma}_{\mathbf{x}}(t) + u_{\mathbf{x}}(t))\; h^{j}(t)
\end{split}
\end{equation}
we need to show that the linear operator
\begin{equation}\label{eq:345}
D_{2} \mathcal{E}^{s} (\zero{\gamma}_{\mathbf{x}},u_{\mathbf{x}}) : T_{u_{\mathbf{x}}} \mathcal{D}^{s} \rightarrow T_{0} \Sigma^{s-2}
\end{equation}
defines an ismorphism of the relevant tangent spaces, \(T_{u_{\mathbf{x}}}\mathcal{D}^{s} \approx \mathcal{D}^{s}\) and \(T_{\mathbf{0}} \Sigma^{s-2} \approx \Sigma^{s-2}\), \(\forall\; s \geq 3\). In other words, for any \(\sigma \in \Sigma^{s-2}\) we need to show that the equation
\begin{equation}\label{eq:346}
D_{2}\mathcal{E}^{s} (\zero{\gamma}_{\mathbf{x}},u_{\mathbf{x}}) \cdot h(\cdot) = \sigma (\cdot)
\end{equation}
admits a unique solution
\begin{equation}\label{eq:347}
h_{\mathbf{x}}(\sigma(\cdot)) \in \mathcal{D}^{s} \approx T_{u_{\mathbf{x}}} \mathcal{D}^{s}.
\end{equation}

A straightforward method for proving this is provided again by the calculus of variations techniques employed above for the nonlinear problem but here specialized to the linear (inhomogeneous) equation given by (\ref{eq:346}). Note that Eq.~(\ref{eq:346}) is the Euler Lagrange equation for the (inhomogeneous, `second variation') action integral, \(\mathcal{J}_{\mathbf{x},\sigma}[h]\), defined by
\begin{equation}\label{eq:348}
\begin{split}
\mathcal{J}_{\mathbf{x},\sigma}[h] &= \frac{1}{2} D_{u}^{2} \mathcal{S}_{\mathbf{x}} [u_{\mathbf{x}}] \cdot (h,h)\\
&\quad + \int_{-\infty}^{0} \sum_{i=1}^{n} \sigma^{i} (t) h^{i} (t)\; dt\\
&= \int_{-\infty}^{0} dt\; \left\lbrace\frac{1}{2} m \sum_{i=1}^{n} \left\lbrack\left(\frac{dh^{i}(t)}{dt}\right)^{2} + \omega_{i}^{2} \left(h^{i}(t)\right)^{2}\right\rbrack\right.\\
&\quad \left.\vphantom{\frac{1}{2} m \sum_{i=1}^{n} \left\lbrack\left(\frac{dh^{i}(t)}{dt}\right)^{2} + \omega_{i}^{2} \left(h^{i}(t)\right)^{2}\right\rbrack}+ \frac{1}{2} \sum_{i,j = 1}^{n} \frac{\partial^{2}A}{\partial x^{i}\partial x^{j}} (\zero{\gamma}_{\mathbf{x}}(t) + u_{\mathbf{x}}(t)) h^{i}(t)h^{j}(t)\right\rbrace + \int_{-\infty}^{0} dt\; \sum_{i = 1}^{n} \sigma^{i}(t)h^{i}(t).
\end{split}
\end{equation}
We can thus prove the needed isomorphism result by showing that \(\mathcal{J}_{\mathbf{x},\sigma}[h]\) always has a smooth minimizer, \(h_{\mathbf{x},\sigma}\), that satisfies the Euler Lagrange equation (\ref{eq:346}) and then showing that such a solution is always unique.

Note, first of all, that the convexity requirement for \(V(x^{1}, \dotsc ,x^{n})\) given by inequality (\ref{eq:322}) implies, when evaluated along an arbitrary minimizer, \(u_{\mathbf{x}} + \zero{\gamma}_{\mathbf{x}}\), that
\begin{equation}\label{eq:349}
\frac{1}{2} m \sum_{i=1}^{n} \chi^{2}_{i} (\xi^{i})^{2} \geq
\frac{1}{2} \sum_{i,j=1}^{n} \frac{\partial^{2}V}{\partial x^{i}\partial x^{j}} (\zero{\gamma}_{\mathbf{x}}(t) + u(t))\; \xi^{i}\xi^{j}
\geq \frac{1}{2} m \sum_{i=1}^{n} \mu_{i}^{2} (\xi^{i})^{2}
\end{equation}
for some constants \(\chi_{i} \geq \mu_{i} > 0\), \(i \in \lbrace 1, \dotsc ,n\rbrace\) and \(\forall\; t \in I\). It follows that the terms in \(\mathcal{J}_{\mathbf{x},\sigma}[h]\) that are quadratic in \(h\) are bounded above and below by (strictly positive) constant multiples of the squared \(H^{1}\)-norm of \(h\). Using H\"{o}lder's inequality on the term linear in \(h\) it is easy to show that \(\mathcal{J}_{\mathbf{x},\sigma}[h]\) is finite and bounded from below \(\forall\; h \in \mathcal{D}\). Coercivity of \(\mathcal{J}_{\mathbf{x},\sigma}[h]\) follows from this same result and the weak lower semicontinuity of this functional is easily proven by an argument that parallels that given above for the nonlinear problem. Thus, applying the fundamental existence theorem cited above, it follows that a minimizer always exists.

For a source of \(\sigma \in \Sigma^{s-2}\), with \(s \geq 3\), one can again apply the regularity arguments given in Ref.~\cite{Buttazzo2005} to prove that the minimizer \(h_{\mathbf{x},\sigma}\) so obtained actually belongs to \(C^{2}(I,\mathbb{R}^{n})\) and satisfies the Euler Lagrange equation (\ref{eq:346}) in the classical sense. By appealing to the smoothness of \(V(x^{1}, \dotsc ,x^{n})\) one can now, upon differentiating Eq.~(\ref{eq:346}) arbitrarily many times, easily show that this minimizer lies in \(C^{k}(I,\mathbb{R}^{n})\) for arbitrary \(k\) and thus is smooth. For an analytic potential function it would further follow, from standard results on ordinary differential equations, that the minimizer is analytic. Finally, from the fact that \(\sigma\) and its derivatives up to order \(s - 2\) are square integrable, it follows from the Euler Lagrange equation and it corresponding derivatives (again appealing to H\"{o}lder's inequality and the Sobolev embedding results) that this minimizer belongs in fact to the Sobolev space \(\mathcal{D}^{s}\).

There is a slight subtlety however concerning uniqueness. We really want to show that the Euler Lagrange equation has a unique \textit{solution} lying in \(\mathcal{D}^{s}\) and not merely that the \textit{minimizer} is unique. However the difference between any two, hypothetically distinct solutions \(h_{1},h_{2} \in \mathcal{D}^{s}\) would satisfy the associated homogeneous equation
\begin{equation}\label{eq:350}
D_{2} \mathcal{E}^{s} (\zero{\gamma}_{\mathbf{x}},u_{\mathbf{x}}) \cdot (h_{1} - h_{2}) (\cdot) = 0.
\end{equation}
Using the explicit formula for \(D_{2} \mathcal{E}^{s} (\zero{\gamma}_{\mathbf{x}},u_{\mathbf{x}}) \cdot (h_{1} - h_{2})\), contracting Eq.~(\ref{eq:350}) with \((h_{1} - h_{2})\) and integrating over \(I\) one easily shows (using the vanishing of elements of \(\mathcal{D}^{s}\) at the endpoints to justify the integration by parts) that
\begin{equation}\label{eq:351}
\begin{split}
\int_{-\infty}^{0} dt\; & \left\lbrace m \sum_{i=1}^{n} \left\lbrack\frac{d}{dt} (h_{1}^{i}(t) - h_{2}^{i}(t))\right\rbrack^{2}\right.\\
&\quad + \left.\vphantom{m \sum_{i=1}^{n} \left\lbrack\frac{d}{dt} (h_{1}^{i}(t) - h_{2}^{i}(t))\right\rbrack^{2}} \sum_{i,j=1} \frac{\partial^{2}V}{\partial x^{i}\partial x^{j}} (\zero{\gamma}_{\mathbf{x}}(t) + u_{\mathbf{x}}(t)) (h_{1}^{i}(t) - h_{2}^{i}(t)) (h_{1}^{j}(t) - h_{2}^{j}(t))\right\rbrace\\
&= 0
\end{split}
\end{equation}
which, in view of (\ref{eq:349}), clearly implies that \(h_{1} = h_{2}\).

Assembling the results derived above we thus conclude that
\begin{equation}\label{eq:352}
S_{(0)}(\mathbf{x}) := \mathcal{S}_{\mathbf{x}}(u_{\mathbf{x}}) = \mathcal{I}_{ip}[\gamma_{\mathbf{x}}]
\end{equation}
is a smooth (i.e., \(C^{\infty}\)) function, globally defined on \(\mathbb{R}^{n}\). To show that \(S_{(0)}(\mathbf{x})\) satisfies the ipze Hamilton-Jacobi equation (\ref{eq:201a}) we need to calculate its gradient. Utilizing the facts that \(\gamma_{\mathbf{x}}(t) := \zero{\gamma}_{\mathbf{x}}(t) + u_{\mathbf{x}}(t)\) satisfies the Euler Lagrange equation and that \(\gamma_{\mathbf{x}}(t)\), together with its derivatives, vanish as \(t \searrow -\infty\), one computes that
\begin{equation}\label{eq:353}
\begin{split}
dS_{(0)}(\mathbf{x}) &= \sum_{i=1}^{n} \frac{\partial S_{(0)}(\mathbf{x})}{\partial x^{i}}\; dx^{i}\\
&= D_{\gamma} \left\lbrack\int_{-\infty}^{0} dt\; \left\lbrace\frac{1}{2} m \sum_{i=1}^{n} (\dot{\gamma}^{i}(t))^{2} + V(\gamma(t))\right\rbrace (\gamma_{\mathbf{x}}(\cdot))\right\rbrack \cdot (d\gamma_{\mathbf{x}}(\cdot))\\
&= \int_{-\infty}^{0} dt\; \left\lbrace\sum_{i=1}^{n} d\gamma_{\mathbf{x}}^{i}(t) \left\lbrack -m \ddot{\gamma}_{\mathbf{x}}^{i}(t) + \frac{\partial V}{\partial x^{i}} (\gamma (t))\right\rbrack\right\rbrace + \lim_{t\nearrow 0}{m} \sum_{i=1}^{n} \dot{\gamma}_{\mathbf{x}}^{i}(t) d\gamma_{\mathbf{x}}^{i}(t)\\
&= \sum_{i=1}^{n} m \dot{\gamma}_{\mathbf{x}}^{i}(0) dx^{i}
\end{split}
\end{equation}
wherein we have used the observation that \(d\; \gamma_{\mathbf{x}}^{i}(0) = d\; \zero{\gamma}_{\mathbf{x}}^{i}(0) = dx^{i}\). Thus, as expected, the gradient of \(S_{(0)}\) at \(\mathbf{x} = (x^{1}, \dotsc ,x^{n})\) is precisely the momentum of the solution curve `starting' at this point,
\begin{equation}\label{eq:354}
\frac{\partial S_{(0)}(\mathbf{x})}{\partial x^{i}} = m\dot{\gamma}_{\mathbf{x}}^{i}(0) := p_{\mathbf{x}}^{i}(0).
\end{equation}
Since we already know that each such solution curve has vanishing ip energy it follows that \(S_{(0)}\) satisfies the ipze Hamilton-Jacobi equation,
\begin{equation}\label{eq:355}
\sum_{i=1}^{n} \frac{1}{2m} \left(\frac{\partial S_{(0)}(\mathbf{x})}{\partial x^{i}}\right)^{2} - V(\mathbf{x}) = 0,
\end{equation}
globally on \(\mathbb{R}^{n}\). These solution curves realize the gradient semi-flow determined by \(S_{(0)}\) in the sense that the solutions of
\begin{equation}\label{eq:356}
\begin{split}
\frac{d\gamma^{i}(t)}{dt} &= \left.\frac{\partial H_{ip}}{\partial p_{i}} \right\vert_{\mathbf{p} = \nabla S_{(0)}(\gamma(t))}\\
&= \frac{1}{m} \frac{\partial S_{(0)}(\gamma(t))}{\partial x^{i}} := \frac{p^{i}(t)}{m}
\end{split}
\end{equation}
automatically satisfy the complementary Hamilton equation
\begin{equation}\label{eq:357}
\begin{split}
\frac{dp_{i}(t)}{dt} &= -\frac{\partial H_{ip}}{\partial x^{i}} (\nabla S_{(0)}(\gamma(t)),\gamma(t))\\
&= \frac{\partial}{\partial x^{i}} V(\gamma(t)).
\end{split}
\end{equation}
This follows directly from noting that
\begin{equation}\label{eq:358}
\begin{split}
\frac{dp_{i}(t)}{dt} &= \sum_{j=1}^{n} \frac{\partial^{2}S_{(0)}(\gamma(t))}{\partial x^{j}\partial x^{i}} \frac{1}{m} \frac{\partial S_{(0)}(\gamma(t))}{\partial x^{j}}\\
&= \frac{1}{2m} \frac{\partial}{\partial x^{i}} \sum_{j=1}^{n} \left(\frac{\partial S_{(0)}(\gamma(t))}{\partial x^{j}}\right)^{2}\\
&= \frac{\partial}{\partial x^{i}} V(\gamma(t))
\end{split}
\end{equation}
where, in the last step, we have appealed to the fact that \(S_{(0)}(\mathbf{x})\) satisfies the ipze Hamilton-Jacobi equation (\ref{eq:355}). The solution curves defined by this gradient semi-flow are, by construction, precisely those which, followed backwards in time, tend asymptotically to the (unstable) equilibrium lying at the peak of the inverted potential. As such they determine the so-called \textit{unstable manifold} of this equilibrium whereas the time-reversed solution curves determine the corresponding \textit{stable manifold}.

One can describe the resulting picture more geometrically in phase space by remarking that the equation
\begin{equation}\label{eq:359}
p_{i} = \frac{\partial S_{(0)}(\mathbf{x})}{\partial x^{i}}
\end{equation}
defines a smooth, global cross section of the cotangent bundle, \(T^{*}\mathbb{R}^{n} \approx \mathbb{R}^{n} \times \mathbb{R}^{n}\), which in fact, is a \textit{Lagrangian submanifold} of this canonical bundle. The last statement follows from the fact that the graph of the gradient of a smooth function on the base manifold, \(\mathbb{R}^{n}\), has the maximal allowed dimension for a Lagrangian submanifold (namely \(n\)) and that the canonical symplectic form,
\begin{equation}\label{eq:360}
\omega := \sum_{i=1}^{n} dx^{i} \land dp_{i},
\end{equation}
pulled back to this submanifold, automatically vanishes:
\begin{equation}\label{eq:361}
\sum_{i=1}^{n} dx^{i} \land d \left(\frac{\partial S_{(0)}(\mathbf{x})}{\partial x^{i}}\right) = \sum_{i,j=1}^{n} \frac{\partial^{2}S_{(0)}(\mathbf{x})}{\partial x^{i}\partial x^{j}}\; dx^{i} \land dx^{j} = 0.
\end{equation}
This Lagrangian submanifold is, of course, foliated by the solution curves (naturally lifted to \(T^{*}\mathbb{R}^{n}\)) defining the unstable manifold of the equilibrium at \((\mathbf{x},\mathbf{p}) = (\mathbf{0},\mathbf{0})\) whereas the `time-reversed' graph defined by
\begin{equation}\label{eq:362}
p_{i} = -\frac{\partial S_{(0)}(\mathbf{x})}{\partial x^{i}}
\end{equation}
is a complementary Lagrangian submanifold of \(T^{*}\mathbb{R}^{n}\) forliated by the solution curves of the stable manifold of this same equilibrium. Note that, since
\begin{equation}\label{eq:363}
\frac{1}{2m} \sum_{i=1}^{n} \left(\frac{\partial S_{(0)}(\mathbf{x})}{\partial x^{i}}\right)^{2} = V(\mathbf{x})
\end{equation}
and that \(V(\mathbf{x})\), by assumption, vanishes only at the origin, these two Lagrangian submanifolds intersect only at the equilibrium \((\mathbf{x},\mathbf{p}) = (\mathbf{0},\mathbf{0})\) which can be regarded as both a (trivial) stable and unstable solution curve.

We conclude this section by analyzing the leading terms in the Taylor expansion of \(S_{(0)}(\mathbf{x})\) about the origin. This will be needed for the study of the transport equations to be developed in the following section. From the definition of \(S_{(0)}(\mathbf{x})\) we clearly have that
\begin{equation}\label{eq:364}
\lim_{\mathbf{x}\rightarrow 0}{S_{(0)}(\mathbf{x})} = 0
\end{equation}
and, from the argument of the preceding paragraph, we also know that
\begin{equation}\label{eq:365}
\lim_{\mathbf{x}\rightarrow 0}{\frac{\partial S_{(0)}(\mathbf{x})}{\partial x^{i}}} = 0.
\end{equation}
Thus the Taylor expansion of \(S_{(0)}(\mathbf{x})\) about the origin begins with the second order term.

To calculate this term precisely recall that \(S_{(0)}(\mathbf{x})\) can be expressed as \begin{equation}\label{eq:366}
S_{(0)}(\mathbf{x}) = \mathcal{S}_{\mathbf{x}}(u_{\mathbf{x}}) = \tilde{\mathcal{I}}_{ip,\mathbf{x}}[u_{\mathbf{x}}]
\end{equation}
which, in view of Eq.~(\ref{eq:316}) results in
\begin{equation}\label{eq:367}
\begin{split}
S_{(0)}(\mathbf{x}) &= \frac{1}{2} m \sum_{i=1}^{n} \omega_{i}(x^{i})^{2}\\
&\quad + \int_{-\infty}^{0} \left\lbrace\frac{1}{2} m \sum_{i=1}^{n} \left\lbrace (\dot{u}_{\mathbf{x}}^{i}(t))^{2} + \omega_{i}^{2} (u_{\mathbf{x}}^{i}(t))^{2}\right\rbrack\right.\\
&\quad \left.\vphantom{\frac{1}{2} m \sum_{i=1}^{n} \left\lbrace (\dot{u}_{\mathbf{x}}^{i}(t))^{2} + \omega_{i}^{2} (u_{\mathbf{x}}^{i}(t))^{2}\right\rbrack} + A \left(\zero{\gamma}_{\mathbf{x}}(t) + u_{\mathbf{x}}(t)\right)\right\rbrace\; dt.
\end{split}
\end{equation}
We shall find that the Taylor expansion of the integral in this formula begins at third order and thus that
\begin{equation}\label{eq:368}
S_{(0)}(\mathbf{x}) = \frac{1}{2} m \sum_{i=1}^{n} \omega_{i}(x^{i})^{2} + O (|\mathbf{x}|^{3}).
\end{equation}

To see this we compute the gradient of expression (\ref{eq:367}) directly, exploiting the facts that \(u_{\mathbf{x}}(t)\) satisfies the Euler-Lagrange equation (\ref{eq:331}), vanishes at the upper endpoint of the domain of integration and has time derivative, \(\dot{u}_{\mathbf{x}}^{i}(t)\), vanishing at the lower endpoint,
\begin{equation}\label{eq:369}
u_{\mathbf{x}}^{i}(0) = 0,\; \lim_{t\searrow -\infty}{\dot{u}_{\mathbf{x}}(t)} = 0.
\end{equation}
The computation gives
\begin{equation}\label{eq:370}
\begin{split}
\partial_{j} S_{(0)}(\mathbf{x}) &= m\omega_{j}x^{j} + \int_{-\infty}^{0} dt\; \left\lbrace\frac{\partial A}{\partial x^{j}} (\zero{\gamma}_{\mathbf{x}}(t) + u_{\mathbf{x}}(t)) e^{\omega_{j}t}\right\rbrace\\
&\quad - \lim_{t\searrow -\infty}{\left\lbrace m \sum_{i=1}^{n}(\dot{u}_{\mathbf{x}}(t) \partial_{j}u_{\mathbf{x}}^{i}(t))\right\rbrace}
\end{split}
\end{equation}
and it will follow that the final term vanishes provided that \(\partial_{j}u_{\mathbf{x}}^{i}(t)\) remains bounded as \(t \rightarrow -\infty\). But our implicit function theorem argument showed that the Euler-Lagrange equation (\ref{eq:331}) (implicitly) determined, for each \(s \geq 1\), a smooth functional
\begin{equation}\label{eq:371}
\chi^{s} : \mathcal{M}^{s} \rightarrow \mathcal{D}^{s}
\end{equation}
such that, reverting to our previous notation,
\begin{equation}\label{eq:372}
u_{\mathbf{x}} = \chi^{s} (\zero{\gamma}_{\mathbf{x}}).
\end{equation}
It follows that
\begin{equation}\label{eq:373}
\partial_{j}u_{\mathbf{x}} = D\chi^{s}(\zero{\gamma}_{\mathbf{x}}) \cdot (\partial_{j}\zero{\gamma}_{\mathbf{x}})
\end{equation}
where \((\partial_{j}\zero{\gamma}_{\mathbf{x}}(t)) = (0, \dotsc , e^{\omega_{j}t}, \dotsc ,0)\). Since \(\partial_{j}\zero{\gamma}_{\mathbf{x}}\) clearly lies in
\begin{equation}\label{eq:374}
T_{\zero{\gamma}_{\mathbf{x}}} \mathcal{M}^{s} = H^{s} (I,\mathbb{R}^{n})
\end{equation}
and since the linear operator \(D\chi^{s}(\zero{\gamma}_{\mathbf{x}})\) yields an isomorphism from this space to
\begin{equation}\label{eq:375}
T_{\chi^{s}(\zero{\gamma}_{\mathbf{x}})} \mathcal{D}^{s} = T_{u_{\mathbf{x}}} \mathcal{D}^{s} \approx \mathcal{D}^{s},
\end{equation}
it follows that \(\partial_{j}u_{\mathbf{x}} \in \mathcal{D}^{s}\) and hence is not only bounded but, in fact, satisfies
\begin{equation}\label{eq:376}
\lim_{t\searrow -\infty}{\partial_{j}u_{\mathbf{x}}^{i}(t)} = 0.
\end{equation}
More explicitly \(\partial_{j}u_{\mathbf{x}}\) is the (unique, smooth) solution to the linearlized Euler-Lagrange equation,
\begin{equation}\label{eq:377}
\begin{split}
m \partial_{j}\ddot{u}_{\mathbf{x}}^{i}(t) &- m \omega_{i}^{2} \partial_{j} u_{\mathbf{x}}^{i}(t)\\
&\quad - \sum_{k=1}^{n} \frac{\partial^{2}A}{\partial x^{i}\partial x^{k}} (\zero{\gamma}_{\mathbf{x}}(t) + u_{\mathbf{x}}(t))\; \partial_{j}u_{\mathbf{x}}^{k}(t)\\
&= \sum_{k=1}^{n} \frac{\partial^{2}A}{\partial x^{i}\partial x^{k}} (\zero{\gamma}_{\mathbf{x}}(t) + u_{\mathbf{x}}(t))\; \partial_{j} \zero{\gamma}_{\mathbf{x}}^{k}(t),
\end{split}
\end{equation}
which is guaranteed to exist by the analysis given previously.

Thus we conclude that the expression for the gradient of \(S_{(0)}\) simplifies to
\begin{equation}\label{eq:378}
\begin{split}
\partial_{j}S_{(0)}(\mathbf{x}) &= m\omega_{j}x^{j}\\
&\quad + \int_{-\infty}^{0} dt\; \left\lbrack\frac{\partial A}{\partial x^{j}} (\zero{\gamma}_{\mathbf{x}}(t) + u_{\mathbf{x}}(t)) e^{\omega_{j}t}\right\rbrack.
\end{split}
\end{equation}
Differentiating again one obtains
\begin{equation}\label{eq:379}
\begin{split}
\partial_{\ell}\partial_{j}S_{(0)}(\mathbf{x}) &= m\omega_{j}\partial^{j}_{\ell}\\
&\quad + \int_{-\infty}^{0} dt\; \left\lbrack\sum_{k=1}^{n} \frac{\partial^{2}A}{\partial x^{k}\partial x^{j}} (\zero{\gamma}_{\mathbf{x}}(t) + u_{\mathbf{x}}(t)) e^{\omega_{j}t} \cdot \frac{\partial}{\partial x^{\ell}} (\zero{\gamma}_{\mathbf{x}}^{k}(t) + u_{\mathbf{x}}^{k}(t))\right\rbrack.
\end{split}
\end{equation}
Recalling that the Taylor expansion of \(A (x^{1}, \dotsc ,x^{n})\) about the origin begins at third order (c.f., Eq.~(\ref{eq:307})) and noting that
\begin{equation}\label{eq:380}
\lim_{\mathbf{x}\rightarrow 0}{(\zero{\gamma}_{\mathbf{x}}(t) + u_{\mathbf{x}}(t))} = 0
\end{equation}
we thus get, upon taking the indicated limit, that
\begin{equation}\label{eq:381}
\lim_{\mathbf{x}\rightarrow 0}{\partial_{\ell}\partial_{j}S_{(0)}(\mathbf{x})} = m\omega_{j}\partial^{j}_{\ell}
\end{equation}
from which Eq.~(\ref{eq:368}) then follows. In the same way one sees that the Taylor expansion of the integral in Eq.~(\ref{eq:370}) begins at second order and thus that
\begin{equation}\label{eq:382}
\partial_{j}S_{(0)}(\mathbf{x}) = m\omega_{j}x^{j} + O(|\mathbf{x}|^{2}).
\end{equation}
The compatibility of these expansions with the ipze Hamilton-Jacobi equation is easily verified.

\section{Integration of the Transport Equations}
\label{sect:integration-transport}

\subsection{Ground State Analysis}
\label{subsec:transport-ground-state}

As discussed at the end of Sect.~(\ref{sec:finite}) one can, for convenience, choose additive constants of integration in such a way that the quantum corrections, \(\lbrace S_{(i)}(\mathbf{x})\; |\; i=1, 2, \dots\rbrace\), to \(S_{(0)}(\mathbf{x})\) all vanish at the origin. Making this choice one finds, from Eqs.~(\ref{eq:216}) and (\ref{eq:217}) that the \(\lbrace S_{(i)} (x^{1}, \dots , x^{n})\rbrace\) must be given by
\begin{subequations}\label{eq:401}
\begin{align}
S_{(1)}(x^{1}, \dots , x^{n}) &= \int_{-\infty}^{0} dt\; \left(\frac{1}{2m} {}^{(n)}\!\Delta S_{(0)} - \zero{\mathcal{E}}_{(0)}\right) \left(\gamma_{\mathbf{x}}(t)\right)\label{eq:401a}\\
S_{(2)}(x^{1}, \dots , x^{n}) &= \int_{-\infty}^{0} dt\; \left(\frac{1}{m} {}^{(n)}\!\Delta S_{(1)} - \frac{1}{m} \nabla S_{(1)} \cdot \nabla S_{(1)} - 2\zero{\mathcal{E}}_{(1)}\right) \left(\gamma_{\mathbf{x}}(t)\right)\label{eq:401b}
\end{align}
\end{subequations}
and, for arbitrary \(k \geq 2\), by
\begin{equation}\label{eq:402}
S_{(k)}(x^{1}, \dots , x^{n}) = \int_{-\infty}^{0} dt\; \left(\frac{k}{2m} {}^{(n)}\!\Delta S_{(k-1)} - \frac{1}{2m} \sum_{j=1}^{k-1} \frac{k!}{j!(k-j)!} \nabla S_{(j)} \cdot \nabla S_{(k-j)} - k\zero{\mathcal{E}}_{(k-1)}\right) \left(\gamma_{\mathbf{x}}(t)\right)
\end{equation}
where, recalling Eqs.~(\ref{eq:312}--\ref{eq:314})
\begin{equation}\label{eq:403}
\begin{split}
\gamma_{\mathbf{x}}(t) &= \zero{\gamma}_{\mathbf{x}}(t) + u_{\mathbf{x}}(t)\\
 &= (x^{1} e^{\omega_{1}t}, \dots , x^{n} e^{\omega_{n}t}) + \left(u_{\mathbf{x}}^{1}(t), \dots , u_{\mathbf{x}}^{n}(t)\right)
\end{split}
\end{equation}
provided that the integrals all converge. Since these integrals extend over semi-infinite ranges along curves that asymptotically approach the origin as \(t \searrow -\infty\) convergence is possible only if the energy coefficients \(\lbrace\zero{\mathcal{E}}_{(0)}, \zero{\mathcal{E}}_{(1)}, \zero{\mathcal{E}}_{(2)}, \dots\rbrace\) are sequentially chosen so that the integrands above all vanish as \(\mathbf{x} \longrightarrow 0\) (i.e., according to Eqs.~(\ref{eq:218a}--\ref{eq:218c})). In particular, in view of Eq.~(\ref{eq:381}), we are forced to choose
\begin{equation}\label{eq:404}
\zero{\mathcal{E}}_{(0)} = \lim_{\mathbf{x} \longrightarrow 0} \frac{{}^{(n)}\!\Delta S_{(0)}}{2m} = \sum_{j=1}^{n} \frac{\omega_{j}}{2}
\end{equation}
so that the integrand \(\left(\frac{1}{2m} {}^{(n)}\!\Delta S_{(0)} - \zero{\mathcal{E}}_{(0)}\right)\) has Taylor expansion (about the origin) beginning with the first order term. If \(S_{(1)}(\mathbf{x})\) exists and is smooth one can compute \(\zero{\mathcal{E}}_{(1)}\) via Eq.~(\ref{eq:218b}) and attempt to evaluate \(S_{(2)}(\mathbf{x})\) through (\ref{eq:401b}). If \(S_{(2)}(\mathbf{x})\) exists and is smooth one can calculate \(\zero{\mathcal{E}}_{(2)}\) and proceed accordingly.

Clearly this inductive procedure will work provided that integrals of the form \(\int_{-\infty}^{0} dt\; \mathcal{G}\left(\gamma_{\mathbf{x}}(t)\right)\) exist and yield smooth functions of \(\mathbf{x} = (x^{1}, \ldots , x^{n})\) whenever \(\mathcal{G}(\mathbf{x})\) is smooth and satisfies \(\mathcal{G}(\mathbf{0}) = 0\). To establish that this is true we shall first need an estimate for the asymptotic behaviors of the curves \(\gamma_{\mathbf{x}}(t)\) as \(t \searrow -\infty\).

Recalling that the curves \(\left\lbrace\gamma_{\mathbf{x}}(t)\; |\; t \in I, \mathbf{x} \in \mathbb{R}^{n}\right\rbrace\) are precisely the integral curves of the gradient (semi-) flow of \(S_{(0)}\) defined by
\begin{equation}\label{eq:405}
\begin{split}
\frac{d\gamma_{\mathbf{x}}^{i}(t)}{dt} &= \frac{1}{m} \frac{\partial S_{(0)}\left(\gamma_{\mathbf{x}}(t)\right)}{\partial x^{i}},\\
\gamma_{\mathbf{x}}^{i}(0) &= x^{i},\qquad i = 1, \ldots , n
\end{split}
\end{equation}
one computes, using the ipze Hamilton-Jacobi equation satisfied by \(S_{(0)}(\mathbf{x})\), that, along this flow, \(S_{(0)}\) obeys the evolution equation
\begin{equation}\label{eq:406}
\begin{split}
\frac{dS_{(0)}}{dt} \left(\gamma_{\mathbf{x}}(t)\right) &= \frac{1}{m} \sum_{i=1}^{n} \frac{\partial S_{(0)}\left(\gamma_{\mathbf{x}}(t)\right)}{\partial x^{i}} \frac{\partial S_{(0)}\left(\gamma_{\mathbf{x}}(t)\right)}{\partial x^{i}}\\
&= 2 V\left(\gamma_{\mathbf{x}}(t)\right)\\
&= m \sum_{i=1}^{n} \omega_{i}^{2} \left(\gamma_{\mathbf{x}}(t)\right)^{2} + 2 A \left(\gamma_{\mathbf{x}}(t)\right).
\end{split}
\end{equation}
On the other hand we know, from Eq.~(\ref{eq:368}) and the definition of \(A(\mathbf{x})\), that
\begin{subequations}\label{eq:407}
\begin{align}
S_{(0)}(\mathbf{x}) &= \frac{1}{2} m \sum_{i=1}^{n} \omega_{i} (x^{i})^{2} + O(|\mathbf{x}|^{3})\label{eq:407a}\\
\intertext{and}
A(\mathbf{x}) &= O(|\mathbf{x}|^{3})\label{eq:407b}
\end{align}
\end{subequations}
It follows that, on a sufficiently small ball centered at the origin and having (Euclidean) radius \(\epsilon\), we have
\begin{equation}\label{eq:408}
\frac{dS_{(0)}}{dt} \left(\gamma_{\mathbf{x}}(t)\right) \geq 2 (\omega_{\mathrm{min}} - C\epsilon) S_{(0)} \left(\gamma_{\mathbf{x}}(t)\right)
\end{equation}
for some constant \(C > 0\) and with \(\omega_{\mathrm{min}} = \mathrm{min} \lbrace\omega_{1}, \ldots , \omega_{n}\rbrace > 0\). Thus, choosing \(\epsilon\) sufficiently small to ensure that \(\omega_{\mathrm{min}} - C\epsilon > 0\), we get that
\begin{subequations}\label{eq:409}
\begin{align}
S_{(0)}\left(\gamma_{\mathbf{x}}(t)\right) \leq K_{\mathbf{x}} e^{2(\omega_{\mathrm{min}} - C\epsilon)t}\label{eq:409a}\\
\intertext{with}
K_{\mathbf{x}} = S_{(0)}\left(\gamma_{\mathbf{x}}(t_{\mathbf{x}}^{\mathbf{\ast}})\right) e^{-2(\omega_{\mathrm{min}} - C\epsilon)t_{\mathbf{x}}^{\mathbf{\ast}}}\label{eq:409b}
\end{align}
\end{subequations}
for all \(t\) in the range \(-\infty < t < t_{\mathbf{x}}^{\mathbf{\ast}} < 0\) where \(t_{\mathbf{x}}^{\mathbf{\ast}}\) is a time prior to which \(\gamma_{\mathbf{x}}(t)\) lies entirely within the chosen ball of radius \(\epsilon\). In view of Eq.~(\ref{eq:407a}) and the fact that \(\epsilon\) can be chosen arbitrarily small it follows from (\ref{eq:409}) that, as \(t \searrow -\infty\), the Euclidean length, \(|\gamma_{\mathbf{x}}(t)|\), of \(\gamma_{\mathbf{x}}(t)\) decays at least exponentially rapidly in the sense that
\begin{equation}\label{eq:410}
\left|\gamma_{\mathbf{x}}(t)\right| \leq \left|\gamma_{\mathbf{x}}(t_{\mathbf{x}}^{\mathbf{\ast}})\right| e^{(\omega_{\mathrm{min}} - C\epsilon) (t - t_{\mathbf{x}}^{\mathbf{\ast}})}
\end{equation}
for all \(t \leq t_{\mathbf{x}}^{\mathbf{\ast}} < 0\).

With this exponentially decaying bound on \(|\gamma_{\mathbf{x}}(t)|\) it is straightforward to verify the convergence of integrals of the form \(\int_{-\infty}^{0} \mathcal{G}(\gamma_{\mathbf{x}}(t))\; dt\) for arbitrary smooth \(\mathcal{G}(\mathbf{x})\) that vanishes at \(\mathbf{x} = \mathbf{0}\). To complete the sequential construction of the \(\lbrace S_{(i)}(\mathbf{x})\; |\; i = 1, 2, \dots\rbrace\) it suffices to show that such integrals are automatically smooth in \(\mathbf{x}\).

For this purpose it is convenient to reexpress these integrals in terms of the Sternberg coordinates for the vector field \(\frac{1}{m} \nabla S_{(0)}\) introduced in Sect.~(\ref{subsec:excited}) and extended to a global diffeomorphism in the Appendix. From Eqs.~(\ref{eq:261}) and (\ref{eq:263}) it is clear that
\begin{equation}\label{eq:411}
\mathcal{G}\left(\gamma_{\mathbf{x}}(t)\right) = \left.\mathcal{G} \circ \mu^{-1}(\mathbf{y})\right|_{\mathbf{y} = \mathbf{y}_{\gamma}(t)}
\end{equation}
where
\begin{equation}\label{eq:412}
\begin{split}
\mathbf{y}_{\gamma}^{i}(t) &= \mu^{i} (x^{1}, \ldots , x^{n}) e^{\omega_{i}t}\\
&= \mathbf{y}^{i} (x^{1}, \ldots x^{n}) e^{\omega_{i}t}
\end{split}
\end{equation}
in which the \(\lbrace\mu^{i}\; |\; i = 1, \ldots , n\rbrace\) are globally smooth functions on \(\mathbb{R}^{n}\) and where \(\mathcal{G} \circ \mu^{-1}(\mathbf{y})\) is smooth throughout the (star-shaped) domain of definition of the \(\lbrace y^{i}\; |\; i = 1, \ldots , n\rbrace\).

It follows that
\begin{equation}\label{eq:413}
\frac{\partial}{\partial x^{\ell}} \int_{-\infty}^{0} \mathcal{G}\left(\gamma_{\mathbf{x}}(t)\right) dt = \int_{-\infty}^{0} \sum_{j=1}^{n} \left.\left(\frac{\partial}{\partial y^{j}} \mathcal{G} \circ \mu^{-1}(\mathbf{y})\right)\right|_{\mathbf{y} = \mathbf{y}_{\gamma}(t)} \frac{\partial\mu^{j} (x^{1}, \ldots , x^{n})}{\partial x^{\ell}} e^{\omega_{j}t} dt
\end{equation}
Since each \(\omega_{j} > 0\) and since the factor multiplying each \(e^{\omega_{j}t}\) is smooth and bounded along \(\gamma_{\mathbf{x}}\) this integral clearly converges for all \(\mathbf{x} \in \mathbb{R}^{n}\). Higher derivatives are readily computed in the same way and it thus follows that \(\int_{-\infty}^{0} \mathcal{G}\left(\gamma_{\mathbf{x}}(t)\right) dt\) is globally smooth.

\subsection{Excited State Analysis}
\label{subsec:transport-excited-state}

In Section (\ref{subsec:excited}) we showed how to construct formal Taylor expansions for the unknown, excited state wave functions \(\lbrace\emm{\phi}_{(k)}\; |\; k = 0, 1, 2, \ldots\rbrace\). By exploiting well-known techniques for constructing globally smooth functions on \(\mathbb{R}^{n}\) that have arbitrarily specified Taylor expansions about the origin \cite{Remizov2010} one can generate a set of functions \(\lbrace\emm{\nu}_{(k)}\; |\; k = 0, 1, 2, \ldots\rbrace\), smooth on \(\mathbb{R}^{n}\), that have the Taylor expansions needed for the unknowns \(\lbrace\emm{\phi}_{(k)}\; |\; k = 0, 1, 2, \ldots\rbrace\). It follows that one can now seek solutions to Eqs.~(\ref{eq:228}--\ref{eq:229}) of the form
\begin{equation}\label{eq:414}
\emm{\phi}_{(k)} = \emm{\nu}_{(k)} + \emm{\chi}_{(k)}
\end{equation}
where the \(\lbrace\emm{\chi}_{(k)}\; |\; k = 0, 1, 2, \ldots\rbrace\) are smooth but vanish to infinite order (i.e., have trivial Taylor expansions) at the origin. Such functions are often referred to as `flat'.

Substituting the decomposition (\ref{eq:414}) into Eqs.~(\ref{eq:228}--\ref{eq:229}) (with \((\ast) \longrightarrow (\mathbf{m})\)) and exploiting the fact that the \(\lbrace\emm{\nu}_{(k)}\rbrace\), by construction, formally satisfy these equations to all orders at the origin, one finds that the \(\lbrace\emm{\chi}_{(k)}\; |\; k = 0, 1, 2, \ldots\rbrace\) must satisfy equations of the form
\begin{equation}\label{eq:415}
\left(\frac{\nabla S_{(0)}}{m}\right) \cdot \nabla\emm{\chi}_{(k)} - \Delta\emm{\mathcal{E}}_{(0)}\emm{\chi}_{(k)} = \emm{\sigma}_{(k)}
\end{equation}
where the `source' terms \(\lbrace\emm{\sigma}_{(k)}\; |\; k = 0, 1, 2, \ldots\rbrace\) are smooth and vanish to infinite order at the origin. Integrating Eq.~(\ref{eq:415}) along the integral curves \(\gamma_{\mathbf{x}}\) of the vector field \(\frac{\nabla S_{(0)}}{m}\) one easily finds that the only functions that could have the desired properties would be given by
\begin{equation}\label{eq:416}
\emm{\chi}_{(k)}(\mathbf{x}) = \int_{-\infty}^{0} e^{-\Delta\emm{\mathcal{E}}_{(0)}t}\;\: \emm{\sigma}_{(k)}\left(\gamma_{\mathbf{x}}(t)\right) dt
\end{equation}
provided that the integral expressions converge to smooth functions that vanish to infinite order. From the exponential decay of \(\gamma_{\mathbf{x}}(t)\) as \(t \searrow -\infty\) (\textit{c.f.}, inequality (\ref{eq:410})) and the fact that each \(\emm{\sigma}_{(k)}(x^{1}, \ldots , x^{n})\) vanishes to infinite order it follows that \(\emm{\sigma}_{(k)}\left(\gamma_{\mathbf{x}}(t)\right) = O(e^{-C|t|})\) for every \(C > 0\) and thus that the integrals above always converge. To verify their smoothness and flatness properties it is convenient again to reexpress the curves \(\gamma_{\mathbf{x}}\) in terms of Sternberg coordinates. Writing \(\emm{\chi}_{(k)}(\mathbf{x})\) in the equivalent form
\begin{equation}\label{eq:417}
\emm{\chi}_{(k)}(\mathbf{x}) = \int_{-\infty}^{0} \left[e^{-\Delta\emm{\mathcal{E}}_{(0)} \cdot t} \left.\left(\emm{\sigma}_{(k)} \circ \mu^{-1}(\mathbf{y})\right)\right|_{\mathbf{y}=\mathbf{y}_{\gamma}(t)}\right] dt
\end{equation}
where
\begin{equation}\label{eq:418}
\begin{split}
y_{\gamma}^{i}(t) &= \mu^{i}(x^{1}, \ldots , x^{n}) e^{\omega_{i}t}\\
 &= y^{i}(x^{1}, \ldots , x^{n}) e^{\omega_{i}t}
\end{split}
\end{equation}
and exploiting the fact that Sternberg coordinates, by construction, satisfy
\begin{equation}\label{eq:419}
y^{i} (x^{1}, \ldots , x^{n}) = x^{i} + O(|\mathbf{x}|^{2})
\end{equation}
it is straightforward to verify that the above expression for \(\emm{\chi}_{(k)}(\mathbf{x})\) vanishes in the limit \(\mathbf{x} \longrightarrow \mathbf{0}\). Partial derivatives of the \(\emm{\chi}_{(k)}(\mathbf{x})\) can be successively computed as in the previous subsection and verified to vanish as above. It follows that the \(\emm{\chi}_{(k)}(\mathbf{x})\) are smooth and vanish to infinite order as desired.

The above argument is essentially equivalent to that given in Sect.~(3) of Ref.~\cite{Dimassi1999}. Our equation for the \(\emm{\chi}_{(k)}\) is slightly simpler than the one dealt with therein however since we have formulated the excited state problem differently by arranging to solve for the energy gaps rather than the actual energies. Our parametrization of the excited state wave functions also differs from that of Ref.~\cite{Dimassi1999} in that we have included all the (ground state) quantum corrections to \(S_{(0)}\) in the exponential factor appearing in our excited state ansatz.

\subsection{Asymptotic Character of the Formal Expansions}
\label{subsec:transport-asymptotic}

The principal difference between our approach and that of the microlocal analysis literature lies in the different methods used to determine the `fundamental solution' \(S_{(0)}(\mathbf{x})\) to the ipze Hamilton-Jacobi equation. But once this function has been obtained the further computation (in our approach) of its quantum corrections \(\lbrace S_{(i)}(\mathbf{x})\; |\; i = 1, 2, \ldots\rbrace\) and, more generally, the excited state wave functions\(\lbrace\emm{\phi}_{(k)}(\mathbf{x})\; |\; k = 0, 1, 2, \ldots\rbrace\) proceeds more or less in parallel to that discussed in \cite{Dimassi1999}, allowance being made for the different parameterizations adopted.

Because of the different parameterizations employed there is no direct, one-to-one correspondence between our quantum corrections and those of \cite{Dimassi1999} but each set clearly determines the other with the same information being simply `packaged' in different ways. Thus the asymptotic nature of these earlier expansions, established in \cite{Dimassi1999} and other references cited therein, applies equally well to our expansions. For that reason we shall simply quote these earlier results as giving a precise meaning to the sense in which Schr\"{o}dinger's equations is being approximately solved.

In Ref.~\cite{Dimassi1999} both ground and excited states are sought in the form
\begin{equation}\label{eq:420}
\emm{\psi}(\mathbf{x}) = \emm{a}(\mathbf{x},\hbar) e^{-S_{(0)}(\mathbf{x})/\hbar}
\end{equation}
and formal expansions (in powers of \(\hbar\)) are developed for the coefficient wave functions \(\emm{a}(\mathbf{x},\hbar)\). But then one can `realize' these formal expansions by actual smooth functions to which the expansions represent asymptotic approximations,
\begin{equation}\label{eq:421}
\emm{a}(\mathbf{x},\hbar) \sim \emm{a}_{(0)}(\mathbf{x}) + \hbar\emm{a}_{(1)}(\mathbf{x}) + \frac{\hbar^{2}}{2!} \emm{a}_{(2)}(\mathbf{x}) + \dots ,
\end{equation}
in the sense that
\begin{equation}\label{eq:422}
\left|\partial_{\mathbf{x}}^{\mathbf{\alpha}}\left(\emm{a} - \sum_{j=0}^{N} \emm{a}_{(j)}(\mathbf{x})\;\: \hbar^{j}\right)\right| \leq C_{K,\alpha,N} \hbar^{N+1},\quad \mathbf{x} \in K
\end{equation}
for all \(K \subset\subset \mathbf{R}^{n}\), \(\alpha \in \mathbb{N}^{n}\), \(N \in \mathbb{N}\). Realizing, in the analogous way, the formal expansions for the energy levels via
\begin{equation}\label{eq:423}
\emm{E}(\hbar) \sim \emm{E}_{(0)} + \hbar \emm{E}_{(1)} + \frac{\hbar^{2}}{2!} \emm{E}_{(2)} + \dots ,
\end{equation}
one proves (\textit{c.f.}, Theorem 3.6 of \cite{Dimassi1999}) that
\begin{equation}\label{eq:424}
\left(\hat{H} - \emm{E}(\hbar)\right) \left(\emm{a}\;\: e^{-S_{(0)}/\hbar}\right) = r\;\: e^{-S_{(0)}/\hbar}
\end{equation}
where \(\left|\partial_{\mathbf{x}}^{\mathbf{\alpha}} r(\mathbf{x},\hbar)\right| \leq C_{K,N,\alpha} \hbar^{N}\), \(\mathbf{x} \in K\) for every \(K \subset\subset \mathbb{R}^{n}\), \(\mathbf{\alpha} \in \mathbb{N}^{n}\), \(N \in \mathbb{N}\). For \textit{analytic potentials} the bound that the error vanishes faster than any integral power of \(\hbar\) can be replaced by the statement that it vanishes faster than \(e^{-\epsilon_{0}/\hbar}\) for some \(\epsilon_{0} > 0\) \cite{Helfer1984}.

There are elements of arbitrariness in the procedure of passing from formal series to their `realizations' alluded to above so one could hardly hope to obtain actual exact solutions to Schr\"{o}dinger's equation in this way. In the section below however we shall investigate some computable examples of our approach for which, through Borel resummation, one may perhaps be able to adduce exact results.

\section{Computable One-Dimensional Examples}
\label{sec:computable-examples}

\subsection{Ground State Calculations}
\label{subsec:ground-state}

As explicitly computable examples of the above framework we here study the family of one-dimensional anharmonic oscillators characterized by potential energies of the form
\begin{equation}\label{eq:501}
V_{\kappa}(x) = \frac{1}{2} m\omega_{o}^{2} x^{2} + gx^{2\kappa},\qquad \kappa = 2,3,4 \dots
\end{equation}
concentrating, for simplicity, on the quartic case (with \(\kappa = 2\)). The ipze Hamilton-Jacobi equation for the quartic oscillator is readily integrated to yield
\begin{equation}\label{eq:502}
S_{(0)}(x) = \frac{m^{2}\omega_{0}^{3}}{6g} \left\lbrack\left( 1 + \frac{2gx^{2}}{m\omega_{0}^{2}}\right)^{3/2} - 1\right\rbrack
\end{equation}
and thus to give a zeroth-order approximation to the ground state wave function of the form
\begin{equation}\label{eq:503}
\zero{\psi} \cong N\; e^{-(S_{(0)}/\hbar)~ +~ \dots}
\end{equation}
where \(N\) is a normalization constant. This approximation already begins to capture the more-rapid-than-gaussian decay at large \(|x|\) expected for the actual ground state solution and reduces to the exact (gaussian) harmonic oscillator result in the limit \(g \searrow 0\).

The transport equation for \(S_{(1)}\) takes the form
\begin{equation}\label{eq:504}
\frac{1}{m} \frac{dS_{(0)}}{dx} \frac{dS_{(1)}}{dx} = -\left(\zero{\mathcal{E}}_{(0)} - \frac{1}{2m} \frac{d^{2}S_{(0)}}{dx^{2}}\right)
\end{equation}
which, in view of the linear behavior of \(\frac{dS_{(0)}}{dx}\) in \(x\) for small \(x\),
\begin{equation}\label{eq:505}
\frac{dS_{(0)}}{dx} = m\omega_{0}x \sqrt{1 + \frac{2gx^{2}}{m\omega_{0}^{2}}},
\end{equation}
generates a logrithmically singular solution for \(S_{(1)}\) unless one fixes \(\zero{\mathcal{E}}_{(0)}\) by demanding that
\begin{equation}\label{eq:506}
\zero{\mathcal{E}}_{(0)} = \lim_{x\rightarrow 0}{\frac{1}{2m} \frac{d^{2}S_{(0)}}{dx^{2}}} = \frac{1}{2} \omega_{0}.
\end{equation}
Substituting this choice for \(\zero{\mathcal{E}}_{(0)}\) into Eq.~(\ref{eq:504}) and integrating yields the (everywhere smooth) result
\begin{equation}\label{eq:507}
S_{(1)} = \frac{1}{2} \ln{\left\lbrace\sqrt{1 + \frac{2gx^{2}}{m\omega_{0}^{2}}} \left(\frac{1 + \sqrt{1 + \frac{2gx^{2}}{m\omega_{0}^{2}}}}{2}\right)\right\rbrace}
\end{equation}
where we have adjusted the arbitrary additive constant of integration to arrange that \(\lim_{x\rightarrow 0}{S_{(1)}} = 0\).

Thus the first order correction to \(\zero{\psi}\), using \(\frac{S_{\hbar}}{\hbar} \cong \frac{S_{(0)}}{\hbar} + S_{(1)} + \dots\) becomes
\begin{equation}\label{eq:508}
\zero{\psi} \cong
\frac{N \cdot e^{-\frac{m^{2}\omega_{0}^{3}}{6g} \left\lbrack
\left(1 + \frac{2gx^{2}}{m\omega_{0}^{2}}\right)^{3/2} - 1\right\rbrack + \dots}}{\left(1 + \frac{2gx^{2}}{m\omega_{0}^{2}}\right)^{1/4}
\left(\frac{1 + \left(1 + \frac{2gx^{2}}{m\omega_{0}^{2}}\right)^{1/2}}{2}\right)^{1/2}}
\end{equation}
and, at this order, the approximation for the ground state energy is simply
\begin{equation}\label{eq:509}
\zero{E}_{\hbar} \cong \frac{1}{2} \hbar\omega_{0} + \dots
\end{equation}
which of course is just the exact result for a \textit{harmonic} oscillator.

Continuing in this way one finds the transport equation for \(S_{(2)}\) to be
\begin{equation}\label{eq:510}
\frac{1}{m} \frac{dS_{(0)}}{dx} \frac{dS_{(2)}}{dx} = -\left\lbrace 2\zero{\mathcal{E}}_{(1)} + \frac{1}{m} \left(\frac{dS_{(1)}}{dx}\right)^{2} - \frac{1}{m} \frac{d^{2}S_{(1)}}{dx^{2}}\right\rbrace
\end{equation}
for which, to avoid a logarithmic singularity in the solution, one is forced to take
\begin{equation}\label{eq:511}
\begin{split}
\zero{\mathcal{E}}_{(1)} &= \lim_{x\rightarrow 0}{\left\lbrace\frac{1}{2m} \frac{d^{2}S_{(1)}}{dx^{2}} - \frac{1}{2m} \left(\frac{dS_{(1)}}{dx}\right)^{2}\right\rbrace}\\
&= \frac{3g}{4m^{2}\omega_{0}^{2}}.
\end{split}
\end{equation}
The corresponding correction for the ground state energy thus takes the form
\begin{equation}\label{eq:512}
\zero{E}_{\hbar} \cong \hbar\omega_{0} \left\lbrace\frac{1}{2} + \frac{3}{4} \frac{g\hbar}{m^{2}\omega_{0}^{3}} + \dots\right\rbrace
\end{equation}
and, upon integration, Eq.~(\ref{eq:510}) then yields
\begin{equation}\label{eq:513}
S_{(2)} = \frac{\left\lbrace 3m^{2}\omega_{0}^{2} \left\lbrack 1 - \sqrt{1 + \frac{2gx^{2}}{m\omega_{0}^{2}}}\right\rbrack + 20gmx^{2} + \frac{18g^{2}x^{4}}{\omega_{0}^{2}}\right\rbrace}
{6x^{2}(m\omega_{0})^{3} \left( 1 + \frac{2gx^{2}}{m\omega_{0}^{2}}\right)^{3/2}}.
\end{equation}
In spite of the factor \(\frac{1}{x^{2}}\), \(S_{(2)}\) is readily verified to be everywhere smooth.

This pattern continues indefinitely in the sense that only differentiations and evaluations of smooth functions at \(x = 0\) are needed to calculate the energy coefficients \(\left\lbrace\zero{\mathcal{E}}_{(0)}, \zero{\mathcal{E}}_{(1)}, \zero{\mathcal{E}}_{(2)}, \dots\right\rbrace\) while elementary integrations suffice to generate the sequence of quantum corrections \(\left\lbrace S_{(1)}, S_{(2)}, S_{(3)}, \dots\right\rbrace\) to \(S_{(0)}\). In particular the transport equation at order \(k\) yields an explicit smooth expression for \(\frac{dS_{(k)}}{dx}\) in terms of previously computed quantities and the corresponding energy coefficient is then given by
\begin{equation}\label{eq:514}
\zero{\mathcal{E}}_{(k)} = \lim_{x\rightarrow 0}{\left\lbrack\frac{1}{2m} \frac{d^{2}S_{(k)}}{dx^{2}} \right\rbrack}.
\end{equation}
We have carried out these calculations explicitly to compute the first 26 terms in each of the expansions
\begin{equation}\label{eq:515}
\zero{E_{\hbar}} = \hbar \left( \zero{\mathcal{E}}_{(0)} + \hbar\zero{\mathcal{E}}_{(1)} + \frac{\hbar^{2}}{2!}\zero{\mathcal{E}}_{(2)} + \frac{\hbar^{3}}{3!}\zero{\mathcal{E}}_{(3)} + \dots\right)
\end{equation}
and
\begin{equation}\label{eq:516}
S_{\hbar} = \left( S_{(0)} + \hbar S_{(1)} + \frac{\hbar^{2}}{2!} S_{(2)} + \frac{\hbar^{3}}{3!}S_{(3)} + \dots\right)
\end{equation}
and find that the energy expansion (\ref{eq:515}) agrees precisely with the conventional result obtained from (Rayleigh-Schr\"{o}dinger) perturbation theory carried out to the corresponding order (c.f., \cite{Mathematica}).

As discussed more fully below we find analogous, precise agreement with the conventional ground state energy expansions (carried out to the same order) for the sectic, octic and dectic oscillators characterized by
\begin{equation}\label{eq:517}
V_{\kappa}(x) = \frac{1}{2} m\omega_{0}^{2}x^{2} + gx^{2\kappa}
\end{equation}
with \(\kappa = 3, 4, 5\). In view of these results it seems plausible that agreement of the energy expansions extends to all orders in \(\hbar\) for each of the oscillators studied and perhaps also to a much larger family of anharmonic oscillators (e.g., those having potential energies \(V_{\kappa}\) with \(\kappa = 6, 7, \dots\) or, more generally, those allowing intermediate terms of the form \(rx^{\alpha}\) with \(2 < \alpha < 2\kappa\)). One might hope to verify these conjectures by deriving suitable recurrence relations for the energy coefficients in the various cases under study and comparing them with corresponding recurrence relations derivable from the conventional (Rayleigh-Schr\"{o}dinger) approach \cite{Kleinert2009} but we shall not pursue that issue here.

For the quartic oscillator in particular the energy expansion can be rewritten as
\begin{equation}\label{eq:518}
\begin{split}
\zero{E}_{\mathrm{quartic}} &= \hbar\omega_{0} \left\lbrace\frac{1}{2} + \frac{3}{4} \left(\frac{g\hbar}{m^{2}\omega_{0}^{3}}\right) - \frac{21}{8} \left(\frac{g\hbar}{m^{2}\omega_{0}^{3}}\right)^{2}\right.\\
 &\left.\vphantom{\frac{1}{2} + \frac{3}{4} \left(\frac{g\hbar}{m^{2}\omega_{0}^{2}}\right) - \frac{21}{8} \left(\frac{g\hbar}{m^{2}\omega_{0}^{3}}\right)^{2}}  + \frac{333}{16} \left(\frac{g\hbar}{m^{2}\omega_{0}^{3}}\right)^{3} - \dots\right\rbrace\\
 &:= \hbar\omega_{0} \left\lbrace\frac{1}{2} + \mathcal{F}_{\mathrm{quartic}} \left(\frac{g\hbar}{m^{2}\omega_{0}^{3}}\right)
\right\rbrace
\end{split}
\end{equation}
where \(\mathcal{F}_{\mathrm{quartic}}\) is a formal power series in the dimensionless quantity \(\frac{g\hbar}{m^{2}\omega_{0}^{3}}\). It is well-known that this series, derived from conventional perturbation theory as a power series in the coupling constant \(g\) (and thus the conjecturally equivalent series derived as above as a power series in \(\hbar\)), is divergent but asymptotic and amenable to Borel resummation. In particular the Borel sum has been proven to equal the exact ground state energy for this oscillator \cite{Galindo1991} and corresponding results are known to hold for the higher order oscillators as well \cite{Simon1982}. The exact expression for this energy (in say the quartic case) is not analytic in \(\mu := \frac{g\hbar}{m^{2}\omega_{0}^{3}}\) about \(\mu = 0\) and thus a formal series expansion for this quantity in powers of \(\mu\) can never converge. Fortunately the asymptotic (and in fact Borel summable) character of this series allows one to extract accurate results from it nevertheless.

For the sectic oscillator (\(\kappa = 3\) in the potential expression) one derives, as above, the formal series
\begin{equation}\label{eq:519}
\begin{split}
\zero{E}_{\mathrm{sectic}} &= \hbar\omega_{0} \left\lbrace\frac{1}{2} + \frac{15}{8} \left(\frac{g\hbar^{2}}{m^{3}\omega_{0}^{4}}\right)\right.\\
&\left.\vphantom{\frac{1}{2} + \frac{15}{8} \left(\frac{g\hbar^{2}}{m^{3}\omega_{0}^{4}}\right)} - \frac{3495}{64} \left(\frac{g\hbar^{2}}{m^{3}\omega_{0}^{4}}\right)^{2} + \frac{1239675}{256} \left(\frac{g\hbar^{2}}{m^{3}\omega_{0}^{4}}\right)^{3} - \dots\right\rbrace
\end{split}
\end{equation}
whereas, for the octic and dectic oscillators one gets
\begin{equation}\label{eq:520}
\begin{split}
\zero{E}_{\mathrm{octic}} &= \hbar\omega_{0} \left\lbrace\frac{1}{2} + \frac{105}{16} \left(\frac{g\hbar^{3}}{m^{4}\omega_{0}^{5}}\right)\right.\\
&\left.\vphantom{\frac{1}{2} + \frac{105}{16} \left(\frac{g\hbar^{3}}{m^{4}\omega_{0}^{5}}\right)} - \frac{67515}{32} \left(\frac{g\hbar^{3}}{m^{4}\omega_{0}^{5}}\right)^{2} + \frac{401548875}{128} \left(\frac{g\hbar^{3}}{m^{4}\omega_{0}^{5}}\right)^{3} - \dots\right\rbrace
\end{split}
\end{equation}
and
\begin{equation}\label{eq:521}
\begin{split}
\zero{E}_{\mathrm{dectic}} &= \hbar\omega_{0} \left\lbrace\frac{1}{2} + \frac{945}{32} \left(\frac{g\hbar^{4}}{m^{5}\omega_{0}^{6}}\right)\right.\\
&\left.\vphantom{\frac{1}{2} + \frac{105}{16} \left(\frac{g\hbar^{4}}{m^{5}\omega_{0}^{6}}\right)} - \frac{140057505}{1024} \left(\frac{g\hbar^{4}}{m^{5}\omega_{0}^{6}}\right)^{2} + \frac{78210463124745}{16384} \left(\frac{g\hbar^{4}}{m^{5}\omega_{0}^{6}}\right)^{3} - \dots\right\rbrace
\end{split}
\end{equation}
respectively. As mentioned above we have computed each of these expansions, by the method developed herein, up to the term of order \(\hbar^{25}\) and find that they each coincide with the corresponding (truncated) series expansions derived via Rayleigh-Schr\"{o}dinger perterbation theory. Notice however that since \(\mu_{\mathrm{sectic}} := \frac{g\hbar^{2}}{m^{3}\omega_{0}^{4}}\), \(\mu_{\mathrm{octic}} := \frac{g\hbar^{3}}{m^{4}\omega_{0}^{5}}\) and \(\mu_{\mathrm{dectic}} := \frac{g\hbar^{4}}{m^{5}\omega_{0}^{6}}\) are the relevant, dimensionless expansion parameters for each of the indicated cases, and that only integral powers of \(g\) occur in each of the formal expansions, there are actually fewer (non-vanishing) terms in each successive expansion. More precisely those energy coefficients that would correspond to non-integral powers of \(g\) were explicitly found to vanish even though the corresponding quantum corrections to \(S_{(0)}\) were non-vanishing.

Remarkably all of the integrals involved in computing the quantum corrections \(\left\lbrace S_{(1)}, S_{(2)}, S_{(3)}, \dots\right\rbrace\) to \(S_{(0)}\) (up to the highest order computed, namely \(S_{(25)}\)) were expressible explicitly in terms of elementary functions for the \textit{quartic} and \textit{sectic} oscillators whereas for the octic and dectic cases some (but not all) of the quantum corrections required, in addition, hypergeometric functions for their evaluation. It seems plausible to conjecture that these patterns persist to all orders in \(\hbar\) and thus, for the quartic and sectic cases in particular, lead to formal expansions for \(S_{\hbar}\) in terms of elementary functions.

The quartic case is especially interesting in that each of the computed functions \(\left\lbrace\frac{dS_{(0)}}{dx}, \dotsc , \frac{dS_{(25)}}{dx}\right\rbrace\) is seen, by inspection, to be odd in \(x\) and to have uniformly definite sign on each of the intervals (\(-\infty ,0\)) and (\(0, \infty\)). The corresponding integrals for \(S_{(0)}\) and \(S_{(1)}\) have already been given in Eqs.~(\ref{eq:502}) and (\ref{eq:507}) respectively and these functions behave exceptionally whereas, for the subsequent quantum corrections \(\left\lbrace S_{(\ell)}(x),\ell = 2, 3, 4, \dotsc , 25\right\rbrace\) one finds that their constants of integrations can always be chosen so that each of the functions \(S_{(\ell)} : \mathbb{R} \rightarrow \mathbb{R}\) has definite sign (alternating with \(\ell\)) throughout its domain and each decays monotonically from its global maximum (or minimum, depending upon \(\ell\)) at \(x = 0\) to the value 0 as \(|x| \rightarrow \infty\). It seems plausible to conjecture that this pattern continues to all orders in \(\hbar\) for the quartic case and, if so, that one might establish such a result mathematically by a suitable inductive argument.

One further finds that the sequence of ratios \(\left\lbrace - \frac{S_{(\ell)}(0)}{\zero{\mathcal{E}}_{(\ell)}}, \ell = 2, 3, \dotsc , 25\right\rbrace\) decreases monotonically with increasing \(\ell\). If this pattern persists and if the aforementioned conjectures for the quartic case can be verified it would follow not only that the formal series \(\sum_{\ell = 2}^{\infty} \left(\frac{S_{(\ell)}(0)\hbar^{\ell}}{\ell!}\right)\) is Borel summable but also that \(\sum_{\ell = 2}^{\infty} \left(\frac{S_{(\ell)}(x)\hbar^{\ell}}{\ell!}\right)\) is uniformly Borel summable for all \(x \in \mathbb{R}\). Under these circumstances the function
\begin{equation}\label{eq:522}
\zero{\psi}(x) := Ne^{-\frac{S_{(0)}(x)}{\hbar} - S_{(1)}(x) - \frac{\mathcal{S}(x)}{\hbar}},
\end{equation}
where \(\mathcal{S}(x)\) is the Borel sum of the formal series \(\sum_{\ell = 2}^{\infty} \frac{S_{(\ell)}(x)\hbar^{\ell}}{\ell!}\) and \(N > 0\) is a normalization constant, would be a natural candidate for the exact ground state wave function for the quartic oscillator.

For the sectic and higher order oscillators studied explicitly the corresponding patterns are more difficult to identify but one should keep in mind that the observed and conjectured simple features of the quartic oscillator were sufficient but not strictly necessary for its Borel summability.

\subsection{Excited States for Quartic Oscillators}
\label{subsec:quartic}

We now turn to the construction of formal expansions for excited state wave functions, concentrating, for simplicity, on the quartic case. Extending our notation
\begin{equation}\label{eq:523}
\zero{E}_{\hbar} := \hbar\zero{\mathcal{E}}_{\hbar} = \hbar \left(\zero{\mathcal{E}}_{(0)} + \hbar\zero{\mathcal{E}}_{(1)} + \frac{\hbar^{2}}{2!} \zero{\mathcal{E}}_{(2)} + \frac{\hbar^{3}}{3!} \zero{\mathcal{E}}_{(3)} + \dots\right)
\end{equation}
for the ground state energy to that for a generic excited state we write
\begin{equation}\label{eq:524}
\starr{E}_{\hbar} := \hbar\starr{\mathcal{E}}_{\hbar} = \hbar\left(\starr{\mathcal{E}}_{(0)} + \hbar\starr{\mathcal{E}}_{(1)} + \frac{\hbar^{2}}{2!}\starr{\mathcal{E}}_{(2)} + \frac{\hbar^{3}}{3!}\starr{\mathcal{E}}_{(3)} + \dots\right).
\end{equation}
We shall soon find however that these states can, as in the case of a pure harmonic oscillator, be naturally labeled by a positive integer \(n\) and thenceforth sharpen and unify the above notation to
\begin{equation}\label{eq:525}
\enn{E}_{\hbar} := \hbar\enn{\mathcal{E}}_{\hbar} = \hbar\left(\enn{\mathcal{E}}_{(0)} + \hbar\enn{\mathcal{E}}_{(1)} + \frac{\hbar^{2}}{2!}\enn{\mathcal{E}}_{(2)} + \frac{\hbar^{3}}{3!}\enn{\mathcal{E}}_{(3)} + \dots\right)
\end{equation}
where \(n = 0, 1, 2, 3, \dots\).

We look for excited states \(\starr{\psi}_{\hbar} (x)\) by setting
\begin{equation}\label{eq:526}
\starr{\psi}_{\hbar} (x) = \starr{\phi}_{\hbar} (x)e^{-S_{\hbar}(x)/\hbar},
\end{equation}
where
\begin{equation}\label{eq:527}
S_{\hbar}(x) = \left(S_{(0)}(x) + \hbar S_{(1)}(x) + \frac{\hbar^{2}}{2!} S_{(2)}(x) + \frac{\hbar^{3}}{3!} S_{(3)}(x) + \dots\right)
\end{equation}
is the formal expansion defined for the ground state wave function in the preceeding section, and deriving the associated Schr\"{o}dinger equation for the factor \(\starr{\phi}_{\hbar}(x)\). Since \(S_{\hbar}(x)\) will remain fixed throughout we refrain from attaching a superfluous overhead `naught' or `star' to it. We shall, of course, only be interested in generating non-constant solutions to the Schr\"{o}dinger equation for \(\starr{\phi}_{\hbar}(x)\) since the trivial, constant solutions merely reproduce multiples of the ground state.

Letting \(\hat{H}\) designate the Schr\"{o}dinger operator for the quartic oscillator,
\begin{equation}\label{eq:528}
\hat{H} = -\frac{\hbar^{2}}{2m} \frac{d^{2}}{dx^{2}} + \frac{1}{2} m\omega_{0}^{2}x^{2} + gx^{4},
\end{equation}
we thus seek solutions to
\begin{equation}\label{eq:529}
\hat{H} \left(\starr{\phi}_{\hbar}e^{-S_{\hbar}/\hbar}\right) = \hbar\starr{\mathcal{E}}_{\hbar} \left(\starr{\phi}_{\hbar}e^{-S_{\hbar}/\hbar}\right)
\end{equation}
in the sense of formal expansions in \(\hbar\) (holding uniformly \(\forall\; x \in \mathbb{R}\)) recalling that, in the foregoing section, we solved
\begin{equation}\label{eq:530}
\hat{H} \left(e^{-S_{\hbar}/\hbar}\right) = \hbar\zero{\mathcal{E}}_{\hbar} \left(e^{-S_{\hbar}/\hbar}\right)
\end{equation}
for \(S_{\hbar}\) and \(\zero{\mathcal{E}}_{\hbar}\) in the analogous way.

Expanding out Eq.~(\ref{eq:529}) and using Eq.~(\ref{eq:530}) to reexpress those terms in which \(\hat{H}\) acts purely on the \(e^{-S_{\hbar}/\hbar}\) factor one readily derives the associated Schr\"{o}dinger equation for \(\starr{\phi}_{\hbar}\),
\begin{equation}\label{eq:531}
\frac{1}{m} \frac{dS_{\hbar}}{dx} \frac{d\starr{\phi}_{\hbar}}{dx} - \frac{\hbar}{2m} \frac{d^{2}\starr{\phi}_{\hbar}}{dx^{2}} = \left(\Delta\starr{\mathcal{E}}_{\hbar}\right)\starr{\phi}_{\hbar}
\end{equation}
where
\begin{equation}\label{eq:532}
\begin{split}
\Delta\starr{\mathcal{E}}_{\hbar} &:= \starr{\mathcal{E}}_{\hbar} - \zero{\mathcal{E}}_{\hbar}\\
 &= \left\lbrack\left(\starr{\mathcal{E}}_{(0)} - \zero{\mathcal{E}}_{(0)}\right) + \hbar \left(\starr{\mathcal{E}}_{(1)} - \zero{\mathcal{E}}_{(1)}\right) + \frac{\hbar^{2}}{2!} \left(\starr{\mathcal{E}}_{(2)} - \zero{\mathcal{E}}_{(2)}\right)\right.\\
 &\left.\vphantom{\left(\starr{\mathcal{E}}_{(0)} - \zero{\mathcal{E}}_{(0)}\right) + \hbar \left(\starr{\mathcal{E}}_{(1)} - \zero{\mathcal{E}}_{(1)}\right) + \frac{\hbar^{2}}{2!} \left(\starr{\mathcal{E}}_{(2)} - \zero{\mathcal{E}}_{(2)}\right)}\; + \frac{\hbar^{3}}{3!} \left(\starr{\mathcal{E}}_{(3)} - \zero{\mathcal{E}}_{(3)}\right) + \dots\right\rbrack\\
 &= \left\lbrack\Delta\starr{\mathcal{E}}_{(0)} + \hbar\Delta\starr{\mathcal{E}}_{(1)} + \frac{\hbar^{2}}{2!}\Delta\starr{\mathcal{E}}_{(2)} + \frac{\hbar^{3}}{3!}\Delta\starr{\mathcal{E}}_{(3)} + \dots\right\rbrack
\end{split}
\end{equation}
designates the `gap' between the sought-for energy coefficients \(\left\lbrace\starr{\mathcal{E}}_{(0)}, \starr{\mathcal{E}}_{(1)}, \starr{\mathcal{E}}_{(2)}, \dots\right\rbrace\) and those previously derived for the ground state \(\left\lbrace\zero{\mathcal{E}}_{(0)}, \zero{\mathcal{E}}_{(1)}, \zero{\mathcal{E}}_{(2)}, + \dots\right\rbrace\).

In the spirit of the preceeding section we now expand \(\starr{\phi}_{\hbar}\) in the analogous fashion, setting
\begin{equation}\label{eq:533}
\starr{\phi}_{\hbar}(x) = \starr{\phi}_{(0)}(x) + \hbar\starr{\phi}_{(1)}(x) + \frac{\hbar^{2}}{2!}\starr{\phi}_{(2)}(x) + \frac{\hbar^{3}}{3!}\starr{\phi}_{(3)}(x) + \dots
\end{equation}
and substitute this together with the previously defined expansions for \(\Delta\starr{\mathcal{E}}_{\hbar}\) and \(S_{\hbar}\) into Eq.~(\ref{eq:531}), requiring the latter to hold, uniformly \(\forall\; x \in \mathbb{R}\), order by order in \(\hbar\).

The transport equation for the zeroth order term \(\starr{\phi}_{(0)}\) is easily seen to be
\begin{equation}\label{eq:534}
\frac{1}{m} \frac{dS_{(0)}}{dx} \frac{d\starr{\phi}_{(0)}}{dx} = \left(\Delta\starr{\mathcal{E}}_{(0)}\right)\starr{\phi}_{(0)}.
\end{equation}
wherein, from the preceeding section, one has
\begin{equation}\label{eq:535}
\frac{dS_{(0)}}{dx} = m\omega_{0}x\; \sqrt{1 + \frac{2gx^{2}}{m\omega_{0}^{2}}}.
\end{equation}
Integrating Eq.~(\ref{eq:534}) one finds the general solution
\begin{equation}\label{eq:536}
\starr{\phi}_{(0)}(x) = \starr{c}_{(0)} \left\lbrack\frac{x}{1 + \sqrt{1 + \frac{2gx^{2}}{m\omega_{0}^{2}}}}\right\rbrack^{\left(\frac{\Delta\starr{\mathcal{E}}_{(0)}}{\omega_{0}}\right)}
\end{equation}
where \(\starr{c}_{(0)}\) is a constant of integration. This expression is everywhere smooth and nontrivial if and only if \(\starr{c}_{(0)} \neq 0\) and \(\frac{\Delta\starr{\mathcal{E}}_{(0)}}{\omega_{0}}\) is a positive integer. Thus we henceforth refine the notation by writing
\begin{equation}\label{eq:537}
\begin{split}
\enn{\phi}_{(0)}(x) &:= \enn{c}_{(0)} \left\lbrack\frac{x}{1 + \sqrt{1 + \frac{2gx^{2}}{m\omega_{0}^{2}}}}\right\rbrack^{n},\\
\Delta\enn{\mathcal{E}}_{(0)} &= n\omega_{0},\qquad \enn{c}_{(0)} = \mathrm{constant} \neq 0,
\end{split}
\end{equation}
with \(n = 1,2,3,\dots\). Note that \(\enn{\phi}_{(0)}\) is bounded \(\forall\; x \in \mathbb{R}\) but reduces, in the limit \(g \searrow 0\), to \(\enn{c}_{(0)} x^{n}\) which is simply a constant multiple of the leading term in the (unbounded) Hermite polynomial \(H_{n} \left(\sqrt{\frac{m\omega_{0}}{\hbar}}x\right)\). Recall as well that, in this same limit, \(e^{-S_{\hbar}(x)/\hbar}\) reduces to the gaussian factor \(e^{-\frac{1}{2}\left(\frac{m\omega_{0}}{\hbar}\right)x^{2}}\).

It is worth nothing at this point that, had we chosen to study the (higher order) oscillators having potential energies given by
\begin{equation}\label{eq:538}
\begin{split}
V_{\kappa}(x) &= \frac{1}{2} m\omega_{0}^{2}x^{2} + gx^{2\kappa},\\
& \kappa = 2, 3, 4, \dots
\end{split}
\end{equation}
we would have obtained the following, more general result
\begin{equation}\label{eq:539}
{}_{\kappa}\enn{\phi}_{(0)}(x) = {}_{\kappa}\enn{c}_{(0)}\; \left\lbrack\frac{x}{\left(1 + \sqrt{1 + \frac{2gx^{2(\kappa - 1)}}{m\omega_{0}^{2}}}\right)^{1/(\kappa - 1)}}\right\rbrack^{n}
\end{equation}
with \(\kappa = 2, 3, 4, 5, \dots\) and \(n = 1, 2, 3, \dots\). Notice that, for each allowed value of \(\kappa\) and \(n\), \({}_{\kappa}\enn{\phi}_{(0)}(x)\) is bounded \(\forall\; x \in \mathbb{R}\) but again reduces to a constant multiple of \(x^{n}\) in the (harmonic) limit \(g \searrow 0\).

Returning to the quartic case and its associated Schr\"{o}dinger equation (\ref{eq:531}) one readily derives the transport equation for the first order `correction' \(\enn{\phi}_{(1)}\)
\begin{equation}\label{eq:540}
\begin{split}
\frac{1}{m} \frac{dS_{(0)}}{dx} \frac{d\enn{\phi}_{(1)}}{dx} &- n\omega_{0}\enn{\phi}_{(1)}\\
&= \Delta\enn{\mathcal{E}}_{(1)}\enn{\phi}_{(0)} + \frac{1}{2m} \frac{d^{2}\enn{\phi}_{(0)}}{dx^{2}} - \frac{1}{m} \frac{dS_{(1)}}{dx} \frac{d\enn{\phi}_{(0)}}{dx}
\end{split}
\end{equation}
wherein, of course, \(\enn{\phi}_{(0)}\) is a smooth, nontrivial solution to the zeroth order equation
\begin{equation}\label{eq:541}
\frac{1}{m} \frac{dS_{(0)}}{dx} \frac{d\enn{\phi}_{(0)}}{dx} - n\omega_{0}\enn{\phi}_{(0)} = 0,
\end{equation}
of the type just discussed, \(S_{(1)}\) is given by Eq.~(\ref{eq:307}) and \(\Delta\enn{\mathcal{E}}_{(1)}\) is an as yet undetermined constant. Comparing Eqs. (\ref{eq:540}) and (\ref{eq:541}) one sees that a natural technique for solving the first of these is the method of `variation of parameters' wherein one seeks a solution of the form
\begin{equation}\label{eq:542}
\enn{\phi}_{(1)}(x) = \enn{u}_{(1)}(x) \enn{\phi}_{(0)}(x).
\end{equation}
Some straightforward calculations, which made use of the preceeding, explicit expressions for \(S_{(0)}\), \(S_{(1)}\) and \(\enn{\phi}_{(0)}\), lead easily to the following equation for \(\enn{u}_{(1)}\):
\begin{equation}\label{eq:543}
\begin{split}
\frac{d\enn{u}_{(1)}(x)}{dx} &= \frac{m}{(xQ(x))^{3}} \left\lbrace\left(\frac{n\omega_{0}}{2}\right) \frac{m\omega_{0}}{Q(x)} [nQ(x) - m\omega_{0}]\right\rbrace\\
&~~ + \frac{m}{(xQ(x))} \left\lbrack\Delta\enn{\mathcal{E}}_{(1)} - \frac{2mg(n\omega_{0})\left(2Q(x) + \frac{3}{2}m\omega_{0}\right)}{Q^{3}(x)(Q(x) + m\omega_{0})}\right\rbrack
\end{split}
\end{equation}
where
\begin{equation}\label{eq:544}
\begin{split}
Q(x) &:= m\omega_{0} \sqrt{1 + \frac{2gx^{2}}{m\omega_{0}^{2}}}\\
&= \frac{1}{x} \frac{dS_{(0)}(x)}{dx}.
\end{split}
\end{equation}
The integral yielding \(\enn{u}_{(1)}(x)\) can be done explicitly but generates a logarithmic singularity at \(x = 0\) unless \(\Delta\enn{\mathcal{E}}_{(1)}\) is chosen to be
\begin{equation}\label{eq:545}
\Delta\enn{\mathcal{E}}_{(1)} = \frac{3}{2} \frac{gn(n+1)}{m^{2}\omega_{0}^{2}}
\end{equation}
in which case the expression for \(\enn{u}_{(1)}\) reduces to:
\begin{equation}\label{eq:546}
\begin{split}
\enn{u}_{(1)}(x) &= \frac{1}{4m\omega_{0}^{3} (xQ(x))^{2}}\; \left\lbrace 2 m^{2}\omega_{0}^{4} \cdot n\right.\\
& \; + gm\omega_{0}^{2}x^{2} (9n + 5n^{2}) + g^{2}x^{4} (18n + 10n^{2})\\
& \left.\vphantom{2 m^{2}\omega_{0}^{4} \cdot n}\; - (m\omega_{0}^{3} + 6g\omega_{0}x^{2}) Q(x) (n + n^{2})\right\rbrace.
\end{split}
\end{equation}
In deriving this we have adjusted the (otherwise arbitrary) constant of integration so as to cancel a constant term in the Laurent expansion for \(\enn{u}_{(1)}\) which now takes the form
\begin{equation}\label{eq:547}
\begin{split}
\enn{u}_{(1)}(x) &\simeq \frac{-n(n-1)}{4m\omega_{0}}\; \frac{1}{x^{2}} + \frac{(25n + 9n^{2})g^{2}x^{2}}{8m^{3}\omega_{0}^{5}}\\
&~~ + \frac{(-45n - 13n^{2}) g^{3}x^{4}}{8m^{4}\omega_{0}^{7}} + O[x^{6}].
\end{split}
\end{equation}
Though \(\enn{u}_{(1)}(x)\) has, for \(n > 1\), a singularity \(\propto \frac{n(n-1)}{x^{2}}\) at the origin, the resulting formula for \(\enn{\phi}_{(1)} = \enn{u}_{(1)}\enn{\phi}_{(0)}\) is easily seen to be smooth, and in fact bounded, \(\forall\; x \in \mathbb{R}\). In the harmonic limit however
\begin{equation}\label{eq:548}
\enn{\phi}_{(1)}(x) \xrightarrow[g \searrow 0]{} -\frac{n(n-1)}{4m\omega_{0}}\; \frac{\enn{c}_{(0)}}{2^{n}}\; x^{n-2}
\end{equation}
so that
\begin{equation}\label{eq:549}
\left(\enn{\phi}_{(0)} + \hbar\enn{\phi}_{(1)}\right)(x) \xrightarrow[g \searrow 0]{} \frac{\enn{c}_{(0)}}{2^{n}}\; \left\lbrace x^{n} - \frac{n(n-1)}{4} \left(\frac{\hbar}{m\omega_{0}}\right) x^{n-2}\right\rbrace
\end{equation}
which (for \(n > 1\)) is simply a constant multiple of the two leading terms in \(H_{n} \left(\sqrt{\frac{m\omega_{0}}{\hbar}} x\right)\).

Combining Eqs.~(\ref{eq:537}) and (\ref{eq:545}) one obtains the first order approximation for the energy gap
\begin{equation}\label{eq:550}
\begin{split}
\Delta\enn{E}_{\hbar} &:= \enn{E}_{\hbar} - \zero{E}_{\hbar}\\
&\simeq \hbar \left(\Delta\enn{\mathcal{E}}_{(0)} + \hbar\Delta\enn{\mathcal{E}}_{(1)} + \dots\right)\\
&\simeq \hbar\omega_{0} \left\lbrack n + \frac{3}{2} \frac{g\hbar}{m^{2}\omega_{0}^{3}} (n^{2} + n) + \dots\right\rbrack.
\end{split}
\end{equation}

The transport equation for the second order `correction' \(\enn{\phi}_{(2)}(x)\) is readily computed to be
\begin{equation}\label{eq:551}
\begin{split}
&\frac{1}{m} \frac{dS_{(0)}}{dx} \frac{d\enn{\phi}_{(2)}}{dx} - n\omega_{0}\enn{\phi}_{(2)}\\
&= -\frac{1}{m} \frac{dS_{(2)}}{dx} \frac{d\enn{\phi}_{(0)}}{dx} - \frac{2}{m} \frac{dS_{(1)}}{dx} \frac{d\enn{\phi}_{(1)}}{dx} + \frac{1}{m} \frac{d^{2}\enn{\phi}_{(1)}}{dx^{2}}\\
&~~ + \Delta\enn{\mathcal{E}}_{(2)}\enn{\phi}_{(0)} + 2\enn{\phi}_{(1)} \left\lbrack\frac{3}{2} \frac{gn(n+1)}{m^{2}\omega_{0}^{2}}\right\rbrack
\end{split}
\end{equation}
wherein \(\left\lbrace\enn{\phi}_{(0)}, \enn{\phi}_{(1)}, S_{(0)}, S_{(1)}\right\rbrace\) are the functions defined above, \(S_{(2)}\) is given by Eq.~(\ref{eq:513}) and \(\Delta\enn{\mathcal{E}}_{(2)}\) is an an yet undetermined constant.

Applying the method of variation of paramters we set
\begin{equation}\label{eq:552}
\enn{\phi}_{(2)}(x) = \enn{u}_{(2)}(x)\enn{\phi}_{(0)}(x)
\end{equation}
and derive, easily, the associated explicit formula for \(\frac{d\enn{u}_{(2)}}{dx}\). The integral yielding \(\enn{u}_{(2)}\) contains a logarithmic singularity at \(x = 0\) unless \(\Delta\enn{\mathcal{E}}_{(2)}\) is chosen to be
\begin{equation}\label{eq:553}
\Delta\enn{\mathcal{E}}_{(2)} = \frac{-g^{2}}{4m^{4}\omega_{0}^{5}} (59n + 51n^{2} + 34n^{3})
\end{equation}
in which case the expression for \(\enn{u}_{(2)}\) reduces to:
\begin{equation}\label{eq:554}
\begin{split}
\enn{u}_{(2)}(x) &= \frac{n}{48m^{3}\omega_{0}^{6}x^{4}(Q(x))^{5}}\; \left\lbrace -2m\omega_{0}\left\lbrack 3m^{4}\omega_{0}^{8} (11 + 5n + 4n^{2})\right.\right.\\
&~~ + 3gm^{3}\omega_{0}^{6}x^{2} (35 + 28n + 32n^{2} + 5n^{3})\\
&~~ + 2g^{2}m^{2}\omega_{0}^{4}x^{4} (-257 - 42n + 104n^{2} + 75n^{3})\\
&~~ + 28g^{3}m\omega_{0}^{2}x^{6} (-59 - 24n + 8n^{2} + 15n^{3})\\
&~~ + \left.\vphantom{3m^{4}\omega_{0}^{8} (11 + 5n + 4n^{2})} 24g^{4}x^{8} (-59 - 24n + 8n^{2} + 15n^{3})\right\rbrack\\
&~~ + Q(x) \left\lbrack 3m^{4}\omega_{0}^{8} (16 + 21n + 2n^{2} + n^{3})\right.\\
&~~ + 6gm^{3}\omega_{0}^{6}x^{2} (12 + 37n + 24n^{2} + 7n^{3})\\
&~~ + g^{2}m^{2}\omega_{0}^{4}x^{4} (-1211 - 351n + 380n^{2} + 282n^{3})\\
&~~ + 4g^{3}m\omega_{0}^{2}x^{6} (-1355 - 837n + 8n^{2} + 156n^{3})\\
&~~ + \left.\left.\vphantom{3m^{4}\omega_{0}^{8} (16 + 21n + 2n^{2} + n^{3})} 4g^{4}x^{8} (-1355 - 891n - 100n^{2} + 102n^{3})\right\rbrack\right\rbrace
\end{split}
\end{equation}
wherein, as above \(Q(x)\) is given by Eq.~(\ref{eq:544}). We have again adjusted the choice of a constant of integration so as to cancel a constant term in the Laurent expansion for \(\enn{u}_{(2)}\) which now takes the form
\begin{equation}\label{eq:555}
\begin{split}
\enn{u}_{(2)}(x) &\simeq \frac{(n-3)(n-2)(n-1)n}{16m^{2}\omega_{0}^{2}} \cdot \frac{1}{x^{4}}\\
&~~ + \frac{g}{m^{3}\omega_{0}^{4}}\; \frac{n^{2}(n-1)}{x^{2}}\\
&~~ + \frac{g^{3}n}{16m^{5}\omega_{0}^{8}}\; (-1098 - 719n - 140n^{2} + 13n^{3}) x^{2}\\
&~~ + \frac{g^{4}n}{64m^{6}\omega_{0}^{10}}\; (12083 + 7217n + 1460n^{2} - 4n^{3}) x^{4}\\
&~~ + O [x^{6}].
\end{split}
\end{equation}
Since the singular term \(\propto \frac{(n-3)(n-2)(n-1)n}{x^{4}}\) is present only for \(n \geq 4\) and that \(\propto \frac{n^{2}(n-1)}{x^{2}}\) only when \(n \geq 2\) it's clear that the product \(\enn{\phi}_{(2)} = \enn{u}_{(2)}\enn{\phi}_{(0)}\) is in fact smooth \(\forall\; x \in \mathbb{R}\). It is easily seen to be bounded as well though, in the harmonic limit it reduces to
\begin{equation}\label{eq:556}
\enn{\phi}_{(2)}(x) \xrightarrow[g\searrow 0]{} \frac{n(n-3)(n-2)(n-1)}{16m^{2}\omega_{0}^{2}}\; \frac{\enn{c}_{(0)}}{2^{n}} x^{n-4}
\end{equation}
so that
\begin{equation}\label{eq:557}
\begin{split}
\left(\enn{\phi}_{(0)} + \hbar\enn{\phi}_{(1)} + \frac{\hbar^{2}}{2!}\enn{\phi}_{(2)}\right) (x) &\xrightarrow[g\searrow 0]{} \frac{\enn{c}_{(0)}}{2^{n}}\; \left\lbrace x^{n} - \frac{n(n-1)}{4} \left(\frac{\hbar}{m\omega_{0}}\right) x^{n-2}\right.\\
 &~~ + \left.\vphantom{x^{n} - \frac{n(n-1)}{4} \left(\frac{\hbar}{m\omega_{0}}\right) x^{n-2}}\frac{n(n-1)(n-2)(n-3)}{32}\; \left(\frac{\hbar}{m\omega_{0}}\right)^{2} x^{n-4}\right\rbrace
 \end{split}
\end{equation}
which (for \(n > 3\)) is a constant multiple of the three leading terms in \(H_{n} \left(\sqrt{\frac{m\omega_{0}}{\hbar}} x\right)\). It is not difficult to prove that, in this limiting case (i.e., setting \(g = 0\) everywhere), our method will simply regenerate the full Hermite polynomials and then terminate thereby reconstructing the exact excited states (after multiplication by the gaussian factor) for the pure harmonic oscillator.

Returning to the quartic oscillator results and combining Eqs.~(\ref{eq:537}), (\ref{eq:545}) and (\ref{eq:553}) we find that our approximation yields the following second order formula for the \(n\)-th excited state's (total) energy
\begin{equation}\label{eq:558}
\begin{split}
\enn{E}_{\hbar} &\simeq \hbar\omega_{0}\; \left\lbrack\left( n + \frac{1}{2}\right) + \frac{3}{2} \left(\frac{g\hbar}{m^{2}\omega_{0}^{3}}\right) \left( n^{2} + n + \frac{1}{2}\right)\right.\\
&~~ \left.\vphantom{\left( n + \frac{1}{2}\right) + \frac{3}{2} \left(\frac{g\hbar}{m^{2}\omega_{0}^{3}}\right) \left( n^{2} + n + \frac{1}{2}\right)}- \frac{1}{8} \left(\frac{g\hbar}{m^{2}\omega_{0}^{3}}\right)^{2} (34n^{3} + 51n^{2} + 59n + 21)\right\rbrack .
\end{split}
\end{equation}
This agrees with the standard result computed using conventional perturbation theory carried out to the corresponding order \cite{Pathak2001}. We conjecture that such agreement will persist to arbitrary order but a full proof of this, if true, would presumably require the development of suitable recurrence relations for the (excited state) energy coefficients. We shall not pursue that issue herein.

It may seem paradoxical that, to the orders that we have computed (and conjecturally to all orders), both ground and excited state energy formulas agree with those of conventional perturbation theory whereas the corresponding wave functions of the two approaches differ dramatically. Our expectation though is that one could use our results to regenerate the conventional approximate wave functions of perturbation theory by expanding our formulas for \(S_{\hbar}(x)\), etc., in powers of \(g\), Taylor expanding the coefficients of the gaussian factor \(e^{-\frac{1}{2}\left(\frac{m\omega_{0}}{\hbar}\right) x^{2}}\) and truncating the results at the desired order in \(g\) and finally reexpressing these truncated coefficients as finite series of Hermite polynomials \(\left\lbrace H_{n} \left(\sqrt{\frac{m\omega_{0}}{\hbar}} x\right)\right\rbrace\).

We conclude this section by pointing out that Eq.~(\ref{eq:539}) for \({}_{\kappa}\enn{\phi}_{(0)}(x)\) simply expresses this quantity as (a constant multiple of) the \(n\)-th power of the corresponding Sternberg coordinate for the associated oscillator. More precisely, the transformation
\begin{equation}\label{eq:559}
y = \frac{2^{1/(\kappa - 1)}x}{(1 + \sqrt{1 + \frac{2gx^{2(\kappa - 1)}}{m\omega_{0}^{2}}})^{1/(\kappa - 1)}}
\end{equation}
with smooth inverse
\begin{equation}\label{eq:560}
x = \frac{y}{(1 - \frac{g}{2m\omega_{0}^{2}} y^{2(\kappa - 1)})^{1/(\kappa - 1)}}
\end{equation}
maps \(\mathbb{R} \ni x\) diffeomorphically to the interval
\begin{equation}\label{eq:561}
\left(-\left(\frac{2m\omega_{0}^{2}}{g}\right)^{1/2(\kappa - 1)}, \left(\frac{2m\omega_{0}^{2}}{g}\right)^{1/2(\kappa - 1)}\right) \ni y
\end{equation}
and transforms the Hamilton-Jacobi flow vector field
\begin{equation}\label{eq:562}
X_{H} := \frac{1}{m} \frac{dS_{(0)}}{dx} \frac{d}{dx} = \omega_{0}x \sqrt{1 + \frac{2gx^{2(\kappa - 1)}}{m\omega_{0}^{2}}} \frac{d}{dx}
\end{equation}
to the Sternberg form
\begin{equation}\label{eq:563}
X_{H} = \omega_{0} y \frac{d}{dy}
\end{equation}
that was discussed, in a more general setting, at the end of Sect.{~\ref{sect:integration-transport}}.

Though we did not exploit this transformation to carry out the foregoing calculations it is quite conceivable that they would have been simplified thereby and, furthermore, that the use of Sternberg coordinates could facilitate the resolution of some of the conjectural issues mentioned above.

\section{Concluding Remarks}
\label{sec:concluding}

For the Lagrangians normally considered in classical mechanics it would not be feasible to define their corresponding action functionals over (semi-)infinite domains, as we have done, since the integrals involved, when evaluated on solutions to the Euler Lagrange equations, would almost never converge. It is only because of the special nature of our problem, with its inverted potential energy function and associated boundary conditions, that we could define a convergent action integral for the class of curves of interest and use this functional to determine corresponding minimizers. Since this problem is of a highly non-standard type within the calculus of variations we have felt obligated to give, in Sect.~(\ref{sec:existence-solution}), a rather complete and self-contained analysis of the existence, uniqueness and smoothness properties of the associated minimizing curves together with a proof of smoothness and asymptotic behavior for the action function, \(S_{(0)}(\mathbf{x})\), computed from these minimizers.

A remarkable feature of our construction, given the hypotheses of convexity and coercivity imposed upon the potential energy \(V(\mathbf{x})\), was that it led to a \textit{globally smooth} solution to the corresponding ipze Hamilton-Jacobi equation. Normally the solutions to a Hamilton-Jacobi equation in mechanics fail to exist globally, even for rather elementary problems, because of the occurrence of caustics in the associated families of solution curves. For our problem however caustics were non-existent for the (semi-)flow generated by the gradient of \(S_{(0)}(\mathbf{x})\). The basic reason for this was the inverted potential character of the forces considered which led to the development of diverging (in the future time direction) solution curves having, in effect, uniformly positive Lyapunov exponents that served to prevent the occurrence of caustics altogether.

By contrast, the more conventional approach (in the physics literature) to semi-classical methods leads instead to the \textit{standard} (non-inverted-potential-non-zero-energy) Hamilton-Jacobi equation for which, especially in higher dimensions, caustics are virtually unavoidable and for which, even in their absence, a non-trivial matching of solutions across the boundary separating classically allowed and classically forbidden regions must be performed. While Maslov and others have developed elegant methods for dealing with these complications \cite{Maslov1981} their techniques and results are more appropriate in the short wavelength limit wherein wave packets of highly excited states are the central objects of interest. On the other hand our approach is aimed at the ground and lower excited states though, in principle, it is not limited thereto.

As we have already mentioned though, our approach is a natural variation of one that has been extensively developed in the microlocal analysis literature but it also differs from this innovative work in fundamental ways that are crucial for our ultimate, intended applications. In the microlocal approach \cite{Dimassi1999,Helfer1984} one begins by analyzing the (classical, inverted potential) dynamics locally, near an equilibrium, by appealing to the stable manifold theorem of mechanics \cite{Abraham1978}. One then shows, by a separate argument, that, for an equilibrium \(p\) (lying in some neighborhood \(U \subset \mathbb{R}^{n}\)) the corresponding stable (\(W^{s}(p) \subset T^{\ast}U\)) and unstable (\(W^{u}(p) \subset T^{\ast}U\)) submanifolds of the associated phase space \(T^{\ast}U\) are in fact Lagrangian submanifolds that can be characterized as the graphs of the (positive and negative) gradients of a smooth function \(\phi : U \longrightarrow \mathbb{R}\):
\begin{equation}\label{eq:425}
\begin{split}
W^{u}(p) &= \left\lbrace (\mathbf{x},\mathbf{p})\; |\; \mathbf{x} \in U,\; \mathbf{p} = \nabla\phi (\mathbf{x})\right\rbrace\\
W^{s}(p) &= \left\lbrace (\mathbf{x},\mathbf{p})\; |\; \mathbf{x} \in U,\; \mathbf{p} = -\nabla\phi (\mathbf{x})\right\rbrace.
\end{split}
\end{equation}
This function is shown to satisfy a certain `eikonal' equation (equivalent to our ipze Hamilton-Jacobi equation restricted to \(U\)) and \(\phi (\mathbf{x})\) itself is, of course, nothing but the (locally defined) analogue of our action function \(S_{(0)}(\mathbf{x})\).

The potential energies, \(V(\mathbf{x})\), dealt with in the microlocal literature often entail multiple local minima, or `wells', for which our global convexity and coercivity hypotheses are not appropriate. Much of the detailed analysis therein involves a careful matching of locally defined approximate solutions (constructed on suitable neighborhoods of each well) to yield global asymptotic approximations to the eigenvalues and eigenfunctions for such problems. Since, however, we are focused primarily on potential energies having single wells (corresponding to unique classical `ground states'), many of the technical features of this elegant analysis are not directly relevant to the issues dealt with herein.

For this case of a single well, however, we have essentially unified and globalized several of the aforementioned, local arguments, replacing them with an integrated study of the properties of the (inverted potential) action functional \(\mathcal{I}_{ip}[\gamma]\) (\textit{c.f.}, Eq.~(\ref{eq:301})). When one turns from finite dimensional problems to field theoretic ones \cite{Marini} this change of analytical strategy begins to play a crucial role. For the typical (relativistic, bosonic) field theories of interest to us in this context, the Euler Lagrange equations for the corresponding, inverted-potential action functionals that arise are the Euclidean signature, elliptic analogues of the Lorentzian signature, hyperbolic field equations that one is endeavoring to quantize. While generalizations of the aforementioned stable manifold theorem do exist for certain types of infinite dimensional dynamical systems, the elliptic field equations of interest to us do not correspond to well-defined dynamical systems \textit{at all}. In particular their associated Canchy initial value problems are never well-posed.

On the other hand certain elliptic boundary value problems for such Euclidean signature action functionals \textit{are} mathematically meaningful and, fortuitously, a number of the most important of these have already been the object of rigorous study \cite{Marini1992,Maitra2007}. In a companion paper to the present one we shall review some of the principal results of this analysis and extend them with the aim of generalizing the arguments given herein to a natural infinite dimensional setting \cite{Marini}.

\section*{Acknowlegements}

Moncrief is grateful to the Albert Einstein Institute in Golm, Germany, the Erwin Schr\"{o}dinger Institute in Vienna, Austria and the University of Vienna for hospitality and support during his work on this paper. Moncrief was supported in part by NSF grant PHY-0963869 to Yale University.

\appendix
\section{Appendix}
\label{sec:appendix}

Sternberg's linearization theorem \cite{Sternberg1957} implies that, at least on a neighborhood \(\mathcal{U}\) of the origin in \(\mathbb{R}^{n}\), there exists a diffeomorphism
\begin{gather}\label{eq:a01}
\mu : \mathcal{U} \longrightarrow \mu (\mathcal{U}) \subset \mathbb{R}^{n} = \left\lbrace (y^{1}, \dots , y^{n})\right\rbrace\nonumber\\
\mathbf{x} \longmapsto \mu(\mathbf{x}) = \left(y^{1}(\mathbf{x}), \dots , y^{n}(\mathbf{x})\right)
\end{gather}
with
\begin{equation}\label{eq:a02}
y^{i}(\mathbf{x}) = x^{i} + O(|x|^{2})
\end{equation}
and
\begin{equation}\label{eq:a03}
\frac{\partial y^{i}}{\partial x^{j}} = \delta_{j}^{i} + O(|x|)
\end{equation}
such that on \(\mu (\mathcal{U})\) the vector field \(\frac{\nabla S_{(0)}}{m}\) takes the form
\begin{equation}\label{eq:a04}
\sum_{i=1}^{n} \frac{1}{m} \frac{\partial S_{(0)}(\mathbf{x})}{\partial x^{i}} \frac{\partial}{\partial x^{i}} \underset{\mu}{\longrightarrow} \sum_{i=1}^{n} \omega_{i}y^{i} \frac{\partial}{\partial y^{i}}.
\end{equation}
By exploiting special features already established for \(\frac{\nabla S_{(0)}}{m}\) however, we shall be able to extend Sternberg coordinates to a global chart defined on a star-shaped domain \(K \subset \mathbb{R}^{n}\) with \(\mu^{-1} (K) \approx \mathbb{R}^{n} = \left\lbrace(x^{1}, \ldots , x^{n})\right\rbrace\), such that
\begin{equation}\label{eq:a05}
\frac{1}{m} \nabla S_{(0)} = \sum_{i=1}^{n} \omega_{i}y^{i} \frac{\partial}{\partial y^{i}}
\end{equation}
everywhere on \(K\). For this purpose we shall first need some further insight into the (semi-) flow generated by \(\frac{1}{m} \nabla S_{(0)}\).

\subsection{Global Properties  of the Gradient (Semi-) Flow of \(S_{(0)}\)}
\label{subsec:global}

We know from Sect.~(\ref{sec:existence-solution}) that \(S_{(0)} : \mathbb{R}^{n} \longrightarrow \mathbb{R}\) is a globally smooth solution to the ipze Hamilton-Jacobi equation
\begin{equation}\label{eq:a06}
\frac{1}{2m} \sum_{i=1}^{n} \left(\frac{\partial S_{(0)}(\mathbf{x})}{\partial x^{i}}\right)^{2} - V(\mathbf{x}) = 0
\end{equation}
with the asymptotic behavior
\begin{equation}\label{eq:a07}
\begin{split}
S_{(0)}(\mathbf{x})  &= \frac{1}{2} m \sum_{i=1}^{n} \omega_{i}(x^{i})^{2} + O(|\mathbf{x}|^{3})\\
\partial_{j} S_{(0)}(\mathbf{x}) &= m \omega_{j}x^{j} + O(|x|^{2}).
\end{split}
\end{equation}
This function was determined by minimizing the ip action functional (\ref{eq:301}) on the affine space of curves \(\mathcal{D}_{\mathbf{x}}\) (\textit{c.f.}, (\ref{eq:302})\:) having (arbitrary) fixed endpoint \(\mathbf{x} \in \mathbb{R}^{n}\), evaluating the action on the minimizing curve, \(\gamma_{\mathbf{x}}\), and setting
\begin{equation}\label{eq:a08}
S_{(0)}(\mathbf{x}) := \mathcal{I}_{ip}[\gamma_{\mathbf{x}}].
\end{equation}

In view of the coercivity condition (\ref{eq:320}) assumed for the potential energy however, it follows that, for any curve \(\gamma(t) = \left(x^{1}(t), \ldots ,  x^{n}(t)\right)\) lying in \(\mathcal{D}_{\mathbf{x}}\), one has
\begin{equation}\label{eq:a09}
\mathcal{I}_{ip}[\gamma] \geq \mathcal{I}_{ip}^{\ast}[\gamma]
\end{equation}
where
\begin{equation}\label{eq:a10}
\mathcal{I}_{ip}^{\ast}[\gamma]  := \int_{-\infty}^{0} \left\lbrace\frac{1}{2} m \sum_{i=1}^{n} \left\lbrace\left(\dot{x}^{i}(t)\right)^{2} + \nu_{i}^{2} \left( x^{i}(t)\right)^{2}\right\rbrace\right\rbrace dt
\end{equation}
with
\begin{equation}\label{eq:a11}
\nu_{i} := \sqrt{\omega_{i}^{2} - \lambda_{i}^{2}} > 0\; \forall\; i \in [1, \ldots , n].
\end{equation}
But the minimizing curves for  \(\mathcal{I}_{ip}^{\ast}\) within \(\mathcal{D}_{\mathbf{x}}\) are simply  given by
\begin{equation}\label{eq:a12}
\gamma_{\mathbf{x}}^{\ast}(t) = (x^{1} e^{\nu_{i}t}, \ldots , x^{n} e^{\nu_{n}t})
\end{equation}
for which one finds that
\begin{equation}\label{eq:a13}
\begin{split}
S_{(0)}^{\ast}(\mathbf{x}) &:= \mathcal{I}_{ip}^{\ast} [\gamma_{\mathbf{x}}^{\ast}]\\
 &= \frac{1}{2} m \sum_{i=1}^{n} \nu_{i}(x^{i})^{2}.
\end{split}
\end{equation}
It follows that, globally on \(\mathbb{R}^{n}\),
\begin{equation}\label{eq:a14}
S_{(0)}(\mathbf{x}) \geq S_{(0)}^{\ast}(\mathbf{x}) = \frac{1}{2} m \sum_{i=1}^{n} \nu_{i} (x^{i})^{2}
\end{equation}
and thus, in particular, that any level set of \(S_{(0)}\) with level value less than \(s > 0\) lies inside the (hyper-) ellipsoid defined by
\begin{equation}\label{eq:a15}
S_{(0)}^{\ast}(\mathbf{x}) = s.
\end{equation}

From Eq.~(\ref{eq:201a}) and the properties assumed for \(V(\mathbf{x})\) it's clear that the gradient of \(S_{(0)}\) only vanishes at the origin which coincides with the (exceptional) level set \(S_{(0)}^{-1}(0)\). For any level value \(s > 0\) the corresponding level set, \(S_{(0)}^{-1}(s)\), is therefore necessarily a smoothly embedded, \(n-1\) dimensional submanifold of \(\mathbb{R}^{n}\). In view of its boundedness, \(S_{(0)}^{-1}(s)\) is compact and, from the asymptotic behavior of \(S_{(0)}\) near the origin (\textit{c.f.} Eq.~(\ref{eq:368})\:), it's clear that the level sets for sufficiently small \(s > 0\) are topological \(n - 1\) spheres surrounding the origin. Since \(\nabla S_{(0)}\) is nowhere vanishing away from the origin it moreover follows, from basic Morse theory, that every level set of \(S_{(0)}\) (corresponding to a level value \(> 0\)) is also a smoothly embedded, topological \(n - 1\) sphere.

In view of the fact that, along the integral curves \(\gamma_{\mathbf{x}}\) of \(\frac{1}{m} \nabla S_{(0)}\), \(S_{(0)}\) obeys
\begin{equation}\label{eq:a16}
\begin{split}
\frac{d}{dt} S_{(0)} \left(\gamma_{\mathbf{x}}(t)\right)  &= \frac{1}{m} \nabla S_{(0)} \cdot \nabla S_{(0)}\; \left(\gamma_{\mathbf{x}}(t)\right)\\
 &= 2 V \left(\gamma_{\mathbf{x}}(t)\right)
\end{split}
\end{equation}
we see, since \(V(\mathbf{x}) > 0\) except at the origin, that \(S_{(0)} \left(\gamma_{\mathbf{x}}(t)\right)\) is monotonically decreasing as \(t \searrow -\infty\) with
\begin{equation}\label{eq:a17}
\lim_{t\searrow -\infty} S_{(0)} \left(\gamma_{\mathbf{x}}(t)\right) = S_{(0)}(0, \ldots , 0) = 0\qquad \forall\; \mathbf{x} \in \mathbb{R}^{n}.
\end{equation}

Let \(\mathcal{B}_{s} \subset \mathbb{R}^{n}\) designate, for any \(s > 0\), the open \(n\)-ball bounded by the (topological) (\(n - 1\))-sphere \(S_{(0)}^{-1} (s)\). From the asymptotic behavior of \(S_{(0)}(\mathbf{x})\) (c.f. Eq.~(\ref{eq:a07})) there clearly exists an \(s^{*} > 0\), sufficiently small, such that \(\mathcal{B}_{s^{*}}\) lies entirely within the neighborhood \(U \subset \mathbb{R}^{n}\) on which Sternberg's theorem ensures the existence of the diffemorphism \(\mu\) described above. From the monotonicity property (\ref{eq:a16}--\ref{eq:a17}) every integral curve of \(\frac{1}{m} \nabla S_{(0)}\), followed in the negative time direction, eventually enters and remains within \(\mathcal{B}_{s^{*}}\), asymptotically approaching the origin as \(t \searrow -\infty\).

In fact, using the monotonicity formula, it is easy to place an upper bound on the time required for any point \(\mathbf{x} \in \mathcal{B}_{s}\) to flow to (and remains inside) the `Stenberg domain' \(\mathcal{B}_{s^{*}}\). For \(s \leq s^{*}\) there is nothing to do so assume \(s > s^{*}\) and let \(\mathcal{A}_{s^{*},s} \subset \mathbb{R}^{n}\) designate the compact, annular solid region defined by
\begin{equation}\label{eq:a18}
\mathcal{A}_{s^{*},s} := \overline{\mathcal{B}_{s}} \backslash \mathcal{B}_{s^{*}}
\end{equation}
and thus bounded inside and out by \(S_{(0)}^{-1} (s^{*})\) and \(S_{(0)}^{-1} (s)\) respectively.

Since the potential energy function \(V (\mathbf{x})\) is continuous and strictly positive on the compact annulus \(\mathcal{A}_{s^{*},s}\) it achieves (strictly positive) maximal and minimal values thereon. Let \(V_{s^{*},s}^{\mathrm{min}} > 0\) designate the corresponding minimum. From (\ref{eq:a16}) it follows that, for any \(t \leq T_{s^{*},s}^{\mathrm{max}} := \frac{s^{*} - s}{2V_{s^{*},s}^{\mathrm{min}}} < 0\), every integral curve \(\gamma\) having \(\gamma (0) \in \mathcal{B}_{s}\) will satisfy \(\gamma (t) \in \mathcal{B}_{s^{*}}\). Thus, along the (semi-) flow generated by \(\frac{1}{m} \nabla S_{(0)}\), \(\mathcal{B}_{s}\) gets mapped to an image lying within \(\mathcal{B}_{s^{*}}\) for all \(t \leq T_{s^{*},s}^{\mathrm{max}} < 0\).

By standard results on the (semi-) flows of smooth vector fields whose integral curves are all complete in the negative \(t\)-direction, it follows that the aforementioned mapping is, in fact, a \(C^{\infty}\) diffeomorphism \cite{Arnold1978}. Thus
\begin{equation}\label{eq:a19}
\gamma^{t} : \mathcal{B}_{s} \longrightarrow \gamma^{t} (\mathcal{B}_{s}) \subset \mathcal{B}_{s^{*}}
\end{equation}
defined, for \(t \leq T_{s^{*},s}^{\mathrm{max}}\), by
\begin{equation}\label{eq:a20}
\gamma^{t} (\mathbf{x}) = \gamma_{\mathbf{x}} (t),
\end{equation}
where \(\gamma_{\mathbf{x}}(\cdot)\) is the integral curve of \(\frac{1}{m} \nabla S_{(0)}\) satisfying \(\gamma_{\mathbf{x}} (0) = \mathbf{x} \in \mathcal{B}_{s}\), is a diffeomorphism taking \(\mathcal{B}_{s}\) to its range lying within the Sternberg domain \(B_{s^{*}}\).

\subsection{Extending the Sternberg Diffeomorphism}
\label{subsec:diffeomorphism}

Let \(\gamma_{\mathbf{x}} (t_{1})\), \(\gamma_{\mathbf{x}} (t_{2})\) designate any two points along an arbitrary integral curve \(\gamma_{\mathbf{x}}\) of \(\frac{1}{m} \nabla S_{(0)}\) that both lie within the Sternberg domain \(\mathcal{B}_{s^{*}}\) and assume \(t_{1} < t_{2} \leq 0\). Note that \(t_{2} = 0\) would be allowed here if \(\mathbf{x} = \gamma_{\mathbf{x}} (0)\) already lies within \(\mathcal{B}_{s^{*}}\). Utilizing the Sternberg coordinate expression for \(\frac{1}{m} \nabla S_{(0)}\), given by Eq.~(\ref{eq:a04}) and valid throughout \(B_{s^{*}}\), one easily shows that
\begin{equation}\label{eq:a21}
\mu^{i} \left(\gamma_{\mathbf{x}} (t_{1})\right) e^{-\omega_{i}t_{1}} = \mu^{i} \left(\gamma_{\mathbf{x}} (t_{2})\right) e^{-\omega_{i}t_{2}}
\end{equation}
\(\forall\; i \in [1, \ldots ,n]\) and for any such choice of \(\gamma_{\mathbf{x}}\), \(t_{1}\) and \(t_{2}\).

Let us tentatively extend the domain of definition of the Sternsberg map \(\mu\) (assumed for simplicity to coincide initially with the \(n\)-ball \(\mathcal{B}_{s^{*}}\)) to \(\mathcal{B}_{s}\) by setting, for any \(\mathbf{x} \in \mathcal{B}_{s}\) and for some \(t \leq T_{s^{*},s}^{\mathrm{max}} < 0\),
\begin{equation}\label{eq:a22}
y^{i} (\mathbf{x}) = \mu^{i} (\mathbf{x}) := \mu^{i} \left(\gamma_{\mathbf{x}} (t)\right) e^{-\omega_{i}t}
\end{equation}
\(\forall\; i \in [1, \ldots ,n]\). In view of Eq.~(\ref{eq:a21}) it is easily verified that the map so defined is independent of the actual value of \(t\) chosen. Furthermore since, for any such \(t\), \(\gamma^{t} (\cdot) = \gamma_{\cdot} (t)\) is a diffeomorphism defined on \(\mathcal{B}_{s}\), we see that the formula
\begin{equation}\label{eq:a23}
y^{i}(\cdot) = \mu^{i} (\cdot) := \mu^{i} \left(\gamma^{t}(\cdot)\right) e^{-\omega_{i}t}
\end{equation}
\(\forall\; i \in [1, \ldots ,n]\), for the extension of \(\mu\) to \(\mathcal{B}_{s}\), is simply a composition of diffeomorphisms. By virtue of (\ref{eq:a21}) it is easily checked that this extended \(\mu\), if restricted to \(\mathcal{B}_{s^{*}}\), coincides with the Sternberg map originally given.

One can repeat this construction for a sequence \(s_{i} \longrightarrow \infty\) whose corresponding domains \(\mathcal{B}_{s_{i}}\) exhaust \(\mathbb{R}^{n}\) as \(i \longrightarrow \infty\) and verify as above that, with each successive enlargement of the domain for \(\mu\), its restriction to the previous domain coincides with the previous definition. In this way one arrives at a resulting diffeomorphism \(\mu\) whose maximal domain of definition exhausts \(\mathbb{R}^{n}\).

To see that this extended \(\mu\) map has the desired property of transforming \(\frac{1}{m} \nabla S_{(0)}\) to the Sternberg form (\ref{eq:a05}) consider an arbitrary segment \begin{equation}\label{eq:a24}
\gamma_{\mathbf{x}} : (-\epsilon, \epsilon) \longrightarrow \mathbb{R}^{n}
\end{equation}
of an arbitrary integral curve \(\gamma_{\mathbf{x}}\) of \(\frac{1}{m} \nabla S_{(0)}\). For any \(\epsilon > 0\) one can choose a \(T < 0\) sufficiently large and negative that
\begin{equation}\label{eq:a25}
\gamma_{\gamma_{\mathbf{x}}(\lambda)} (T) : (-\epsilon ,\epsilon) \longrightarrow \mathbb{R}^{n}
\end{equation}
will be entirely within \(\mathcal{B}_{s^{*}}\). By virtue of the autonomous character of the vector field \(\frac{1}{m} \nabla S_{(0)}\) it follows that
\begin{equation}\label{eq:a26}
\begin{split}
\gamma_{\gamma_{\mathbf{x}}(\lambda)} (T) &= \gamma_{\gamma_{\mathbf{x}}(0)} (T+\lambda)\\
 &= \gamma_{\mathbf{x}} (T+\lambda) \in \mathcal{B}_{s^{*}}
\end{split}
\end{equation}
\(\forall\; \lambda \in (-\epsilon,\epsilon)\). Computing the Sternberg coordinate form for this curve segment we get
\begin{equation}\label{eq:a27}
\begin{split}
y^{i}(\lambda) &:= y^{i} \left(\gamma_{\mathbf{x}}(\lambda)\right) = \mu^{i} \left(\gamma_{\gamma_{\mathbf{x}}(\lambda)}(T)\right) e^{-\omega_{i}T}\\
 &= \mu^{i} \left(\gamma_{\mathbf{x}}(T+\lambda)\right) e^{-\omega_{i}T}\\
 &= \mu^{i} \left(\gamma_{\mathbf{x}}(T+\lambda)\right) e^{-\omega_{i}(T+\lambda)} e^{\omega_{i}\lambda}\\
 &= \mu^{i} \left(\gamma_{\mathbf{x}}(T)\right) e^{-\omega_{i}T} e^{\omega_{i}\lambda}\\
 &= \mu^{i} (\mathbf{x}) e^{\omega_{i}\lambda}\qquad\qquad \forall\; i \in [1, \ldots ,n]
\end{split}
\end{equation}
Thus
\begin{equation}\label{eq:a28}
\begin{split}
y^{i} (\lambda) &:= y^{i} \left(\gamma_{\mathbf{x}}(\lambda)\right)\\
 &= \mu^{i} (\mathbf{x}) e^{\omega_{i}\lambda}
\end{split}
\end{equation}
which gives immediately that
\begin{equation}\label{eq:a29}
\frac{dy^{i} (\lambda)}{d\lambda} = \omega_{i} y^{i} (\lambda)
\end{equation}
along the arbitrary curve segment. Since \(\mu\) is a globally defined diffeomorphism on \(\mathbb{R}^{n}\) it follows that it transforms \(\frac{1}{m} \nabla S_{(0)}\) to the Sternberg form (\ref{eq:a05}). From the standard definition (c.f., p. 23 of \cite{Dimassi1999}) both \(\mathbb{R}^{n}\) and its image under \(\mu\) are \textit{star-shaped domains} for the (semi-) flow generated by \(\frac{1}{m} \nabla S_{(0)}\).

We conclude this appendix with a brief discussion of how the Jacobian determinant of the transformation to Sternberg coordinates varies along the integral curves of \(\frac{1}{m} \nabla S_{(0)}\). Writing \(y^{i} = \mu^{i} (x^{1}, \ldots , x^{n})\) and \(\left\lbrace x^{a}\right\rbrace = \left\lbrace x^{1}, \ldots , x^{n}\right\rbrace\) we have,
\begin{equation}\label{eq:a30}
\frac{1}{\sqrt{\mathrm{det}g_{ij}}} = \mathrm{det}\left(\frac{\partial\mu^{i}}{\partial x^{a}}\right)
\end{equation}
where
\begin{equation}\label{eq:a31}
g_{ij} \left(\mu (\mathbf{x})\right) \frac{\partial\mu^{i}}{\partial x^{a}}\; \frac{\partial\mu^{j}}{\partial x^{b}} = \delta_{ab}
\end{equation}
is the usual transformation relating the Euclidean metric \(e\) in Sternberg coordinates,
\begin{equation}\label{eq:a32}
e = g_{ij} (\mathbf{y})\; dy^{i} \otimes dy^{j}
\end{equation}
to its Cartesian form
\begin{equation}\label{eq:a33}
e = \delta_{ab}\; dx^{a} \otimes dx^{b}.
\end{equation}
A straightforward calculation results in
\begin{equation}\label{eq:a34}
\begin{split}
-\frac{d}{dt} \ln{\left(\sqrt{\mathrm{det}g_{**}}\right)} &= \frac{d}{dt}\; \ln{\mathrm{det}\left(\frac{\partial\mu^{i}}{\partial x^{a}}\right)}\\
 &= -\frac{1}{m} {}^{(n)}\!\Delta S_{(0)} + \sum_{i = 1}^{n} \omega_{i}.
\end{split}
\end{equation}

\end{document}